\documentclass[a4paper, 12pt]{article}

\usepackage[utf8]{inputenc}
\usepackage{amssymb,amsmath,amsfonts,bbm,eurosym,geometry,ulem,graphicx,caption,subcaption,color,setspace,sectsty,comment,footmisc,caption,pdflscape,array,adjustbox,ragged2e,booktabs}
\usepackage[T1]{fontenc}    
\usepackage{mathpazo}  

\usepackage[capposition=top]{floatrow}
\usepackage{hyperref}
\hypersetup{colorlinks,linkcolor={blue},citecolor={blue},urlcolor={blue}}

\DeclareCaptionLabelFormat{bold}{\textbf{(#2)}}
\usepackage[subrefformat=bold]{subcaption}

\makeatletter
\def\@mb@citenamelist{cite,citep,citet,citealp,citealt,citepalias,citetalias,nocite}
\makeatother
\usepackage{tabularx} 	
\usepackage{threeparttable}
\usepackage{longtable}
\usepackage{threeparttablex}
\usepackage{dcolumn}
\usepackage{array}
\usepackage{tabularray}
\UseTblrLibrary{booktabs}
\usepackage{afterpage}
\usepackage{csvsimple}
\usepackage{pgf}
\usepackage{siunitx}
\usepackage{makecell}
\newcommand{\CI}[1]{\SI[round-precision=0]{#1}{\percent} CIs}

\newcolumntype{L}[1]{>{\raggedright\let\newline\\arraybackslash\hspace{0pt}}m{#1}}
\newcolumntype{C}[1]{>{\centering\let\newline\\arraybackslash\hspace{0pt}}m{#1}}
\newcolumntype{R}[1]{>{\raggedleft\let\newline\\arraybackslash\hspace{0pt}}m{#1}}  
\newlength{\defaultcolsep}
\usepackage{array} 
\newcolumntype{H}{>{\setbox0=\hbox\bgroup}c<{\egroup}@{}}

\usepackage{authblk}  
\usepackage{changepage}  

\PassOptionsToPackage{
  style=numeric-comp,   
  sorting=none,         
  backend=bibtex,
  natbib,
  giveninits=true,
  autocite=superscript, 
  maxbibnames=99,
  useprefix=true
}{biblatex}
\usepackage{biblatex}

\let\cite=\autocite
\let\citep=\autocite
\let\citet=\autocite

\AtEveryBibitem{%
	\clearlist{language}  
	\clearfield{month}    
	\clearfield{day}      
}

\renewbibmacro{in:}{}                            
\DeclareFieldFormat{pages}{#1}                  
\DeclareFieldFormat[report]{title}{\mkbibquote{#1}}  

\renewenvironment{abstract}
 {\small
  \begin{center}
  \bfseries \abstractname\vspace{-.5em}\vspace{0pt}
  \end{center}
  \list{}{
    \setlength{\leftmargin}{1.3cm}%
    \setlength{\rightmargin}{\leftmargin}%
  }%
  \item\relax}
 {\endlist}

\newcommand{\version}{02-NHB-Andrea}

\def\finalNindTrioPld{12871}

\def\finalNindTrioGtpPld{4586}

\def\finalNindPld{51056}

\def\finalNindTrioSec{5063}

\def\finalNindSec{18692}

\def\finalNindTrioTer{7808}

\def\finalNindTrioGtpTer{2704}

\def\finalNindTer{32364}

\def\MeanAgeTFifteenTer{39.72839632801682}
\def\RsqIncrementalYEduPredictedPGIEA{.0708784409994058}
\def\RegThetaPGIEANPld{31866}

\def\medianDPVPld{294962.71875}
\def\predMargDPVEstpTenPld{309659.3341678228}
\def\predMargDPVEstpNinetyPld{350417.8582156428}
\def\predMargDPVLowpTenPld{307099.2776980281}
\def\predMargDPVUpppTenPld{312219.3906376176}
\def\predMargDPVLowpNinetyPld{347299.8489790345}
\def\predMargDPVUpppNinetyPld{353535.8674522511}
\def\predMargDPVNPld{51056}
\def\predMargDPVEstGapPld{40758.52404781996}

\def\predMargIncTTenEstMeanPld{24931.51889833254}

\def\predMargIncTTenEstGapPld{3262.058719545843}
\def\predMargIncTTenLowGapPld{2740.843689871333}
\def\predMargIncTTenUppGapPld{3783.273749220353}

\def\predMargIncTTwentyFiveEstMeanPld{34858.0508737832}

\def\predMargIncTTwentyFiveEstGapPld{7534.262284419867}
\def\predMargIncTTwentyFiveLowGapPld{6381.648963374741}
\def\predMargIncTTwentyFiveUppGapPld{8686.875605464993}

\def\predMargCCINPld{50963}

\def\predMargDPVNSec{18692}

\def\predMargDPVFamNSec{5063}

\def\predMargEmpSpellNSec{17432}

\def\predMargCCINSec{18669}
\def\predMargPsiNSec{15579}
\def\medianDPVTer{333648.671875}
\def\predMargDPVEstpTenTer{346193.9322398672}

\def\predMargDPVNTer{32364}
\def\predMargDPVEstGapTer{45391.53919397341}

\def\predMargDPVFamNTer{7808}

\def\predMargDPVParsEstGapTer{12990.95995613164}

\def\predMargDPVParsEstpTenFTer{330699.1068861979}
\def\predMargDPVParsEstpNinetyFTer{360945.700214089}
\def\predMargDPVParsLowpTenFTer{321958.4726406912}
\def\predMargDPVParsUpppTenFTer{339439.7411317046}
\def\predMargDPVParsLowpNinetyFTer{350897.3673085127}
\def\predMargDPVParsUpppNinetyFTer{370994.0331196653}

\def\predMargEmpSpellEstpTenTer{4.277913622364768}
\def\predMargEmpSpellEstpNinetyTer{4.450278640983013}

\def\predMargEmpSpellNTer{31015}

\def\predMargEmpSpellPvalGapTer{.0071058401134779}

\def\predMargCCINTer{32294}
\def\predMargPsiNTer{22733}
\def\RegYEduPGIEAEstSlopePld{.2825883301317226}

\def\RegYEduPGIEAParsEstSlopePld{.212030341600102}

\def\maxTime{25}

\def\finalNobsAll{963715}

\def\finalNindAll{51056}
\def\finalNindThl{27135}
\def\finalNindBdb{23921}
\def\AKMnobsPTone{16586748}
\def\AKMnobsPTtwo{15060995}
\def\AKMnindFePTone{1881715}
\def\AKMnindFePTtwo{1842564}
\def\AKMnfirmFePTone{126605}
\def\AKMnfirmFePTtwo{50430}
\def\RegThetaPGIEAEstSlopeSec{0.017}
\def\RegThetaPGIEAEstSlopeTer{0.115}

\def\RegThetaPGIEALowSlopeSec{0.00328025210821962}
\def\RegThetaPGIEALowSlopeTer{0.10128025210822}

\def\RegThetaPGIEAUppSlopeSec{0.0307197478917804}
\def\RegThetaPGIEAUppSlopeTer{0.12871974789178}

\def\RegDPVPGIEAParsEstSlopeTer{5047.80770576166}

\def\RegDPVPGIEAParsLowSlopeTer{-2465.5825264482}
\def\RegDPVPGIEAParsUppSlopeTer{12561.1979379715}

\def\RegDPVPGIEAParsEstSlopeFTer{12112.9072308129}

\def\RegDPVPGIEAParsLowSlopeFTer{5508.66742474824}
\def\RegDPVPGIEAParsUppSlopeFTer{18717.1470368775}

\newcommand{\SentenceBlueRed}{The blue line corresponds to 10th and the red - to 90th percentile of EA-PGI distribution. }
\newcommand{\SentenceIncomeReg}{Average income estimated from a regression of annual income on EA-PGI fully interacted with indicators measuring years since graduation and controlling for first ten genetic principal components, gender, year of birth, calendar year, and biobank indicators. }
\newcommand{\SentenceIncomeRegFamp}{Average income estimated from a regression of annual income on the set of own, maternal and paternal EA-PGI, all fully interacted with indicators measuring years since graduation, and controlling for first ten genetic principal components, gender, year of birth, calendar year, and biobank indicators. }
\newcommand{\SentenceHealthReg}{Average health index estimated from a regression of Charlson Comorbidity Index on EA-PGI fully interacted with indicators measuring years since graduation and controlling for first ten genetic principal components, gender, year of birth, calendar year, and biobank indicators. }
\newcommand{\SentenceHealthRegFamp}{Average health index estimated from a regression of Charlson Comorbidity Index on the set of own, maternal and paternal EA-PGI, all fully interacted with indicators measuring years since graduation, and controlling for first ten genetic principal components, gender, year of birth, calendar year, and biobank indicators. }
\newcommand{\SentenceDPVReg}{Average lifetime income adjusted by regressing cumulated income on EA-PGI and controlling for first ten genetic principal components, gender, year of birth, calendar year, and biobank indicators. }
\newcommand{\SentenceDPVRegFamp}{Average lifetime income adjusted by regressing cumulated income on the set of own, maternal and paternal EA-PGI and controlling for first ten genetic principal components, gender, year of birth, calendar year, and biobank indicators. }
\newcommand{\SentenceDPVMethod}{Income discounted to obtain its present value upon graduation (see Section \ref{sec:measurement} for additional information). }
\newcommand{\SentenceCI}[1]{The shaded areas correspond to \CI{#1}. }
\newcommand{\SentenceSE}{Standard errors reported in parentheses. }

\begin{document}


\begin{titlepage}

\title{\vspace{-35pt}
Effects of Genetic Propensity for Education on Labor Market and Health Trajectories across the Working Life%
\thanks{
Corresponding authors: Andrea Ganna (\href{mailto:andrea.ganna@helsinki.fi}{andrea.ganna@helsinki.fi}), Stefano Lombardi (\href{mailto:stefano.lombardi@vatt.fi}{stefano.lombardi@vatt.fi}).
}
}

\renewcommand\Authfont{\normalsize}
\renewcommand\Affilfont{\footnotesize}  
\setlength{\affilsep}{1em} 

\author[1,2,3]{Stefano Lombardi\thanks{
Co-first authors.}
}
\author[4]{Nurfatima Jandarova\textsuperscript{† }}
\author[5]{Kristina Zguro\textsuperscript{† }} 
\author[4]{Jarkko Harju}
\author[6]{Aldo Rustichini}
\author[5,7,8]{Andrea Ganna}

{\footnotesize
\affil[1]{VATT Institute for Economic Research, Helsinki, Finland, 00100}
\affil[2]{IFAU and Uppsala Center for Labor Studies, Uppsala, Sweden, 75105}
\affil[3]{IZA, Bonn, Germany, 53113, and Rockwool Foundation, Berlin, Germany, 10119}
\affil[4]{Economics department at Tampere University, and FIT, Tampere, Finland, 33100}
\affil[5]{Institute for Molecular Medicine Finland (FIMM), HiLIFE, University of Helsinki, Finland, 00290}
\affil[6]{Economics department at University of Minnesota, MN, USA, 55455}
\affil[7]{Broad Institute of MIT and Harvard, MA, USA, 02142}
\affil[8]{Analytic and Translational Genetics Unit, Massachusetts General Hospital, MA, USA, 02114}
}

\date{}

\maketitle

\vspace{-40pt}
\singlespacing

\begin{abstract}
\begin{adjustwidth}{-1cm}{-1cm}%
\singlespacing
\vspace{-20pt}
\noindent 
Education is a major source of inequality in income and health. Polygenic indices for educational attainment (EA-PGI) capture both direct and indirect genetic influences on education, but their effects on income and health remain unclear.
Using Finnish registry data on \num[round-precision=0]{\finalNindAll} graduates followed annually since graduation for up to \num[round-precision=0]{\maxTime} years, we report three findings.
First, higher EA-PGI strongly predicts income growth, but only among higher-educated people: tertiary-educated graduates at the 90th percentile earn €\num[round-precision=0]{\predMargDPVEstGapTer} (\SI{\fpeval{\predMargDPVEstGapTer / \predMargDPVEstpTenTer * 100}}{\percent}) higher discounted lifetime income than those at the 10th percentile. This effect is not mediated by overall health and is entirely absent for the secondary (high school)-educated workers, who do not beneift from higher EA-PGI levels.
Second, EA-PGI does not predict income differences at labor market entry or the quality of the first employer, but rather a higher job-to-job mobility toward higher-quality firms that drives the long-run income divergence.
Third, controlling for parental EA-PGI in \num[round-precision=0]{\finalNindTrioPld} parent–offspring trios reduces the discounted lifetime income gap by \SI[round-precision=0]{\fpeval{100 - \predMargDPVParsEstGapTer / \predMargDPVEstGapTer * 100}}{\percent}, and the effect of paternal (but not maternal) EA-PGI on offspring income exceeds that of the offspring’s own EA-PGI. These findings suggest that genetic factors associated with educational attainment predict income trajectories primarily through faster and more frequent changes to higher-paying employers. However, much of this association reflects indirect paternal genetic effects, consistent with enduring paternal patterns of intergenerational job and income transmission.
\end{adjustwidth}
\end{abstract}

\noindent Classification: Social Sciences, Economics; Biological Sciences, Genetics

\bigskip 


\bigskip

\end{titlepage}


\pagebreak \newpage
\noindent
\onehalfspacing
\clearpage


\setcounter{page}{2}  
\section*{Significance Statement}\label{sec:significance}

Genetic differences are increasingly studied as contributors to socioeconomic inequality, yet little is known about how their influence unfolds over the working life. Using Finnish administrative records linked to molecular genetic data, we follow individuals’ earnings trajectories from labor market entry over 25 years. Genetic predispositions are unrelated to income at entry, but substantial differences emerge later in the career. These gaps are concentrated among university graduates and arise primarily through differential sorting into higher-paying firms. Parental—especially paternal—genetic endowments also account for a meaningful share of these differences. Our findings provide new evidence on how genetic and social processes interact over the life course to shape income inequality.


\newpage


\section*{Introduction}\label{sec:introduction}

Understanding the origins and persistence of inequalities in health, education, and income—collectively referred to as socioeconomic status—is central to economic and social policy. While individual effort and choices contribute to these disparities, socioeconomic outcomes are also shaped by circumstances beyond individual control, including family background and genetic predispositions. The extent to which such circumstances influence life outcomes has important implications for equality of opportunity in society \citep{roemer_equality_2016}.

Social scientists and epidemiologists have long sought to identify the drivers of socioeconomic inequalities\citep{ashenfelter1994estimates, chetty2014land, chetty2016association, bundy2023social}, recognizing that genetic and environmental factors interact over the life cycle to shape individual outcomes\citep{cunha_technology_2007, biroli2025economics}.
The recent development of polygenic indices (PGIs), which summarize genetic predispositions for complex traits, has enabled researchers to directly examine how genetic endowments relate to education and other dimensions of socioeconomic status \citep{harden_using_2020, abdellaoui202315, barth_genetic_2020, papageorge2020genes}.

Recent advances in this literature have focused on distinguishing direct genetic effects—arising from genetic variants randomly transmitted at conception—from indirect genetic influences, such as environmentally mediated effects of parental genotypes (“genetic nurture”) \citep{kong2018nature, okbay2022, tan2024family}. Identifying direct genetic effects makes it possible to study how environmental processes amplify or mitigate genetic predispositions and to investigate the mechanisms through which genetic differences translate into socioeconomic outcomes such as income \citep{rustichini2023educational, carvalho2025genetics}.

Despite the advances in this growing literature, evidence on how genetic predispositions shape income over the working life remains limited. The existing studies that use within-family designs provide static estimates that do not capture how income differences accumulate over the life cycle. They also do not examine potentially relevant mediating channels such as employer sorting, and often rely on coarse, self-reported measures of income and health. Complementary work has examined outcomes such as occupational prestige but without directly analyzing income\citep{akimova2025polygenic}.

This paper examines when and through which mechanisms genetic predispositions translate into income differences over the working life. Using Finnish administrative registers on income, health, and education, we follow individuals’ labor income for 25 years after labor market entry. The data tracks workers in and out of employment and their employers over time, enabling us to analyze how individuals’ own and parental genetic endowments shape income trajectories and the roles of education, firm quality, and health in generating them. We find that genetic predispositions for education are unrelated to income at labor market entry, but substantial differences emerge and widen over the career. These gaps are concentrated among university graduates and arise through differential sorting into higher-paying firms.

The analysis combines polygenic indices for educational attainment (EA-PGIs) with matched employer–employee registers. Linking workers to their employers over their careers allows us to apply the Abowd–Kramarz–Margolis (AKM) framework to estimate worker productivity and firm wage premia \citet{abowd_high_1999, kline_firm_2025}.
The worker productivity index captures worker-specific earnings capacity that is portable across employers, while firm pay premia measure systematic differences in firm wage policies after accounting for worker productivity. This setting allows us to shed light on the labor market processes generating unequal income trajectories across genetic groups: differences in worker productivity, mobility to higher-paying firms, and wage growth within and across firms.

Our empirical strategy exploits the random transmission of genetic variants at conception and implements a within-family design in which outcomes are regressed on the EA-PGI of the individual and those of both parents. This approach addresses population stratification and other forms of genetic confounding, allowing us to estimate direct genetic effects on income. Because the labor market returns to skills differ substantially by education level, the effects of genetic predispositions on income may vary across educational groups. We therefore present results separately for individuals with college and high school education.

Our analysis advances the literature in two ways. First, by studying income trajectories rather than static outcomes, we distinguish between two fundamentally different sources of long-run income inequality: differences present at labor market entry and differences arising from divergent income growth during the career. Similar income disparities later in life can emerge either from large initial gaps or from similar starting points followed by diverging trajectories. These two patterns correspond to very different mechanisms. Initial differences point to inequalities generated before labor market entry, for example through education or early skill formation. Divergent trajectories instead reflect processes operating during the career, such as job mobility, promotions, or health dynamics. Distinguishing between these mechanisms is therefore crucial for understanding how genetic predispositions contribute to the production and accumulation of income inequality over the life cycle.

Second, within this dynamic framework, we analyze a key channel through which genetics may shape income trajectories: job mobility and sorting into firms with different wage-setting quality. Firms play an important role in explaining wage inequality among workers with similar observable skills \citep{card2013workplace, kline2024firm}. Estimating firm pay premia—capturing how much firms pay relative to other firms after controlling for worker productivity—allows us to move beyond documenting income gaps and identify the labor market mechanisms linking genetic predispositions to income trajectories.


\section*{Results}

\subsection*{\normalsize EA-PGI predicts lifetime income trajectories, but only among individuals with tertiary education}

The study includes \num[round-precision=0]{\finalNindAll} individuals with genome-wide genetic data linked to longitudinal health and socioeconomic information. The data covers employment histories (i.e., employee-employer links with job spells length), annual income from labor, and educational records (education level, field and school/university identifiers). 
This information spans thirty years (1987--2019) and all individuals are followed from graduation up to \num[round-precision=0]{\maxTime} years later. We estimate a dynamic model to analyze the relation between EA-PGI and individual income over time, adjusting for calendar year, birth year, gender, the first ten genetic principal components, and biobank indicator. 

Individuals included in this study are generally better educated and with a higher share of women compared to the nationwide population of fresh graduates in the same calendar period (Supplementary Table \ref{subtab:descriptives-genotyped}). Reassuringly, when we reweight the sample to reflect the characteristics in the population, the results are virtually unaffected (Supplementary Table  \ref{tab:predmarg-dpvinc-pgiea-weighted}).

\begin{figure}[htbp!]
    \centering
    \begin{subfigure}[b]{0.70\linewidth}
        \caption{Pooled}
        \includegraphics[width = \linewidth]{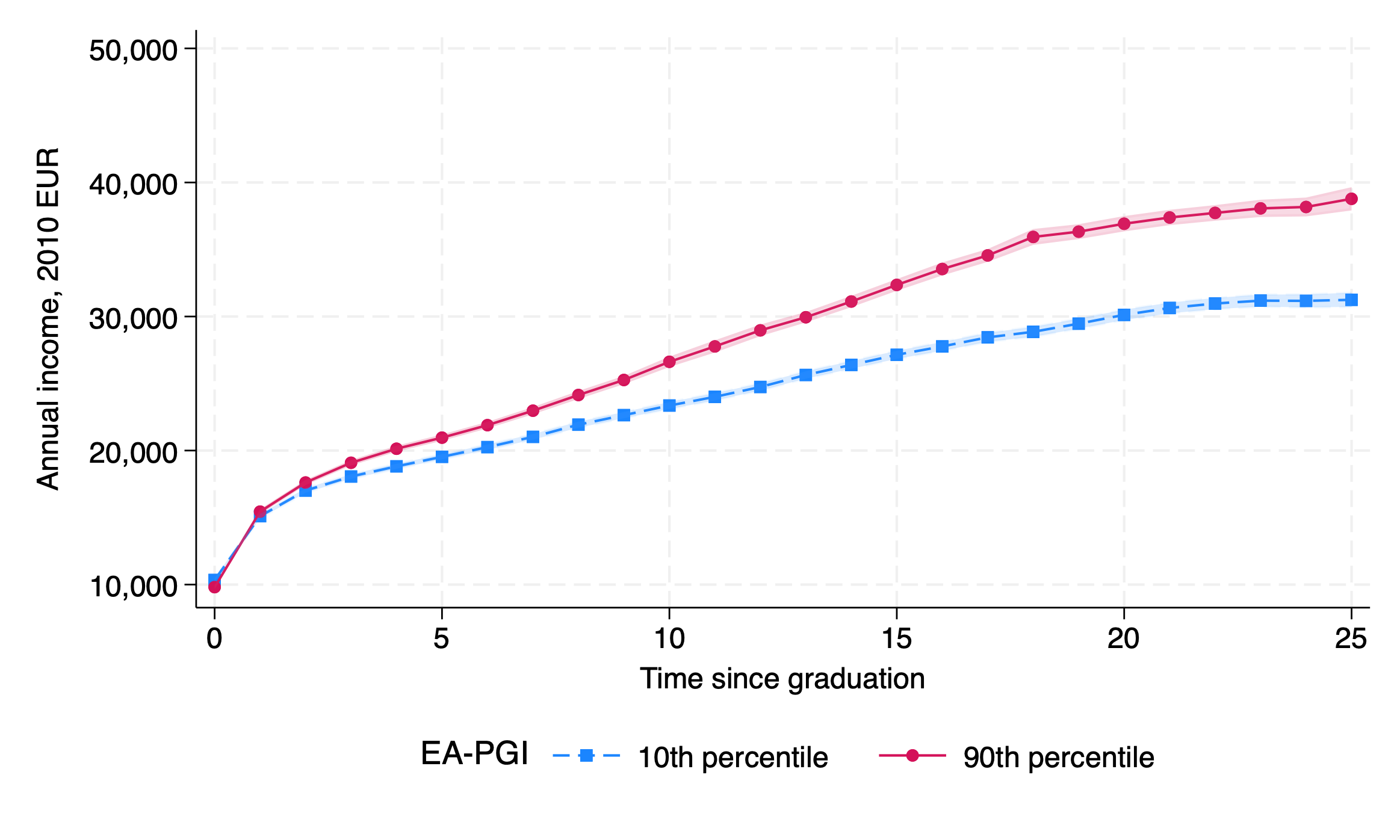}
        \label{subfig:predmarg-inc-pgiea-time-pld}
    \end{subfigure}%
    
    \begin{subfigure}[b]{0.49\linewidth}
        \caption{Secondary}
        \includegraphics[width = \linewidth]{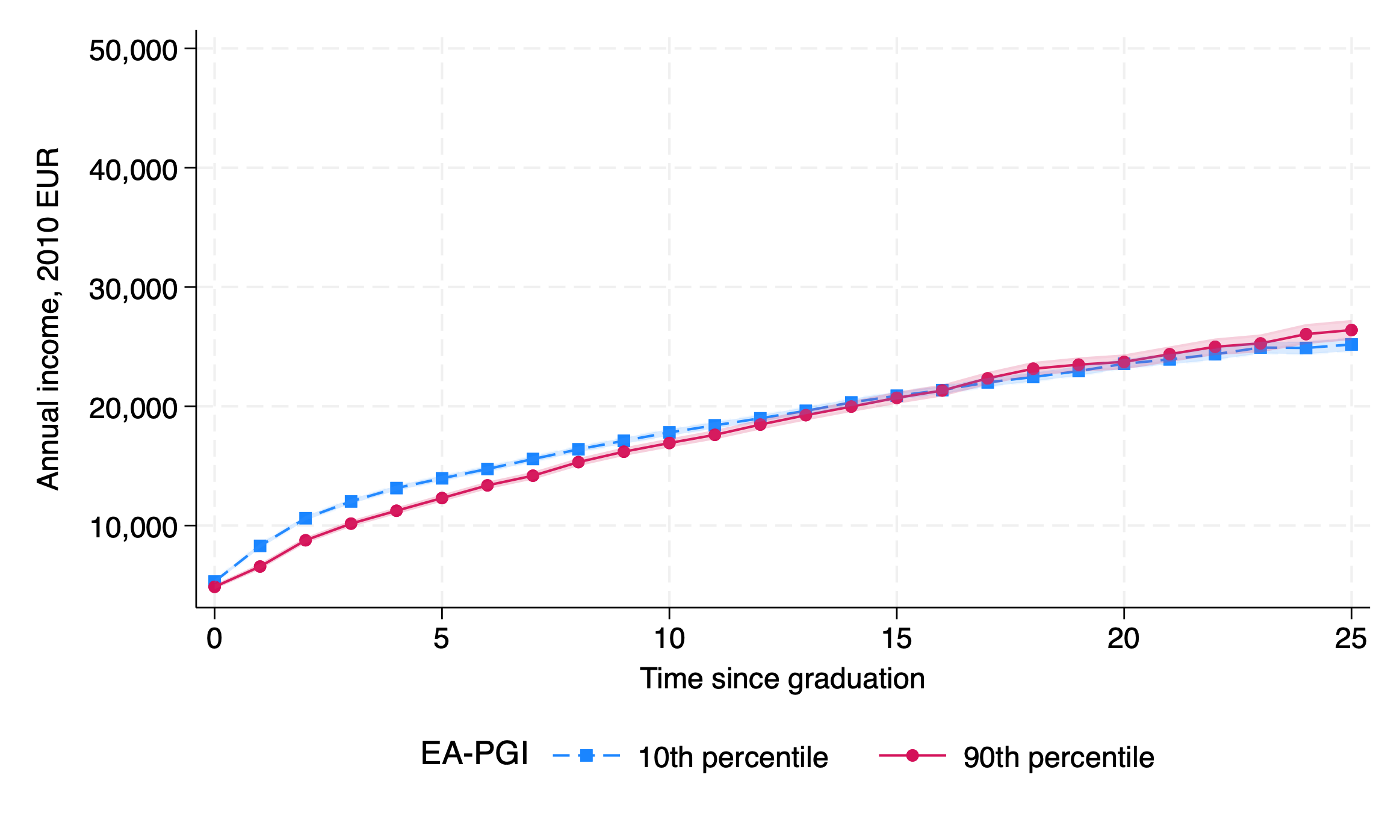}
        \label{subfig:predmarg-inc-pgiea-time-sec}
    \end{subfigure}%
    \begin{subfigure}[b]{0.49\linewidth}
        \caption{Tertiary}
        \includegraphics[width = \linewidth]{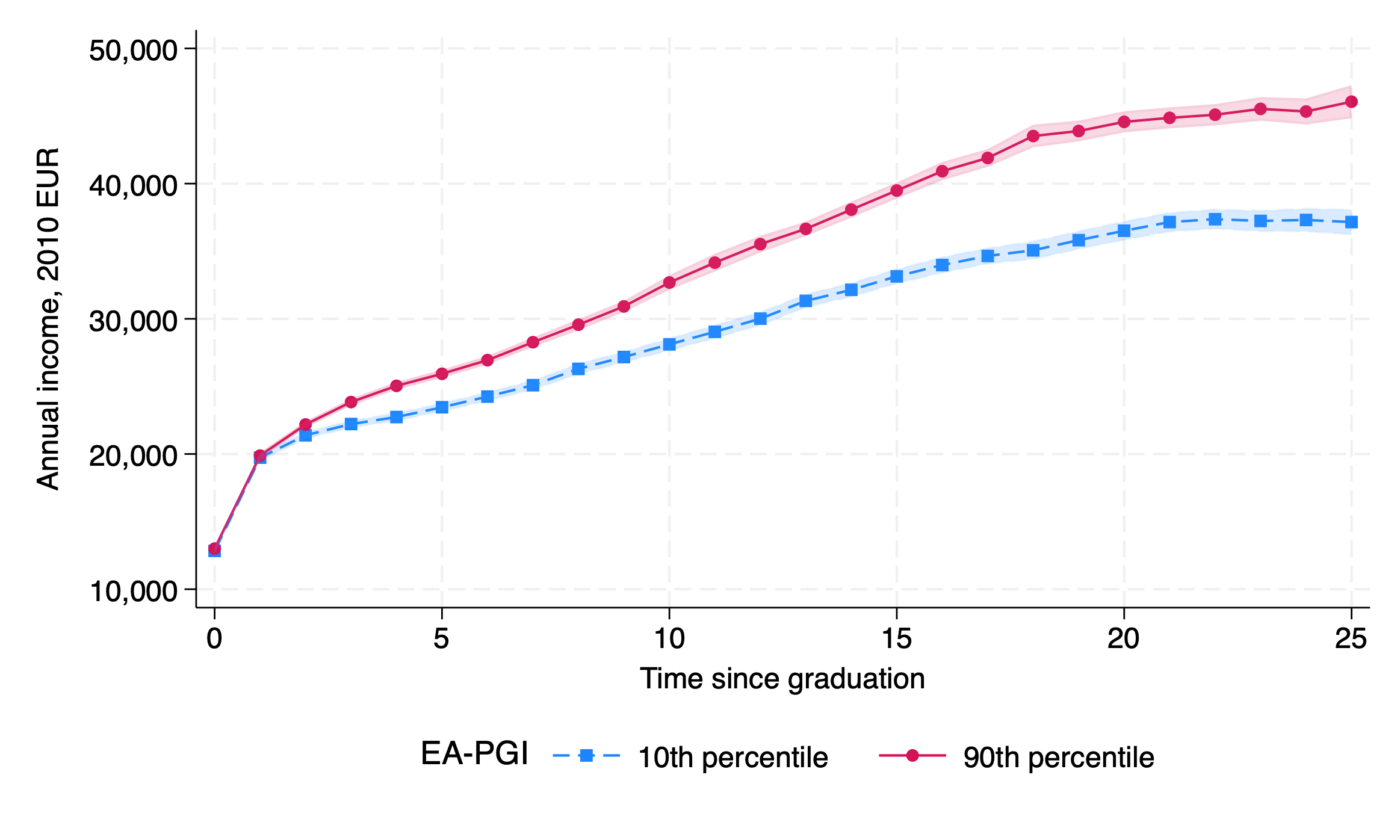}
        \label{subfig:predmarg-inc-pgiea-time-ter}
    \end{subfigure}
    \caption{\textbf{Average annual income by EA-PGI level, over time and by education}\\\footnotesize Panel \subref{subfig:predmarg-inc-pgiea-time-pld} uses full analysis sample ($N=$ \num[round-precision=0]{\predMargDPVNPld}), while Panels \subref{subfig:predmarg-inc-pgiea-time-sec} and \subref{subfig:predmarg-inc-pgiea-time-ter} use subset of workers based on their highest qualification being either secondary ($N=$ \num[round-precision=0]{\predMargDPVNSec}) or tertiary degree ($N=$ \num[round-precision=0]{\predMargDPVNTer}), respectively. \SentenceBlueRed\SentenceIncomeReg\SentenceCI{95}}
    \label{fig:predmarg-inc-pgiea-time-byedu}
\end{figure}

We first confirm that EA-PGI is significantly associated with years of educational attainment, with results comparable to what previously reported in the literature (R2=\SI{\fpeval{\RsqIncrementalYEduPredictedPGIEA * 100}}{\percent},  Supplementary Table \ref{tab:predicted-ea-pgi-increm-r2}) \citep{wang2021robust}.
Next, we provide novel estimates on income over the life cycle. Comparing earned income trajectories between the 10th and 90th percentile of the EA-PGI distribution over \num[round-precision=0]{\maxTime} years since graduation, we show that income levels are initially very similar across groups (Figure \ref{subfig:predmarg-inc-pgiea-time-pld}). However, trajectories begin to diverge substantially over time. 
The gap in average annual income between 90th and 10th percentile of EA-PGI 10 years from graduation is €\num[round-precision=0]{\predMargIncTTenEstGapPld} [CI €\num[round-precision=0]{\predMargIncTTenLowGapPld} - €\num[round-precision=0]{\predMargIncTTenUppGapPld}] and widens further to €\num[round-precision=0]{\predMargIncTTwentyFiveEstGapPld} [CI €\num[round-precision=0]{\predMargIncTTwentyFiveLowGapPld} - €\num[round-precision=0]{\predMargIncTTwentyFiveUppGapPld}] by 25 years from graduation.
To put this in context, the average yearly earnings in the sample is €\num[round-precision=0]{\predMargIncTTenEstMeanPld} and €\num[round-precision=0]{\predMargIncTTwentyFiveEstMeanPld} at 10 and 25 years after graduation, respectively. 
The cumulated income over the 25 years after graduation, discounted to obtain its present value at graduation, is €\num[round-precision=0]{\predMargDPVEstpTenPld} [CI €\num[round-precision=0]{\predMargDPVLowpTenPld} - €\num[round-precision=0]{\predMargDPVUpppTenPld}] for individuals at the 10th percentile and €\num[round-precision=0]{\predMargDPVEstpNinetyPld} [CI €\num[round-precision=0]{\predMargDPVLowpNinetyPld} - €\num[round-precision=0]{\predMargDPVUpppNinetyPld}] for those at the 90th percentile of the EA-PGI distribution (Table \ref{tab:predmarg-dpvinc-pgiea}).
The gap in cumulated income between the 10th and 90th percentile is €\num[round-precision=0]{\predMargDPVEstGapPld}, equivalent to \SI{\fpeval{\predMargDPVEstGapTer / \predMargDPVEstpTenTer * 100}}{\percent} and corresponding to \SI[round-precision=0]{\fpeval{\predMargDPVEstGapPld / \medianDPVPld * 100}}{\percent} of the median income over the same period. 

Strikingly, this genetic gradient in income varies markedly by educational attainment. The income gap by genetics is driven entirely by individuals with tertiary education (Figures \ref{subfig:predmarg-inc-pgiea-time-sec} and \ref{subfig:predmarg-inc-pgiea-time-ter}), while no systematic differences are observed among the people with secondary education (\SI{\fpeval{100 * \finalNindTer / \finalNindPld}}{\percent} of the graduates has tertiary degree). Among the tertiary-educated, differences in income trajectories widen up during the first 15 years after graduation and then stabilize, at which point individuals are, on average, \num[round-precision=0]{\MeanAgeTFifteenTer} years old and  approaching peak labor market attachment.

Such a differential result likely reflects two forces: individuals with relatively high EA-PGI experience higher economic returns to ability, and EA-PGI influences educational sorting, shaping the types and levels of education they pursue. Although we cannot fully separate sorting from the economic returns to higher EA-PGI, we next analyze how EA-PGI correlates with a measure of ability and examine the role of employers in explaining the observed income differences.

\begin{table}[htbp!]
    \centering
    \caption{\textbf{Cumulated lifetime income by EA-PGI level}\\\footnotesize The table reports adjusted lifetime income (up to \num[round-precision=0]{\maxTime} years since graduation) by EA-PGI percentiles. Column (1) uses full analysis sample ($N =$ \num[round-precision=0]{\predMargDPVNPld}), while columns (2) and (3) use subset of workers based on their highest qualification being either secondary ($N =$ \num[round-precision=0]{\predMargDPVNSec}) or tertiary degree ($N =$ \num[round-precision=0]{\predMargDPVNTer}), respectively. \SentenceDPVReg\SentenceDPVMethod\SentenceSE}
    \label{tab:predmarg-dpvinc-pgiea}
    \begin{threeparttable}
        \begin{tabular}{llll}\toprule
& \multicolumn{3}{c}{Dependent variable: Cumulated income} \\\cmidrule(lr){2-4}
& \multicolumn{1}{c}{Pooled} & \multicolumn{1}{c}{Secondary} & \multicolumn{1}{c}{Tertiary} \\
& \multicolumn{1}{c}{(1)} & \multicolumn{1}{c}{(2)} & \multicolumn{1}{c}{(3)} \\ \midrule
\multicolumn{4}{l}{EA-PGI percentiles} \\
\hspace{1em}10th & 309 659 & 262 386 & 346 194 \\
& (1 306) & (1 429) & (1 944) \\
\hspace{1em}20th & 316 476 & 260 040 & 353 786 \\
& (1 037) & (1 153) & (1 539) \\
\hspace{1em}30th & 321 505 & 258 309 & 359 386 \\
& (  898) & (1 046) & (1 304) \\
\hspace{1em}40th & 325 804 & 256 829 & 364 174 \\
& (  844) & (1 042) & (1 175) \\
\hspace{1em}50th & 329 893 & 255 422 & 368 728 \\
& (  857) & (1 116) & (1 137) \\
\hspace{1em}60th & 333 908 & 254 041 & 373 198 \\
& (  930) & (1 248) & (1 188) \\
\hspace{1em}70th & 338 222 & 252 556 & 378 004 \\
& (1 061) & (1 439) & (1 328) \\
\hspace{1em}80th & 343 389 & 250 777 & 383 758 \\
& (1 265) & (1 709) & (1 579) \\
\hspace{1em}90th & 350 418 & 248 358 & 391 585 \\
& (1 591) & (2 120) & (2 006) \\
\midrule Obs. & 51 056 & 18 692 & 32 364 \\\bottomrule
\end{tabular}

    \end{threeparttable}
\end{table}

\subsection*{\normalsize EA-PGI is associated with the tertiary-educated workers' productivity
}

To understand the mechanisms through which the EA-PGI influences income trajectories, we first examine its relationship with both employee and employer productivity in the labor market. To quantify productivity, we use full population data within an Abowd–Kramarz–Margolis (AKM) framework, the workhorse model in economics for decomposing wage variation into worker and firm components \citep{abowd_high_1999, kline2024firm}.

The model exploits repeated measurements of worker's wage, relating it to individual- and firm-specific indicators.
After model estimation, each worker is assigned a corresponding estimated intercept, which we refer to as \textit{worker productivity index}, as it captures the persistent component of the worker's wage that is portable when switching jobs across firms, net of the effect of firm-quality and time-varying worker's characteristics.

\begin{figure}[htbp!]
    \centering
    \includegraphics[width = \linewidth]{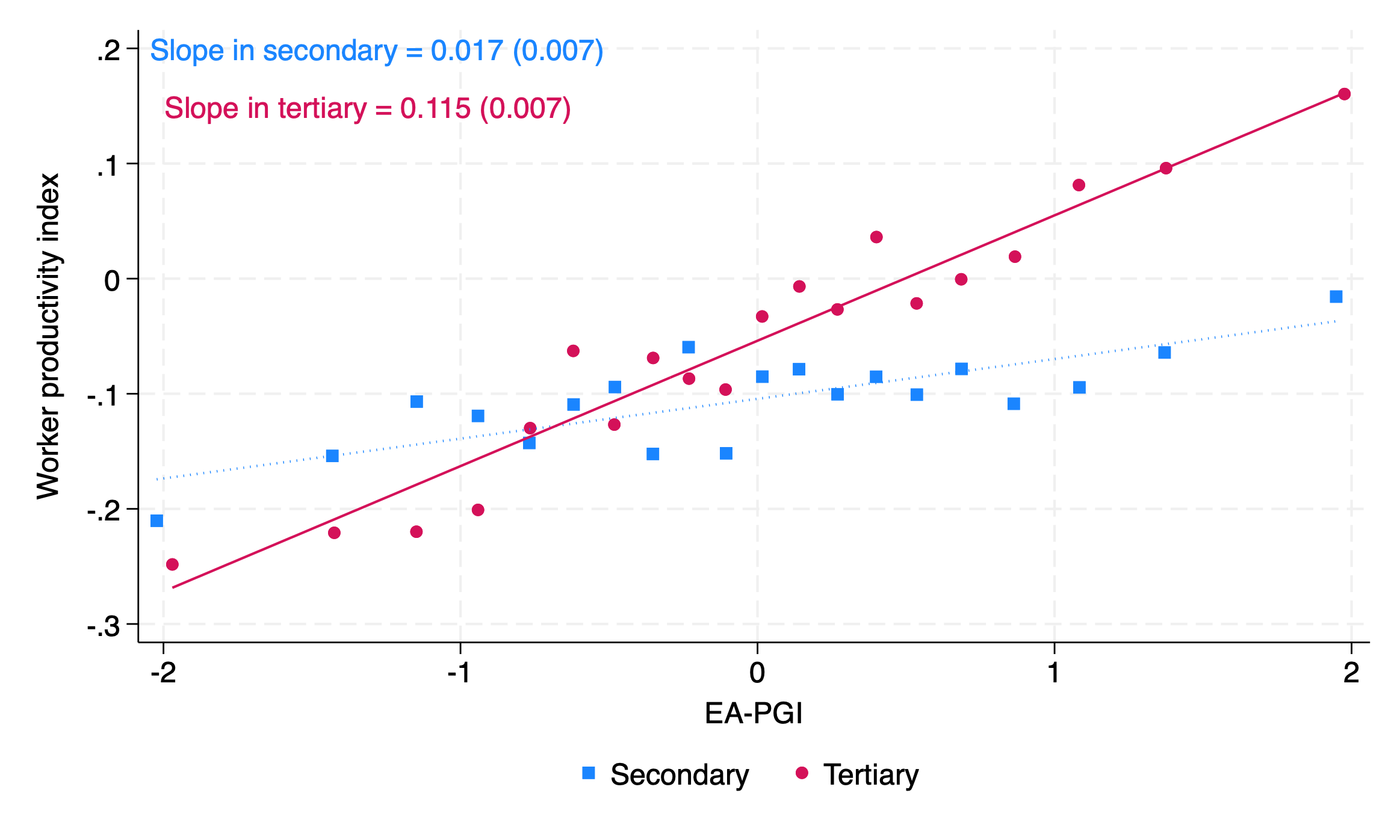}
    \caption{\textbf{Worker productivity index and EA-PGI level by education}\\\footnotesize The vertical axis reports a worker productivity index estimated via an AKM regression (details reported in Section \ref{sec:measurement}); the horizontal axis reports ventiles of the EA-PGI distribution. The binscatter plot shows the relation between these two quantities standardized to have mean 0 and standard deviation 1 and after having residualized them with respect to gender, year of birth indicators, and first 10 genetic principal components (PCs). The data used to obtain the plot is a cross-section of \num[round-precision=0]{\RegThetaPGIEANPld} individuals in the analysis sample with non-missing worker productivity index. The lines correspond to sub-samples based on highest education achieved. The figure also reports the estimated slopes (and standard errors in parentheses) from the corresponding linear regression of worker productivity on EA-PGI controlling for gender, year of birth, calendar year and biobank indicators, and first ten genetic PCs.}
    \label{fig:workerFE_pgsEA}
\end{figure}

We find a statistically significant correlation between the worker productivity index estimated via the AKM model and EA-PGI (both standardized to have mean 0 and standard deviation 1), but primarily for the tertiary-educated workers (Figure \ref{fig:workerFE_pgsEA}). 
For this group, a one standard deviation increase in EA-PGI is associated with a \num[round-precision=3]{\RegThetaPGIEAEstSlopeTer} [CI \num[round-precision=3]{\RegThetaPGIEALowSlopeTer} - \num[round-precision=3]{\RegThetaPGIEAUppSlopeTer}] standard deviation higher worker productivity, compared to a \num[round-precision=3]{\RegThetaPGIEAEstSlopeSec} [CI \num[round-precision=3]{\RegThetaPGIEALowSlopeSec} - \num[round-precision=3]{\RegThetaPGIEAUppSlopeSec}] increase among the secondary-educated individuals. 
Hence, individuals with high EA-PGI tend to be highly productive on the labor market (and therefore are on average paid higher wages), as long as they obtain a tertiary education degree. 

We further examine whether the correlation between EA-PGI and worker productivity persists across education groups after adjusting for fine-grained education-related characteristics (education field, and academic institution). Although the association between EA-PGI and worker productivity attenuates after adjustment (Supplementary Table \ref{tab:workerFE_pgiEA_reg}), it remains statistically significant among tertiary-educated workers and is significantly stronger than among those with secondary education.

\subsection*{\normalsize High EA-PGI individuals transition more rapidly and more frequently to higher-quality firms}

\begin{figure}[htbp!]
\centering
\caption{\textbf{Employer mobility and quality of tertiary-educated individuals by EA-PGI level, over time}\\\footnotesize Panel \subref{subfig:njobs-time} plots the average number of employment spells, and Panel \subref{subfig:akm-fe-time} - the average firm quality index over time since graduation. \SentenceBlueRed Both panels are estimated from a regression of respective outcome on EA-PGI fully interacted with indicators measuring years since graduation and controlling for first ten genetic principal components, gender, year of birth, calendar year, and biobank indicators. The estimation sample is restricted to tertiary-educated workers with non-missing firm identifiers ($N=$ \num[round-precision=0]{\predMargEmpSpellNTer}) in panel \subref{subfig:njobs-time} and non-missing firm quality index ($N=$ \num[round-precision=0]{\predMargPsiNTer}) in panel \subref{subfig:akm-fe-time}. \SentenceCI{95}}
\label{fig:akm-fe-njobs-time}
\begin{subfigure}{1\textwidth}
    \subcaption{\textbf{Cumulative number of jobs since graduation}}
    \centering
    \includegraphics[width = 0.8\linewidth]{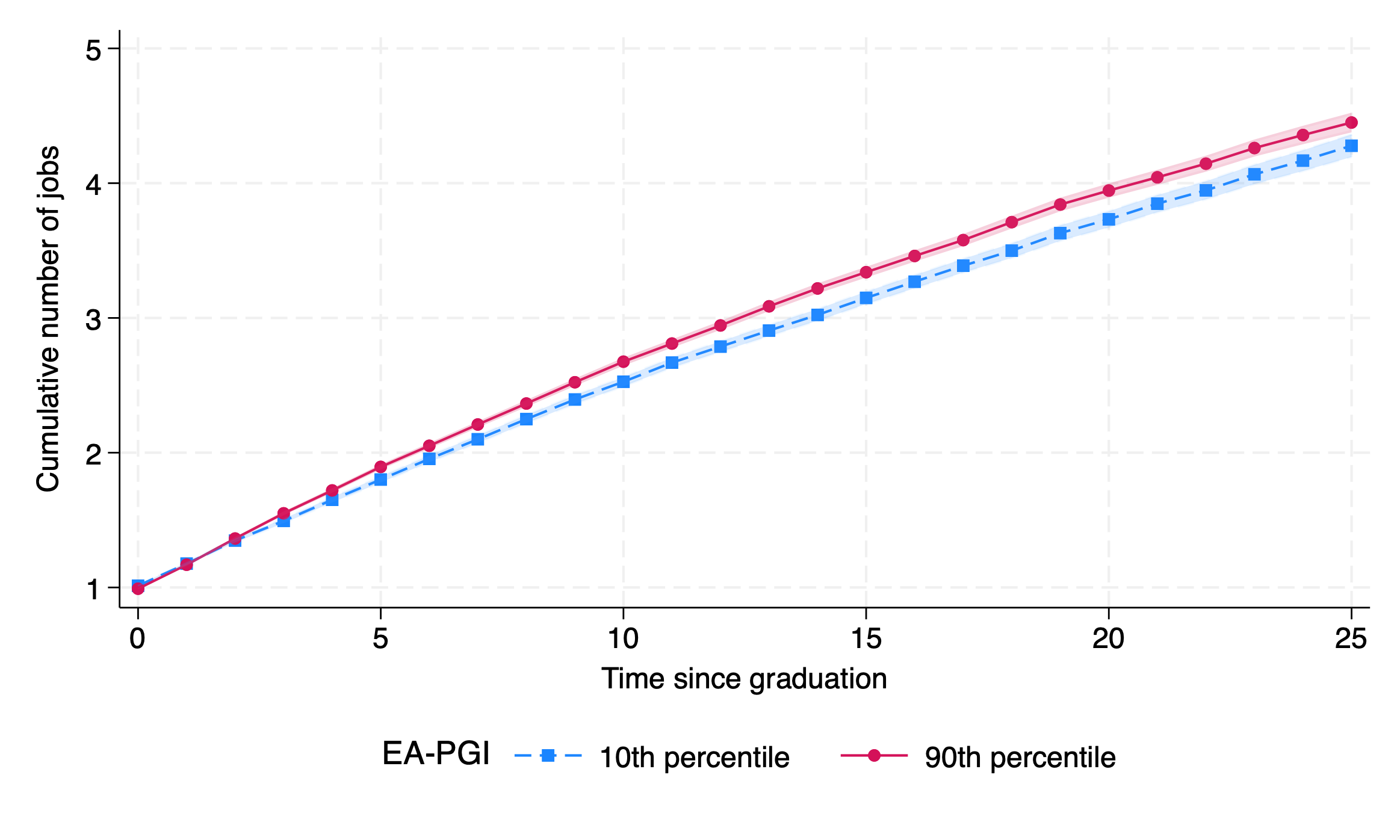}
    \label{subfig:njobs-time}
\end{subfigure}
\begin{subfigure}{1\textwidth}
    \centering
    \subcaption{\textbf{Firm quality since graduation}}
    \includegraphics[width = 0.8\linewidth]{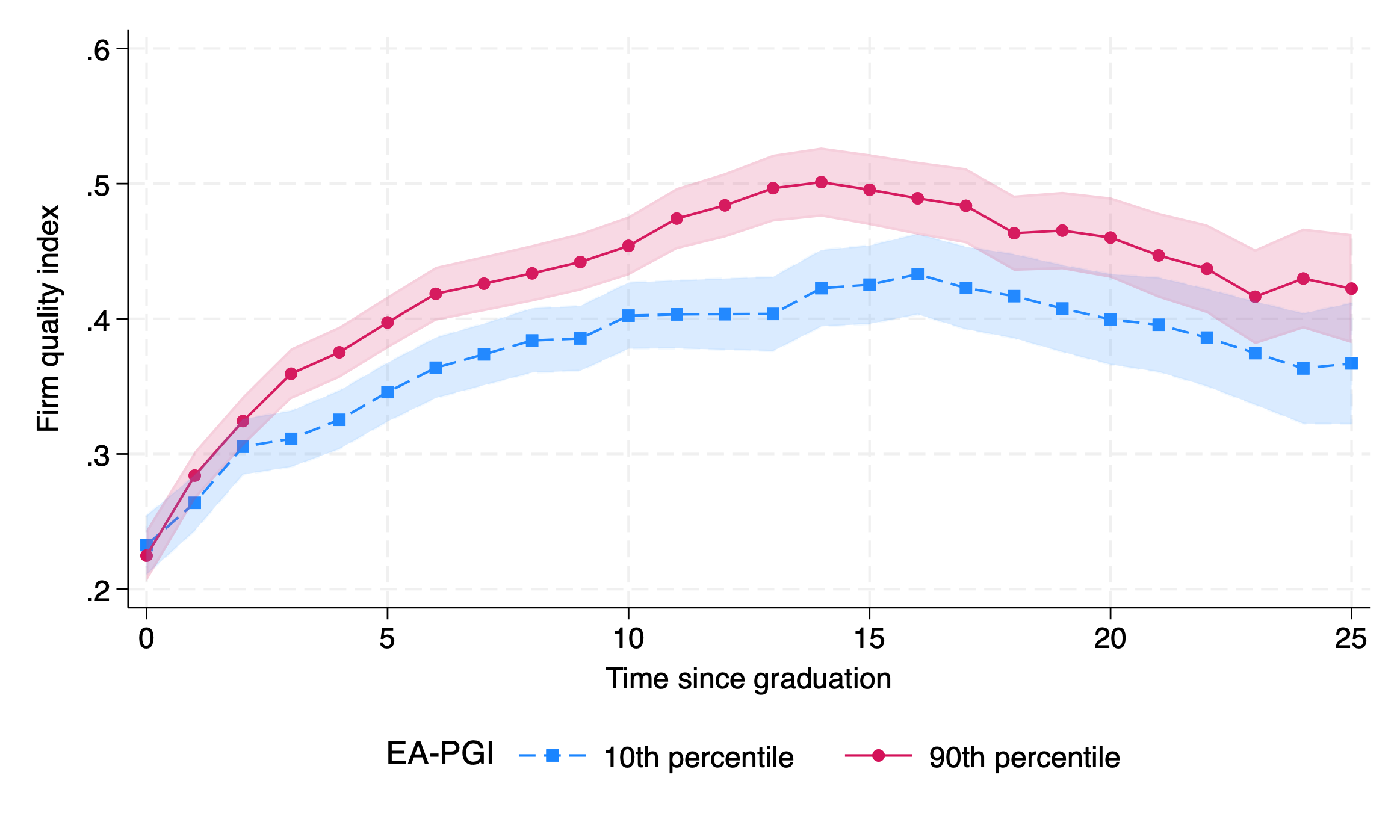}
    \label{subfig:akm-fe-time}
\end{subfigure}
\end{figure}

We next investigate the role of firms in shaping income differences across levels of EA-PGI among tertiary-educated individuals. 
Workers with higher EA-PGI change employers slightly more frequently (\num{\predMargEmpSpellEstpTenTer} jobs vs \num{\predMargEmpSpellEstpNinetyTer} jobs over \num[round-precision=0]{\maxTime} years, P-value=\num[round-precision=3]{\predMargEmpSpellPvalGapTer} for the 10th and 90th percentile EA-PGI, respectively; Figure \ref{subfig:njobs-time}). To assess the quality dimension of these moves, we use the AKM regression to estimate a \textit{firm quality index} which captures the firm-specific component of paid wage that is common to all workers at that firm, after accounting for worker ability and observed characteristics.
We use the firm quality index to rank all firms in which workers in the sample are employed, and use this as an outcome in our trajectories model. 

We show that, on average, higher EA-PGI individuals transition to significantly higher-quality firms (Figure \ref{subfig:akm-fe-time}). Interestingly, the quality of their first employer is not statistically different across the EA-PGI distribution, but the firm quality gap widens as early as three years after graduation. 
This pattern suggests that individuals do not initially self-select into higher- or lower-quality firms based on their EA-PGI, but sorting by EA-PGI appears to begin early in the career, with higher EA-PGI individuals progressively moving toward better-quality firms.

In contrast, among individuals with secondary education, firm quality trajectories appear similar across EA-PGI levels (Supplementary Figure \ref{fig:akm-fe-njobs-time-sec}).
These findings suggest that the higher income trajectories observed among tertiary-educated individuals with high EA-PGI are largely driven by greater access to higher-quality, higher-paying firms over time, rather than by difference in educational attainment per se, initial labor market entry, or the frequency of employment transitions alone.

\subsection*{\normalsize Income growth disparities between high and low EA-PGI  are attributable to differences in mobility between rather than within jobs}

Having established that tertiary-educated individuals with higher EA-PGI tend to transition more rapidly to higher-quality firms, we next examine how much of the income differences between EA-PGI groups can be attributed to mobility within versus between firms. To do so, we decompose changes in labor income between consecutive years since graduation into within-firm wage growth (stayers), employer-to-employer mobility, and transitions into and out of non-employment \citep{hahn2021}. 

\begin{figure}[htbp!]
    \centering
    \begin{subfigure}[b]{0.5\linewidth}
        \caption{\textbf{Stayers}}
        \label{subfig:log-earn-growth-decomp-stayer}
        \includegraphics[width = \linewidth]{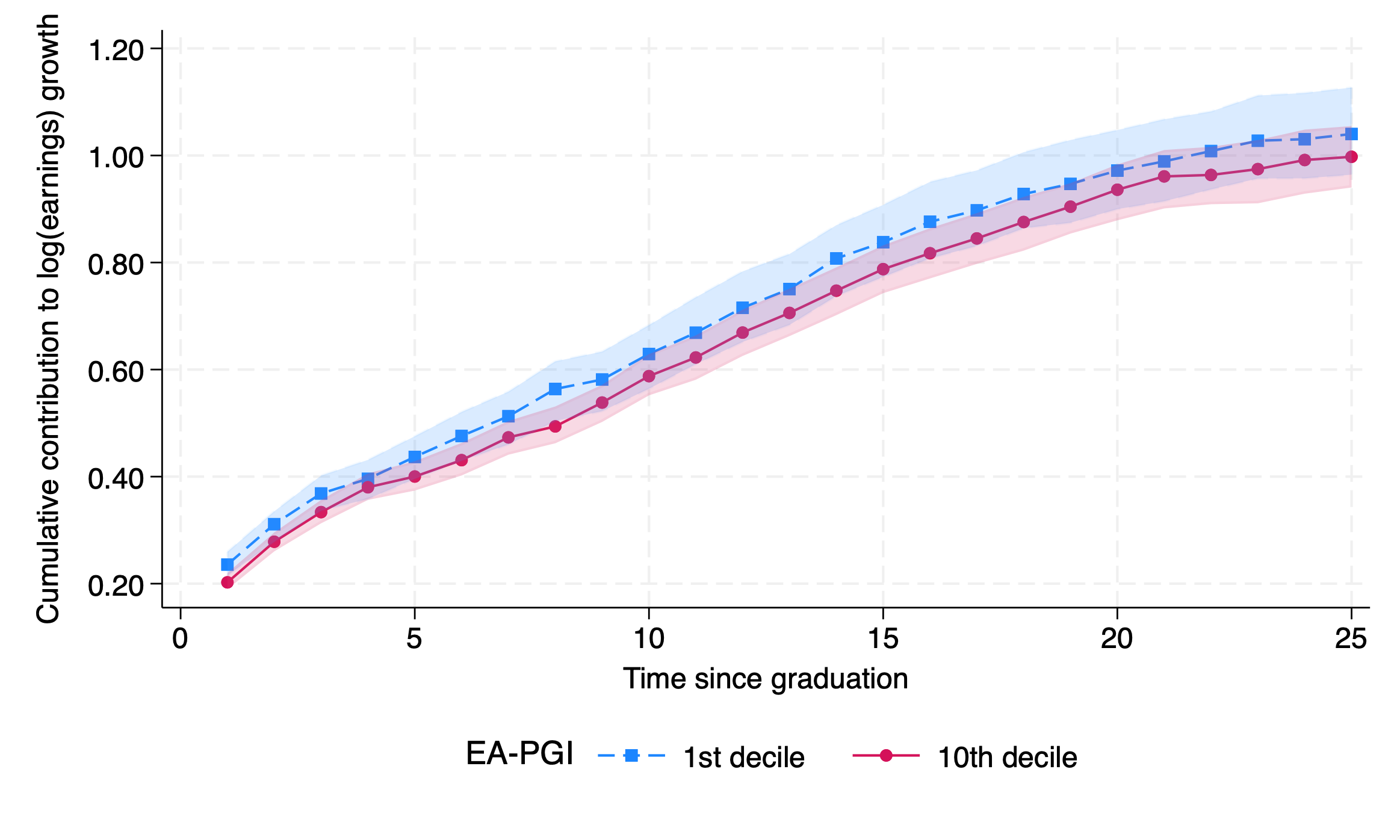}
    \end{subfigure}%
    \begin{subfigure}[b]{0.5\linewidth}
        \caption{\textbf{Employer-to-employer movers}}
        \label{subfig:log-earn-growth-decomp-mover}
        \includegraphics[width = \linewidth]{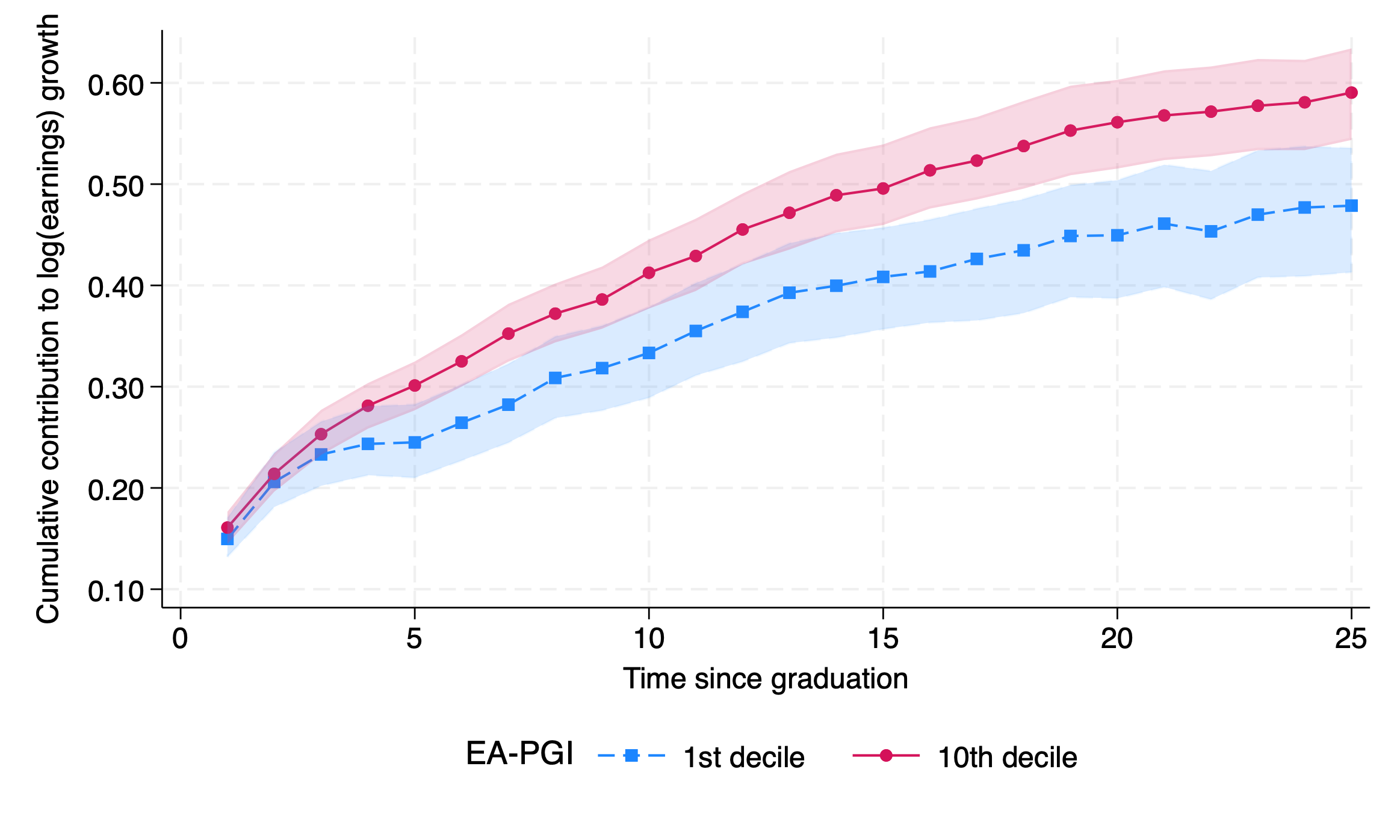}
    \end{subfigure}
    
    \caption{\textbf{Decomposition of cumulative growth in log annual income of tertiary-educated individuals by EA-PGI level, over time}\\\footnotesize Panel \subref{subfig:log-earn-growth-decomp-stayer} reports cumulative contribution of within-firm earnings growth and Panel \subref{subfig:log-earn-growth-decomp-mover} - of employer-to-employer mobility to overall cumulative log earnings growth over time since graduation. The blue line corresponds to 1st and the red - to 10th decile of EA-PGI distribution. The decomposition follows \citet{hahn2021}. The sample for the decomposition is restricted to tertiary-educated workers ($N =$ \num[round-precision=0]{\finalNindTer}). The shaded areas correspond to \CI{95} obtained via 500 block bootstrap iterations, where at each iteration we sample with re-immission \num[round-precision=0]{\finalNindTer} whole income histories from the pool of tertiary graduates.}

    \label{fig:log-earn-growth-decomp}

\end{figure}

Overall, within-firm wage growth contributes most to cumulative income gains over time and does so similarly for individuals at the 10th and 90th percentiles of EA-PGI (Figure \ref{subfig:log-earn-growth-decomp-stayer}). However, the contribution of between-firm mobility to earnings growth becomes increasingly important over time for those at the 90th percentile compared to those at the 10th percentile (Figure \ref{subfig:log-earn-growth-decomp-mover}). 
After 25 years, the cumulative job-to-job contribution to earnings growth differs by approximately 0.1 log-points (about \SI[round-precision=0]{10}{\percent}) between the two groups, suggesting that a substantial portion of the income divergence is driven by firm-to-firm mobility. Transitions into or out of non-employment account for only a negligible share of these disparities (Supplementary Figure \ref{fig:log-earn-growth-decomp-detailed}).

\subsection*{\normalsize Overall health contributes similarly to the EA-PGI–income association for secondary- and tertiary-educated individuals}

Health is a well-established determinant of educational and income disparities,\cite{pallesen2024educational, newman2025income} making it a potential intermediate channel between EA-PGI and income. To examine whether the correlation between EA-PGI and income is mediated by disease burden, we calculate the Charlson Comorbidity Index (CCI), which captures the first occurrence of 17 major chronic conditions over the life course\cite{charlson1987comorbidity, deyo1992icd9} and use it as an outcome in our trajectories regression.

\begin{figure}[htbp!]
    \centering
    \begin{subfigure}[b]{0.49\linewidth}
        \caption{\textbf{Secondary}}
        \includegraphics[width = \linewidth]{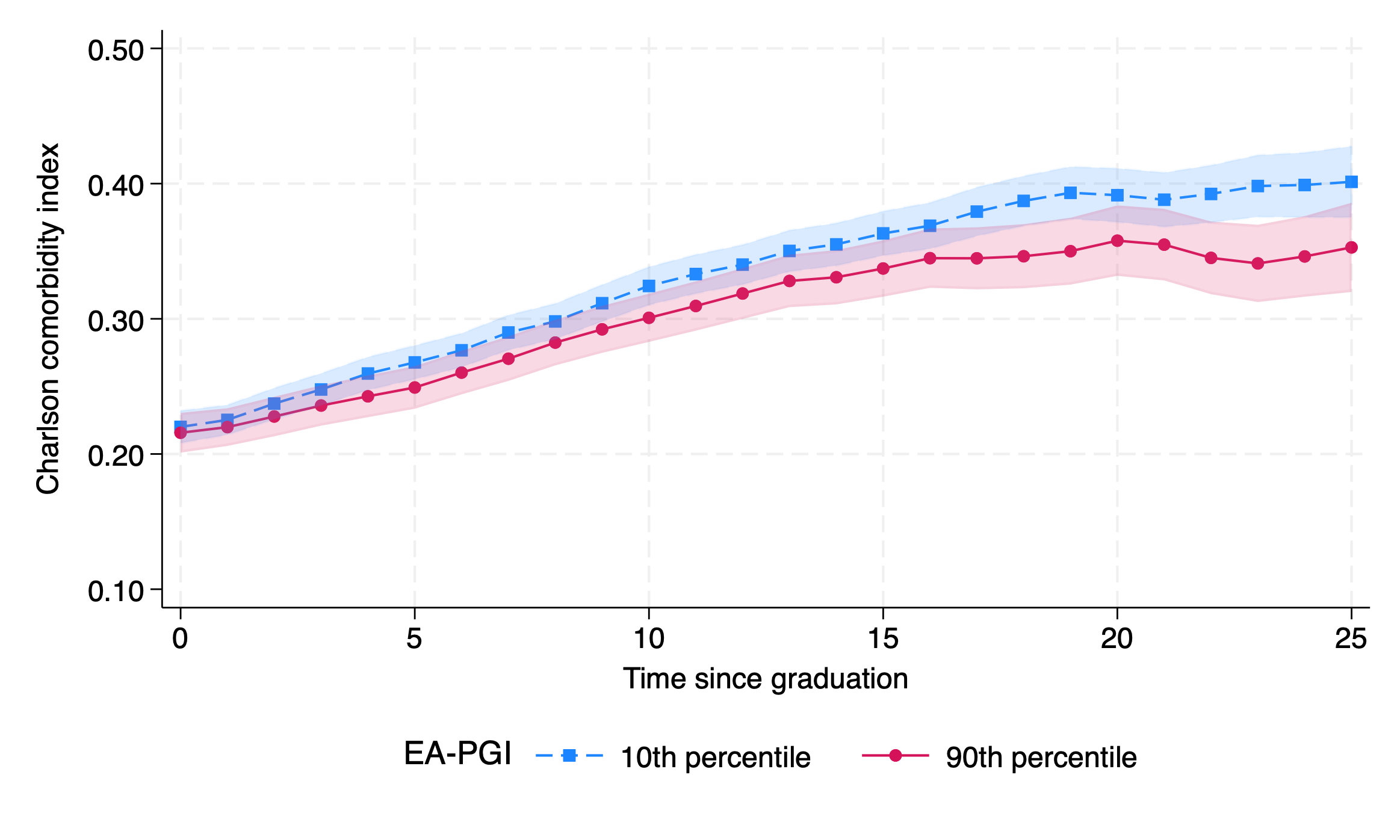}
        \label{subfig:predmarg-cci-pgiea-time-sec}
    \end{subfigure}%
    \begin{subfigure}[b]{0.49\linewidth}
        \caption{\textbf{Tertiary}}
        \includegraphics[width = \linewidth]{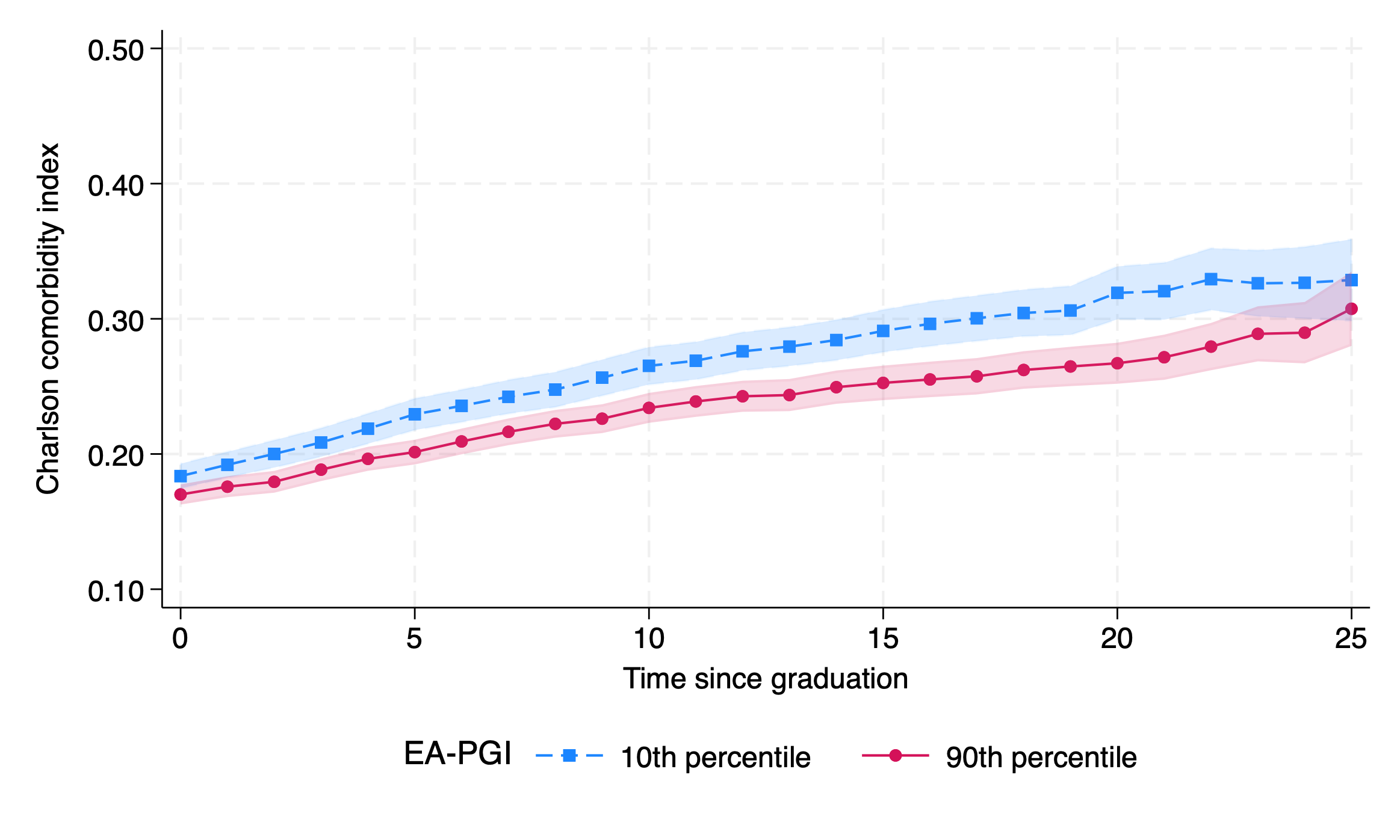}
        \label{subfig:predmarg-cci-pgiea-time-ter}
    \end{subfigure}
    \caption{\textbf{Average health index by EA-PGI level, over time and by education}\\\footnotesize Panels \subref{subfig:predmarg-cci-pgiea-time-sec} and \subref{subfig:predmarg-cci-pgiea-time-ter} report the results for secondary- ($N =$ \num[round-precision=0]{\predMargCCINSec}) and tertiary-educated workers ($N =$ \num[round-precision=0]{\predMargCCINTer}) with non-missing Charlson Comorbidity Index, respectively. \SentenceBlueRed\SentenceHealthReg\SentenceCI{95}}
    \label{fig:predmarg-health-pgiea-age-byedu}
\end{figure}

CCI is, on average, lower among individuals with tertiary compared to secondary education (Figure \ref{fig:predmarg-health-pgiea-age-byedu}), reflecting a well-established lower incidence of major chronic diseases among high-educated individuals\cite{agardh2011type, tillmann2017education, vaccarella2023socioeconomic}. 
Consistent with this pattern, higher EA-PGI is significantly associated with a lower cumulative disease burden as measured by the CCI. The magnitude of this association is highly similar across education groups, suggesting that the stronger association between EA-PGI and income observed among tertiary-educated individuals is not primarily explained by differences in disease burden. This result is confirmed when controlling for parental EA-PGI (Supplementary Figure \ref{fig:predmarg-health-pgiea-age-byedu-famp}).

\subsection*{\normalsize Parental EA-PGI, in particular that of fathers, predicts the income trajectories among tertiary-educated people}

EA-PGI captures both direct genetic effects and indirect effects, such as environmentally mediated influences from parental genetics (i.e., genetic nurture), as well as population stratification and assortative mating \citep{kong2018nature}. 
Indeed, the people at the bottom decile of EA-PGI show remarkably lower socioeconomic status (parental education) than those at the top decile (Supplementary Table \ref{tab:cross-sumstats-pgiea-dec}).
To better isolate direct genetic effects, we leverage parental genetic data. Specifically, we calculate maternal and paternal EA-PGI using SNIPAR for \num[round-precision=0]{\finalNindTrioPld} parent–offspring trios.\citep{young2022mendelian} Of these, \num[round-precision=0]{\finalNindTrioGtpPld} were directly genotyped; the rest were imputed based on 4,482 duos and 3,803 sibling pairs.

We first confirm that the association between offspring EA-PGI and years of education is attenuated by approximately \SI[round-precision=0]{\fpeval{100 - \RegYEduPGIEAParsEstSlopePld / \RegYEduPGIEAEstSlopePld * 100}}{\percent} after accounting for both maternal and paternal EA-PGIs\cite{young2022mendelian}, and that both parental EA-PGIs are significantly associated with offspring education (Supplementary Table \ref{tab:reg-yedu-predicted-pgiea-byspec-pld}), with effects of similar magnitude.

\begin{figure}[htbp!]
    \centering
    \begin{subfigure}[b]{0.5\linewidth}
        \caption{\textbf{Baseline without parental PGI (secondary)}}
        \includegraphics[width = \linewidth]{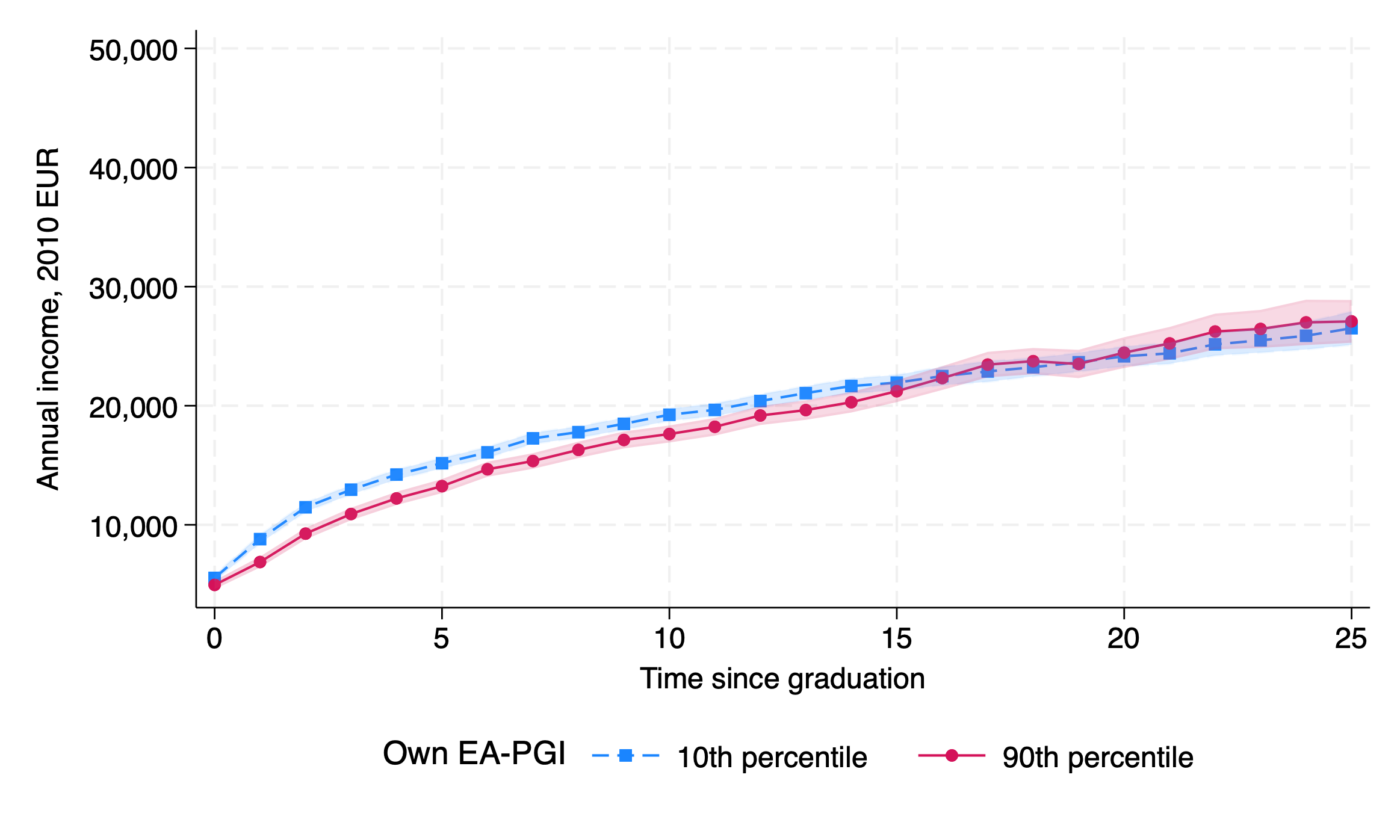}
        \label{subfig:predmarg-inc-pgiea-base-sec}
    \end{subfigure}%
    \begin{subfigure}[b]{0.5\linewidth}
        \caption{\textbf{Baseline without parental PGI (tertiary)}}
        \includegraphics[width = \linewidth]{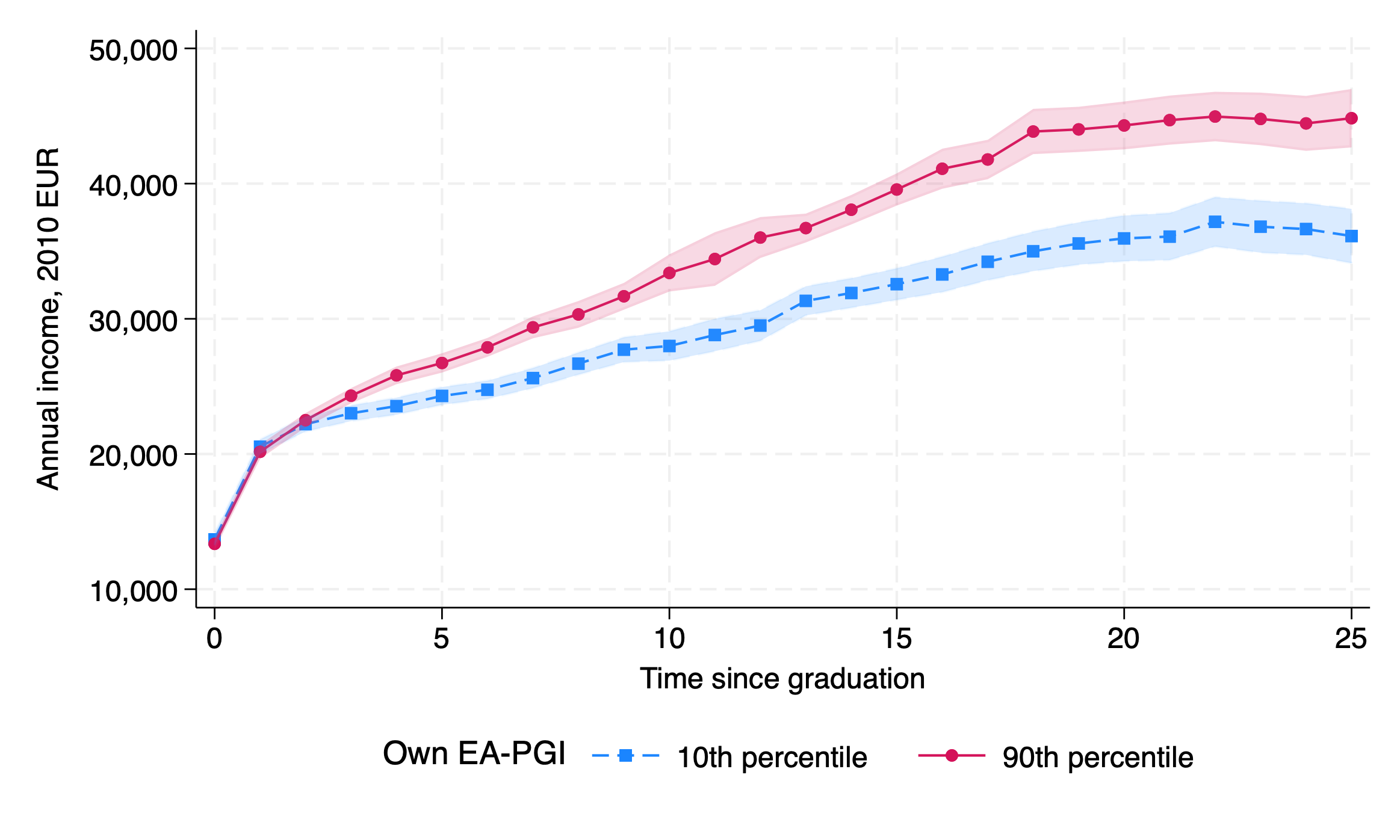}
        \label{subfig:predmarg-inc-pgiea-base-ter}
    \end{subfigure}
    \begin{subfigure}[b]{0.5\linewidth}
        \caption{\textbf{Controlling for parental PGI (secondary)}}
        \includegraphics[width = \linewidth]{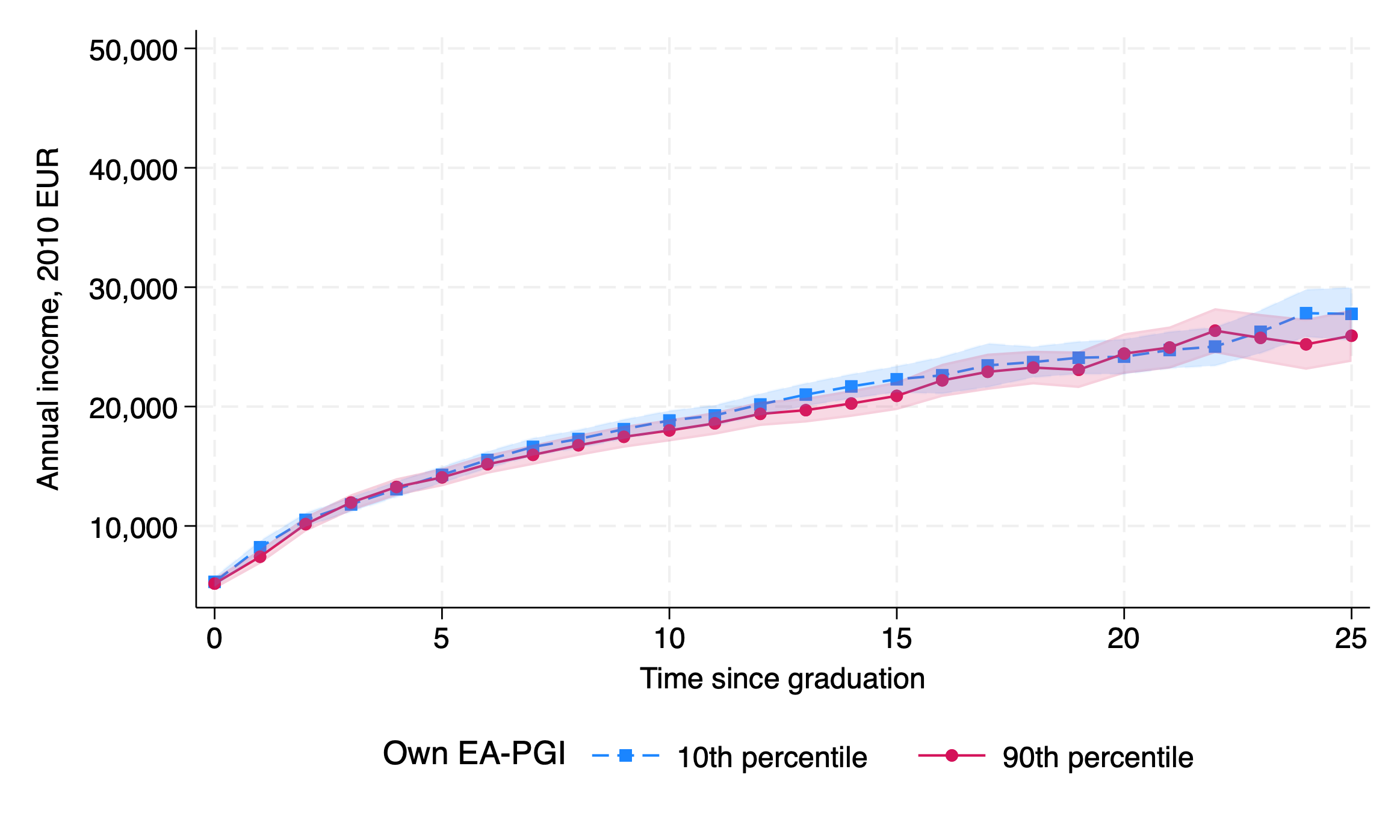}
        \label{subfig:predmarg-inc-pgiea-famp-sec}
    \end{subfigure}%
    \begin{subfigure}[b]{0.5\linewidth}
        \caption{\textbf{Controlling for parental PGI (tertiary)}}
        \includegraphics[width = \linewidth]{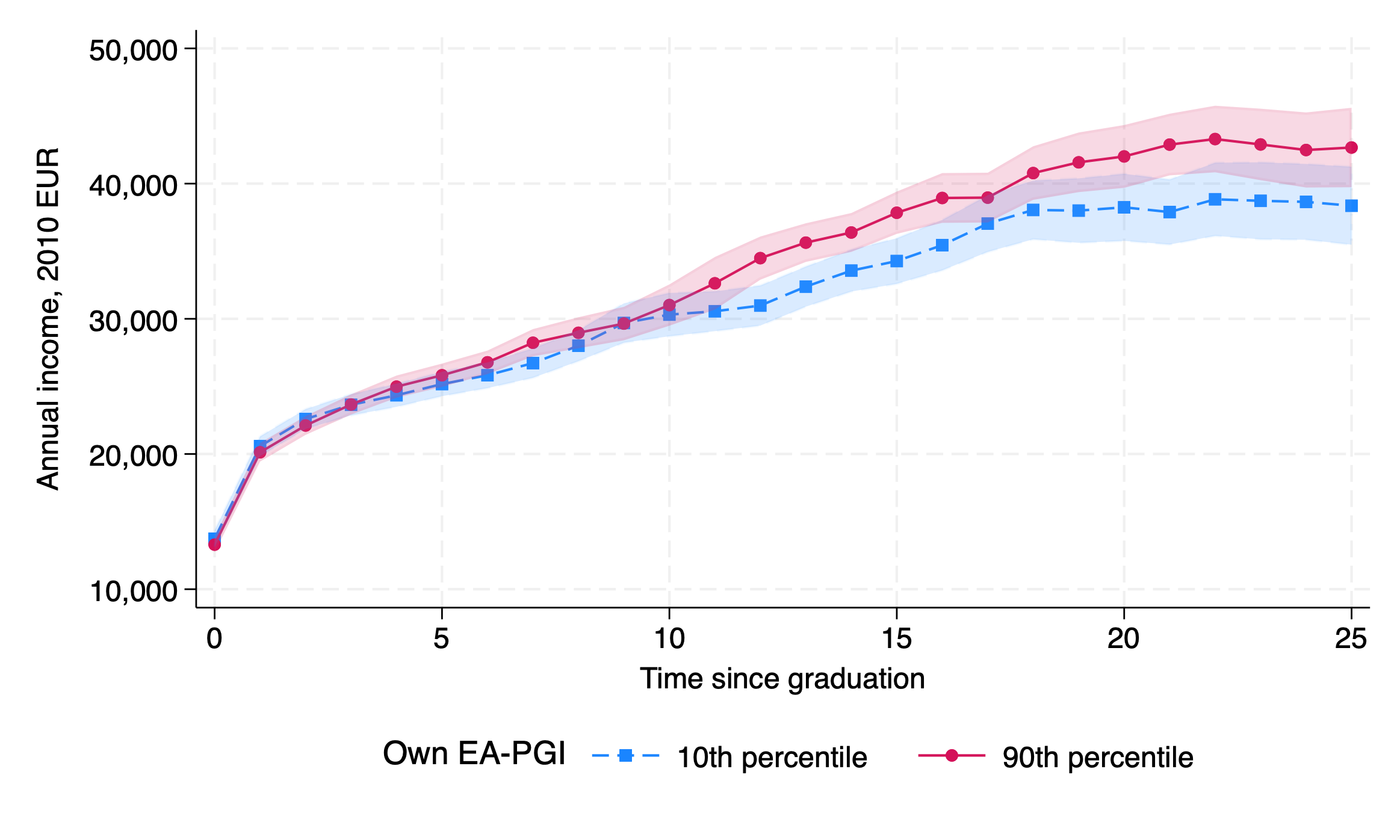}
        \label{subfig:predmarg-inc-pgiea-famp-ter}
    \end{subfigure}
    \caption{\textbf{Average annual income by EA-PGI level, over time and by education, unconditional and conditional on parental EA-PGI among parent-offspring trios}\\\footnotesize Panels \subref{subfig:predmarg-inc-pgiea-base-sec} and \subref{subfig:predmarg-inc-pgiea-base-ter} plot baseline trajectories among secondary- and tertiary-educated workers, respectively, without controlling for parental EA-PGI. Panels \subref{subfig:predmarg-inc-pgiea-famp-sec} and \subref{subfig:predmarg-inc-pgiea-famp-ter} plot the trajectories among secondary- and tertiary-educated workers after controlling for parental EA-PGI fully interacted with time since graduation. The estimation sample in Panels \subref{subfig:predmarg-inc-pgiea-base-sec} and \subref{subfig:predmarg-inc-pgiea-famp-sec} is restricted to secondary- ($N =$ \num[round-precision=0]{\finalNindTrioSec}) and in Panels \subref{subfig:predmarg-inc-pgiea-base-ter} and \subref{subfig:predmarg-inc-pgiea-famp-ter} - to tertiary-educated ($N =$ \num[round-precision=0]{\finalNindTrioTer}) workers with non-missing parental EA-PGI. \SentenceBlueRed\SentenceIncomeReg\SentenceCI{95}}
    \label{fig:predmarg-inc-pgiea-byspec-byedu}
\end{figure}

\begin{table}[htbp!]
    \centering
    \begin{threeparttable}
        \caption{\textbf{Cumulated lifetime income of tertiary-educated individuals by EA-PGI level, conditional on parental EA-PGI among parent-offspring trios}\\\footnotesize The table reports adjusted lifetime income (up to \num[round-precision=0]{\maxTime} years since graduation) by EA-PGI percentiles. The estimation sample is restricted to parent-offspring trios with tertiary-educated offspring ($N =$ \num[round-precision=0]{\predMargDPVFamNTer}). Column (1) reports the estimates in the baseline specification without controlling for parental EA-PGI. Columns (2)-(4) report average cumulated lifetime income by own, maternal and paternal EA-PGI percentiles, respectively, conditional on parental EA-PGI. \SentenceDPVReg\SentenceDPVMethod\SentenceSE}
        \label{tab:predmarg-dpvinc-pgiea-famp}
        \begin{tabular}{lrrrr}
\toprule
& \multicolumn{4}{c}{Dependent variable: Cumulated income} \\\cmidrule(lr){2-5}
& Baseline & \multicolumn{3}{c}{Controlling for parental EA-PGI} \\\cmidrule(lr){2-2}\cmidrule(lr){3-5}
& Own EA-PGI & Own EA-PGI  & Mother EA-PGI & Father EA-PGI \\
 & \multicolumn{1}{c}{(1)} & \multicolumn{1}{c}{(2)} & \multicolumn{1}{c}{(3)} & \multicolumn{1}{c}{(4)} \\ \midrule
\multicolumn{5}{l}{\textit{EA-PGI percentiles}} \\
\hspace{1em}10th & 325 796 & 339 181 & 339 879 & 330 699 \\
& (4 015) & (5 690) & (4 311) & (4 459) \\
\hspace{1em}20th & 332 743 & 341 428 & 341 893 & 335 844 \\
& (3 160) & (4 186) & (3 286) & (3 313) \\
\hspace{1em}30th & 337 737 & 343 044 & 343 306 & 339 552 \\
& (2 677) & (3 232) & (2 713) & (2 664) \\
\hspace{1em}40th & 341 952 & 344 408 & 344 571 & 342 722 \\
& (2 413) & (2 611) & (2 388) & (2 338) \\
\hspace{1em}50th & 345 756 & 345 639 & 345 677 & 345 561 \\
& (2 328) & (2 320) & (2 313) & (2 309) \\
\hspace{1em}60th & 349 747 & 346 930 & 346 834 & 348 447 \\
& (2 413) & (2 399) & (2 466) & (2 543) \\
\hspace{1em}70th & 354 053 & 348 323 & 348 007 & 351 537 \\
& (2 685) & (2 884) & (2 823) & (3 019) \\
\hspace{1em}80th & 359 280 & 350 014 & 349 535 & 355 397 \\
& (3 197) & (3 799) & (3 487) & (3 808) \\
\hspace{1em}90th & 365 948 & 352 172 & 351 466 & 360 946 \\
& (4 022) & (5 198) & (4 500) & (5 126) \\
\midrule Obs. & 7 808 & 7 808 & 7 808 & 7 808 \\\bottomrule
\end{tabular}

    \end{threeparttable}
\end{table}

Next, we estimate the impact of offspring EA-PGI on income trajectories while adjusting for both maternal and paternal EA-PGIs. Controlling for parental EA-PGI accounts for income variation attributable to parental genetics as well as family background characteristics unevenly distributed across EA-PGI groups \citep{rustichini2023educational}.
Among secondary-educated individuals, the income gap across EA-PGI percentiles remains negligible, consistent with our earlier findings (Figure \ref{subfig:predmarg-inc-pgiea-base-sec}).
In contrast, among tertiary-educated individuals, adjusting for parental EA-PGI substantially reduces the income gap between the 90th and 10th percentiles of offspring EA-PGI by approximately \SI[round-precision=0]{\fpeval{100 - \predMargDPVParsEstGapTer / \predMargDPVEstGapTer * 100}}{\percent} (Figure \ref{subfig:predmarg-inc-pgiea-base-ter} vs \ref{subfig:predmarg-inc-pgiea-famp-ter}; Table \ref{tab:predmarg-dpvinc-pgiea-famp}; Supplementary Tables \ref{tab:reg-dpvinc-pgiea-family} and \ref{tab:predmarg-dpvinc-pgiea-fam-sec}).
        
However, unlike what was observed for years of education, only paternal (and not maternal) EA-PGI is significantly associated with offspring income trajectories (Supplementary Table \ref{tab:reg-dpvinc-pgiea-family} and Supplementary Figure \ref{fig:predmarg-inc-pgiea-famp-ter-own-par}). Strikingly, the effect of paternal EA-PGI on offspring income is larger than that of the offspring’s own EA-PGI (€\num[round-precision=0]{\RegDPVPGIEAParsEstSlopeFTer} [CI €\num[round-precision=0]{\RegDPVPGIEAParsLowSlopeFTer} - €\num[round-precision=0]{\RegDPVPGIEAParsUppSlopeFTer}] vs €\num[round-precision=0]{\RegDPVPGIEAParsEstSlopeTer} [CI €\num[round-precision=0]{\RegDPVPGIEAParsLowSlopeTer} - €\num[round-precision=0]{\RegDPVPGIEAParsUppSlopeTer}] per one standard deviation in respective EA-PGI). To contextualize, regardless of offspring EA-PGI, fathers in the 10th percentile of EA-PGI are associated with a cumulative offspring income \num[round-precision=0]{\maxTime} years after graduation of €\num[round-precision=0]{\predMargDPVParsEstpTenFTer} [CI €\num[round-precision=0]{\predMargDPVParsLowpTenFTer} - €\num[round-precision=0]{\predMargDPVParsUpppTenFTer}], compared with €\num[round-precision=0]{\predMargDPVParsEstpNinetyFTer} [CI €\num[round-precision=0]{\predMargDPVParsLowpNinetyFTer} - €\num[round-precision=0]{\predMargDPVParsUpppNinetyFTer}] among fathers in the 90th percentile. These effects are consistent across both directly genotyped and imputed trios (Supplementary Table \ref{tab:predmarg-dpvinc-pgiea-family}).


\section*{Discussion}\label{sec:discussion}

In this study, we examine the mechanisms through which known genetic factors associated with educational attainment affect socioeconomic inequalities. To our knowledge, this is the first large study to investigate direct and indirect (parental) EA-PGI effects on longitudinal income trajectories; prior work has used small samples or cross-sectional designs, which limits our understanding of how inequality accumulates over the life cycle and is transmitted.
It is important to recognize that these studies have contributed to an aversion, anger and even fear of studying genetics in social stratification research \citep{martschenko2019genetics, mills2022sociogenomics}. We include a document with Frequently Asked Questions (FAQs)  to offer an accessible explanation of what our study does and, importantly, does not, find and how the results should be interpreted.

Labor income is the focus of our analysis because it provides a well-defined, downstream indicator of socioeconomic status, reflecting differences in education, health, and skills. Together with government transfers, labor income constitutes the main source of monetary inflows accumulated over the life course and is therefore a fundamental determinant of wealth \citep{black2025importance}. Moreover, income inequality is a central concern in both economics \citep{kline2024firm} and epidemiology \citep{lynch2000income,pickett2015income,schwandt2021inequality}, which allows us to incorporate genetics into well-established analytical frameworks.

We show that workers earn similar income upon labor market entry, irrespective of their EA-PGI level. After entry, however—and only among tertiary-educated individuals—income differences between those at the 90th and 10th percentiles of EA-PGI widen steadily over time, culminating in a \SI[round-precision=0]{\fpeval{\predMargDPVEstGapTer / \medianDPVTer * 100}}{\percent} cumulative income gap 25 years after graduation. 
The fact that EA-PGI predicts income trajectories only among tertiary-educated individuals reflects two forces. First, individuals with relatively high EA-PGI experience higher economic returns to ability. 
Second, EA-PGI influences educational sorting itself, shaping the types and levels of education pursued. While we cannot quantify exactly the relative importance of the two channels (sorting and economic returns to higher EA-PGI), our analysis based on the AKM model estimation \citep{abowd_high_1999} offers additional insights on the role of workers' ability and of employers in explaining the observed income differences. 

We show that a measure of labor market productivity (or ability) is indeed positively associated with EA-PGI, but again only for tertiary-educated workers. This relationship persists even after controlling for detailed education-related characteristics (education field and type of academic institution). The finding that EA-PGI relates to labor market productivity only among highly educated individuals raises the question of whether some high-ability individuals with only secondary education might benefit from additional educational support or incentives to pursue tertiary studies. While we cannot provide a definitive answer to this, we note that any such discussion must consider the role of family socioeconomic background \citep{ichino2024college} and  return to the importance of family background later.

Our trajectory framework also allows us to document a new mechanism through which genetic endowments affect income: employer sorting. We show that there is no evidence of sorting into the first employer on the basis of EA-PGI. This is consistent with labor market entrants having limited information about firms and match quality, and with firms similarly knowing little about the productivity of new hires. Despite the absence of initial sorting, we show that EA-PGI subsequently matters for job mobility among university graduates, who are more likely to move to higher-paying employers. This provides novel evidence that employer learning about workers’ productivity and job mobility \citep{farber1996learning} are important channels underlying the genetic gradient in income, but only for jobs accessed through higher-education degrees.
This result complements a literature in economics that shows that employer-specific characteristics and pay policies play a crucial role in explaining wage inequality beyond occupation alone (Supplementary Figure \ref{fig:wage_dispersion})\cite{card2013workplace, kline2024firm}.

Because health is a well-documented determinant of educational and income disparities, we also examine whether it mediates the EA-PGI–income relationship. Previous studies have shown that EA-PGI is associated with multiple diseases \citep{okbay2022}. We find that the association between EA-PGI and cumulative disease burden is similar across education groups. This suggests that differences in health do not explain the stronger EA-PGI–income association among the tertiary-educated.

By Mendel’s laws, parents of children with higher EA-PGI must have higher EA-PGI themselves, and higher EA-PGIs result in higher parental education. This pattern is confirmed in our data: individuals in the 10th percentile of EA-PGI come from substantially lower socioeconomic backgrounds than those in the 90th percentile (Supplementary Table \ref{tab:cross-sumstats-pgiea-dec}). We therefore examine the role of parental EA-PGIs in accounting for the observed income gap across offspring’s EA-PGI levels. Adjusting for parental EA-PGI reduced the offspring income gap by approximately \SI[round-precision=0]{\fpeval{abs(\predMargDPVParsEstGapTer / \predMargDPVEstGapTer * 100 - 100)}}{\percent}, indicating that indirect effects play an important role in shaping income development.

Notably, most of this attenuation is driven by the father’s EA-PGI rather than the mother’s, a pattern not observed for educational attainment itself. This may reflect differential mechanisms: maternal resources, such as time and cognitive engagement, are particularly relevant for early-life development and educational choices, whereas paternal resources may become more influential during labor-market entry, through direct income support, occupational networks, or knowledge transmission about high-quality firms \citep{del2014household}. These results are especially striking given Finland’s relatively high level of gender equality in employment opportunities \citep{WEF2024GlobalGenderGap}.

This study advances the literature in several ways. First, we analyze a substantially larger genotyped sample, including \num[round-precision=0]{\finalNindTrioPld} parent–offspring trios, than previous studies, linked to decades of high-quality administrative data. This allows us to produce precise estimates of the genetic drivers of socioeconomic disparities over the life cycle. Second, by exploiting repeated measurements of workers linked to employers, we apply panel-data methods that account for worker and firm heterogeneity, which in turns allows us to provide novel evidence on the role of genetics in explaining the production of inequality in the labor markets. This allows us to decompose income variation into worker- and firm-level components, bridging genetics and modern labor economics \citep{card_firms_2018, song_firming_2018, kline_firm_2025}.
Lastly, we leverage continuous, rich information from individual tax records, avoiding self-report bias and measurement error that may affect survey-based measures of income. 

Our study also has limitations. First, our sample consists solely of Finnish individuals who tend to be positively selected compared to the general population. This limits the generalizability to other ancestries or institutional settings. To address the fact that genotyped individuals are not fully representative of the Finnish population, we applied inverse-probability weighting to make the sample representative of all fresh graduates in the same calendar period \citep{davies_causal_2018}.
Weighted analyses produced results highly consistent with our main estimates (Supplementary Table \ref{tab:predmarg-dpvinc-pgiea-weighted}). 
Second, EA-PGI should not be interpreted as a pure “genetic endowment for educational attainment,” as it may capture pathways correlated with educational attainment, such as lifestyle or social factors, that are modifiable and context-dependent. As such, changes in these underlying environments or social conditions could alter the observed relationship between EA-PGI and income. 
Lastly, as noted, the large attenuation we observe when controlling for parental EA-PGI indicates that indirect effects—particularly those operating through fathers—play an important role in shaping income differences. However, our setting does not allow us to disentangle the (parental) genetic and alternative environmental components underlying this result, which we leave to future work. 

Taken together, our results indicate that EA-PGI is a robust predictor of income across the working life course, but only among tertiary-educated individuals. These effects operate primarily through individuals with higher EA-PGI experiencing faster and more frequent labor-market transitions to higher-quality employers, which in turn offer higher wages and thereby further amplify income inequality. Moreover, a substantial share of the EA-PGI–income association arises through indirect paternal genetic effects, suggesting that in the relatively recent generations studied (average graduation year 2000), fathers continued to play an important role in shaping offspring labor-market trajectories through untransmitted genetic factors.

    
\section*{Materials and Methods}\label{subsec:dynamic}

\setcounter{section}{1} 

\subsection{Data Sources}\label{sec:data}

\paragraph{Genetic data.} The genotyped sample was obtained from Finnish biobanks and consists of individuals who provided consent for research use of their blood samples. Participants in our dataset were drawn from several population-based epidemiological cohorts: \num[round-precision=0]{\finalNindThl} from the THL and \num[round-precision=0]{\finalNindBdb} from the Blood Donor study.\footnote{
THL biobank website: 
\url{http://www.thl.fi/biobank}. 
Blood service biobank website: \url{https://www.veripalvelu.fi/en/biobank/}.
}

To quantify the genetic contribution to educational attainment, we constructed a polygenic index for years of education (EA-PGI) based on the largest genome-wide association study of educational attainment to date \citep{okbay2022}, excluding Finnish cohorts to avoid overfitting. In our analysis sample, the PGI explained \SI{\fpeval{100 * \RsqIncrementalYEduPredictedPGIEA}}{\percent}\footnote{Computed as incremental $R^2$, following \citet{okbay2022}, and controlling for gender fully interacted with year of birth indicators, biobank indicator and first 10 genetic principal components} of the variance in years of education (Supplementary Table \ref{tab:predicted-ea-pgi-increm-r2})—slightly lower than estimates from previous studies \cite{lee2018gene}, possibly due to differences in cohort composition or educational classification.

Our sample tends to be positively selected (younger, better educated, with a higher share of women) compared to the population of fresh graduates. To analyze whether sample composition is a likely driver of our results, we apply an inverse probability weighting approach to make the sample representative of the graduates population \citep[see e.g.][]{davies_causal_2018}.
Supplementary Table \ref{tab:descriptives} provides details on the re-weighting procedure and shows that it is effective in making the sample representative of the general population along the central dimensions that are initially unbalanced (including entry income, which we do not re-weight in our routine).   
Using these weights when estimating the income trajectories  yields results that are qualitatively very similar to those in our main analysis, both when using the full genotyped sample and when using the family trios (Supplementary Table \ref{tab:predmarg-dpvinc-pgiea-weighted}).

\paragraph{Register data.} We link the genotyped data to administrative registers from Statistics Finland (FOLK databases), covering the years 1987--2019. In addition, we utilize these register data independently, as they include the entire population of individuals permanently residing in Finland at the end of each year. The registers provide detailed information on employment histories, which allows us to identify the main employer at the end of each calendar year. They also contain income by source, from which we identify yearly labor income. 
We include people with zero income in the income analysis, thereby avoiding conditioning on employment. Given the long time series used in the analysis, all monetary values are deflated to 2010 EUR to account for inflation and ensure comparability across time. The choice of the reference year is arbitrary; we choose 2010 because it lies midway between 2000 -- the average graduation year in the sample -- and the last follow-up year, 2019.

The registers further include demographic variables (gender, year of birth) and detailed educational information (highest degree, 4-digit field, school or institution ID, vocational vs. academic track), occupation, and industry codes. 

\paragraph{Health registers.} Health outcomes are obtained from two nationwide registers maintained by the Finnish Institute for Health and Welfare: the Care Register for Health Care (Hilmo) and the Register of Primary Health Care Visits (Avohilmo). In this study, Hilmo covers inpatient visits, operations, and specialized outpatient visits for the period to 1987--2024, when diagnoses follow ICD-9 and ICD-10 coding. Avohilmo, which uses ICD-10, covers primary care outpatient visits since 2011. For individuals absent from Hilmo, Avohilmo is used    to complement the coverage. Both registers contain patient identifiers, care episode details, and one or more discharge diagnoses.

\subsection{Methods}\label{sec:measurement}

\subsubsection*{Polygenic Indices} We construct polygenic indices (PGIs) by aggregating single nucleotide polymorphisms (SNPs), common genetic variants identified in Genome-Wide Association Studies (GWAS) as predictive of years of education and health-related outcomes (see e.g. \cite{biroli2025economics}). SNPs are linearly combined using GWAS-derived effect sizes as weights, producing out-of-sample PGIs predictive of each trait of interest.

Our primary measure is the PGI for educational attainment (EA-PGI), standardized to mean zero and unit variance. Its distribution across education groups is shown in Supplementary Figure \ref{fig:density_pgi_edulev_edutrack}. To control for ancestry and population stratification, we  compute the first ten principal components of the genetic data and include them as covariates in all analyses.

\subsubsection*{Worker Ability and Firm Quality Measurement}

To operationalize worker ability (or productivity) on the labor market $\theta_i$, and firm quality $\psi_{J}$ we estimate an (AKM) regression\cite{abowd_high_1999, kline2024firm}:

\begin{equation}
    \label{eq:akm}
    y_{it} = \mathbf{X}_{it}\beta + \psi_{J(i, t)} + \theta_i + \varepsilon_{it}
\end{equation}

where $y_{it}$ denotes log monthly labor income\footnote{The outcome in \eqref{eq:akm} is monthly income, defined as annual earnings divided by number of months worked. Results are robust to using hourly wages derived from the Structure of Earnings Register (SES), which covers the whole public sector and a sample of about half of the private sector. To maximize sample size and coverage, we use monthly earnings as our baseline income measure.} for individual $i$ in year $t$; $\mathbf{X}_{it}$ includes education fully interacted with calendar year and cubic age polynomial; $\psi_{J(i, t)}$ represents firm fixed effects; and $\theta_i$ are worker fixed effects. Estimated $\hat{\theta}_i$ provides a measure of worker productivity (unobserved heterogeneity) and $\hat{\psi}_{J(i, t)}$ offers firm-specific wage premia, interpreted as firm quality. 
Workers with higher fixed effects earn more across firms relative to a reference worker, holding observables constant. Similarly, firms with higher AKM fixed effects pay higher average wages, consistent with higher productivity serving as a proxy firm quality, as they capture persistent wage differences across firms after accounting for worker characteristics.

Except for the additive separability of firm and worker fixed effects, no functional form assumptions are made on either of the fixed effects, which are estimated non-parametrically.
The estimation sample includes all full-time employees aged 20--60. The main employer is defined as the highest-paying ongoing job at year-end. To ensure labor market attachment, we restrict to workers earning at least 50\% of the national median monthly income. 
the model is estimated by using the sample of firms linked by workers job-to-job transitions, also called connected set \citep{kline_firm_2025}. Firms that do not experience hires or separations during their observation period are not part of the connected set and do not have an associated firm fixed effect.  

For the estimation of the model, employment spells in very small firms ($<$5 employees) or shorter than four months are excluded. The resulting panel comprises \SI[fixed-exponent=6, table-omit-exponent]{\fpeval{\AKMnindFePTone + \AKMnindFePTtwo}} million workers, \num[fixed-exponent=6, table-omit-exponent]{\fpeval{\AKMnobsPTone + \AKMnobsPTtwo}} million person-year observations, and \num[fixed-exponent=3, table-omit-exponent]{\fpeval{\AKMnfirmFePTone + \AKMnfirmFePTtwo}} thousand firms. The AKM estimation is performed separately for two periods (1987--2003 and 2004--2019) due to computational constraints. Correlations of worker and firm fixed effects across the two periods are reasonably high (Supplementary Figure \ref{fig:akm-8703-0419-binscatter}). 

We standardize worker and firm fixed effects to have zero mean and unit variance, and link them to the genotyped sample. Supplementary Tables \ref{tab:sumstats-akm-gtp} and \ref{tab:akm-sumstats-vardec} provide summary statistics for the sample used in estimating AKM, and standard statistics and income variance decomposition following AKM estimations.

\subsubsection*{Income and Health Trajectory Model} 

To study how genetic predispositions affect income over the life cycle, we estimate: 

\begin{equation}
    \label{eq:income-gap}
    y_{icmt} = \alpha + \tau_c + \tau_m + \beta_{t} PGI_{i} + \gamma X_{i} + \varepsilon_{icmt}
\end{equation}

where $y_{icmt}$ is labor income of individual $i$ or the Charlson Comorbidity Index (see below), in birth cohort $c$, calendar year $m$, and number of years since graduation $t$. $PGI_{i}$ is standardised EA-PGI; $\tau_c$ and $\tau_m$ are cohort and year effects; and $X_i$ includes ten genetic principal components, gender, and biobank indicator (THL or Blood Donors). 
The coefficients of interest are $\beta_{t}$, which capture the income differential for EA-PGI over time since graduation. Since PGIs are randomly assigned at conception, $\beta_{t}$ has a causal interpretation, conditional on adequate control for population stratification via $X_i$. Any residual stratification would lead the coefficients to capture a compound effect of own genetics and other environmental factors. While assigning sign to the bias is not immediate, we believe that it is reasonable to consider our estimates as conservative lower bounds of true causal effects of own EA-PGI.\footnote{The PGI captures only the contribution of common genetic variants identified in external GWAS, not the full genetic architecture of education.}

Our model income trajectories specifications include continue EA-PGI evaluated at 10th and 90th percentiles of EA-PGI distribution. The results are similar when specifying EA-PGI via deciles (Supplementary Figure \ref{fig:trajectory_10groups}). Finally, when  computing cumulated lifetime income we discount income by a 3\% interest rate, computing its present value upon graduation. 
In the computation, we sum over all discounted income rows (including zeros) between graduation year and up to \num[round-precision=0]{\maxTime} years later.

\subsubsection*{Income decomposition by EA-PGI group and over time since graduation}

We implement the approach by \citet{hahn2021} to decompose the log-income growth separately by EA-PGI group (1st and 10th deciles).

Each year since graduation $t$, workers are partitioned into one of four groups: \textit{stayers} (workers who stay with the same employer); \textit{employer-to-employer transitions} (workers who change firm); \textit{entrants from non-employment} (hires from nonemployment); \textit{exiters to non-employment} (incumbent workers separating to nonemployment). 
The average income growth between $t-1$ and $t$ is decomposed into four weighted contributions based on the four worker types (weighted by the share of workers in each worker type). 
Since entrants and exiters move between employment and non-employment, their contribution is obtained by comparing their average income to that of the workers who are continuously employed in the time period. 

In line with \citet{hahn2021}, and confirmed by our analysis, job-to-job movers' transitions are associated with large earnings gains for individuals. Moreover, the entrants from nonemployment earn substantially less than the continuously-employed workers to which their salary is compared to. Hence, their entry into employment lessens (subtracts from) the average earnings and their contribution to the average earnings growth is negative. The opposite occurs for exiters to nonemployment: they also tend to earn  less than the continuously-employed workers, but  because these low-paying jobs dissolve, this contributes positively to the earnings growth. 

We present results by cumulating the income growth components over $t = 1 ,..., 25.$

\subsubsection*{Charlson Comorbidity Index}

The Charlson Comorbidity Index (CCI) \citep{charlson1987comorbidity,deyo1992icd9} assigns fixed weights to comorbid conditions associated with higher mortality risk. The index is the weighted sum of an individual’s comorbidities, with weights derived from Cox regression models. 

We compute CCI scores using the ICCI R package \citep{ICCI}, which implements the ICD-9 and ICD-10 coding \citep{quan2005coding} via R's comorbidity package \citet{gasparini2018comorbidity}.\footnote{The package accommodates multiple ICD versions. Source code:  \href{https://github.com/dsgelab/ICCI}{ICCI GitHub repository}.} For each individual, we compute cumulative CCI scores by age, recalculating the index at successive cutoffs (0–19, 0–20, …, up to 0–50 years).

\paragraph{Definitions and Sample Restrictions.} For the trajectory-based analyses, we construct a panel of all genotyped adults with either secondary or tertiary qualification, followed from year of graduation onward between 1987 and 2019. To ensure that individuals in our sample have completed their education phase, we remove those with secondary degree that have not been observed past age 30; people that obtain tertiary degree before age 30 are retained in the sample. The final sample includes \num[round-precision=0]{\finalNindAll} individuals and \num[round-precision=0]{\finalNobsAll} person-year observations (see Supplementary Table \ref{tab:trajectory-restrictions-samplecounts}).

\section*{Acknowledgments}

We would like to thank  Adrian Adermon, Mikko Arvas, Pietro Biroli, Rosa Cheesman, Lena Hensvik, Paul Hufe, Andrea Ichino, Francis Kramarz, Martin Nybom, Gerard van den Berg, and participants to the European Social Science Genetics Network Conference 2023 (Bologna), Nordic Summer Institute 2025 (Uppsala), Workshop on Inequality and education 2025 (Modena), Workshop in Wages, Employment, and Inequality 2025 (Helsinki), for useful comments and discussions. 

We thank all study participants for their generous participation in biobank research, and we are grateful to Blood Service Biobank and to THL Biobank for issuing the data permits to utilize the genetic data (BSB permit no. 005-2020 and THLBB2023\_47). The project fall under IRB ethical approval HUS/1311/2020, Statistics Finland data permit TK/125/07.03.00/2020, and Findata permit THL/4014/14.02.00/2022. Lombardi gratefully acknowledges funding from the Research Council of Finland (decision no. 350399) and Yrjö Jahnsson Foundation (decision no. 20247803).

\newpage


\setstretch{0.5}
\printbibliography

\onehalfspacing
\restoregeometry
\newpage


\clearpage
\appendix
\pagenumbering{roman}
\setcounter{page}{1}
\section{Supplementary Figures and Tables}\label{sec:appendix}

\counterwithin{figure}{section}
\setcounter{figure}{0}

\counterwithin{table}{section}
\setcounter{table}{0}

\subsection{Supplementary Figures}\label{subsec:appendix_figs}

\begin{figure}[htbp!]
    \centering
    \includegraphics[width = \linewidth]{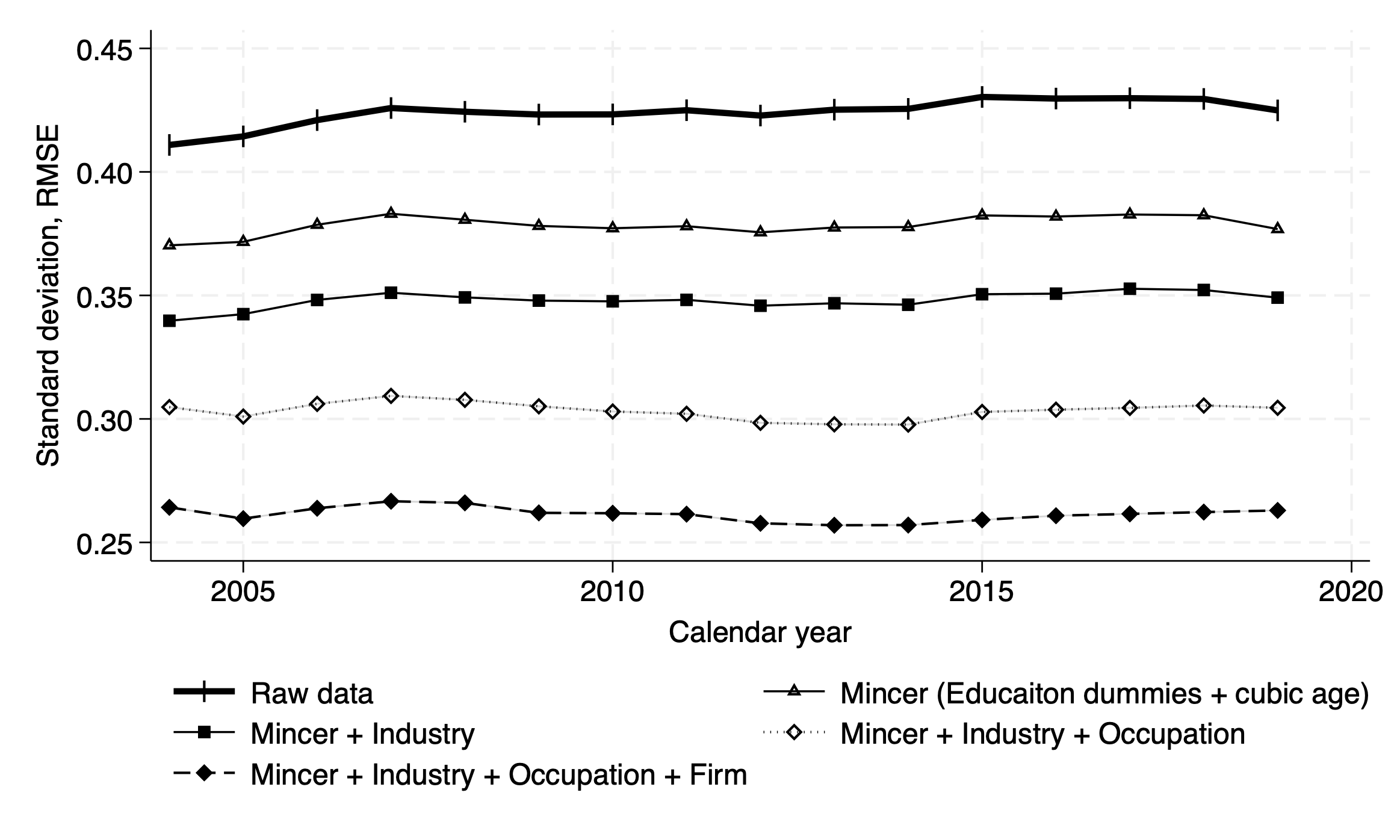}
    \caption{\textbf{Standard deviation in wages from alternative wage models}\\ \footnotesize Measures of dispersion in actual and residualized wages for full-time workers, using the sample described in “Worker Ability and Firm Quality Measurement” section. Residual wages are obtained by sequentially adjusting worker-level wages for the covariates listed in the legend using linear models. The Mincer specification adjusts for education indicators and a cubic age profile. All models are estimated separately by calendar year. 95\% confidence intervals.
    }
    \label{fig:wage_dispersion}
\end{figure}

\begin{figure}[htbp!]
    \centering
    \includegraphics[width = \linewidth]{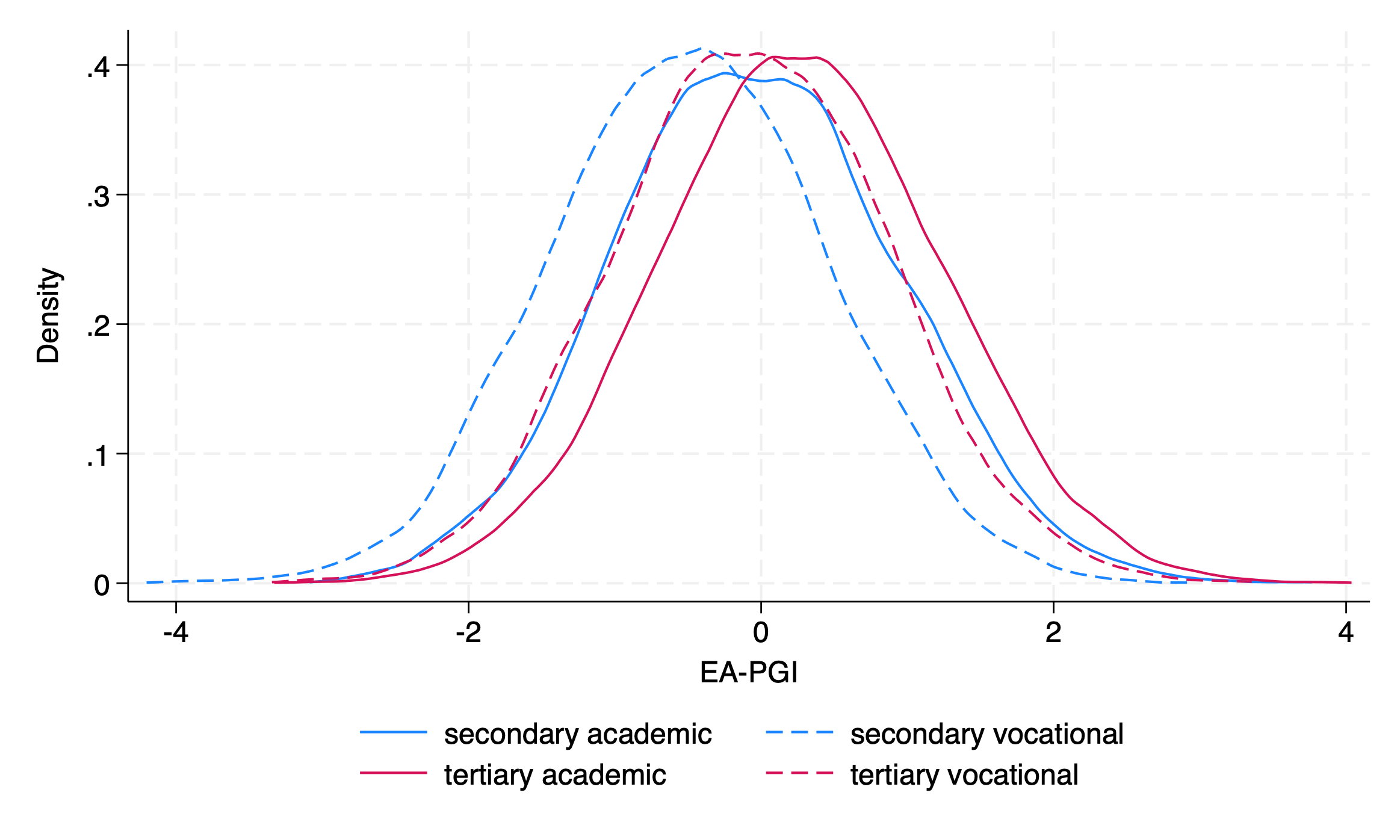}
    \caption{\textbf{Density of EA-PGI by highest education level and track}\\\footnotesize The density plot uses full analysis sample with non-missing education track information ($N = 50,940$). The blue lines correspond to secondary, and the red - to tertiary degree level. Solid lines correspond to academic, and the dashed lines - to vocational education tracks.}
    \label{fig:density_pgi_edulev_edutrack}
\end{figure}

\begin{figure}[htbp!]
    \centering
    \includegraphics[width = \linewidth]{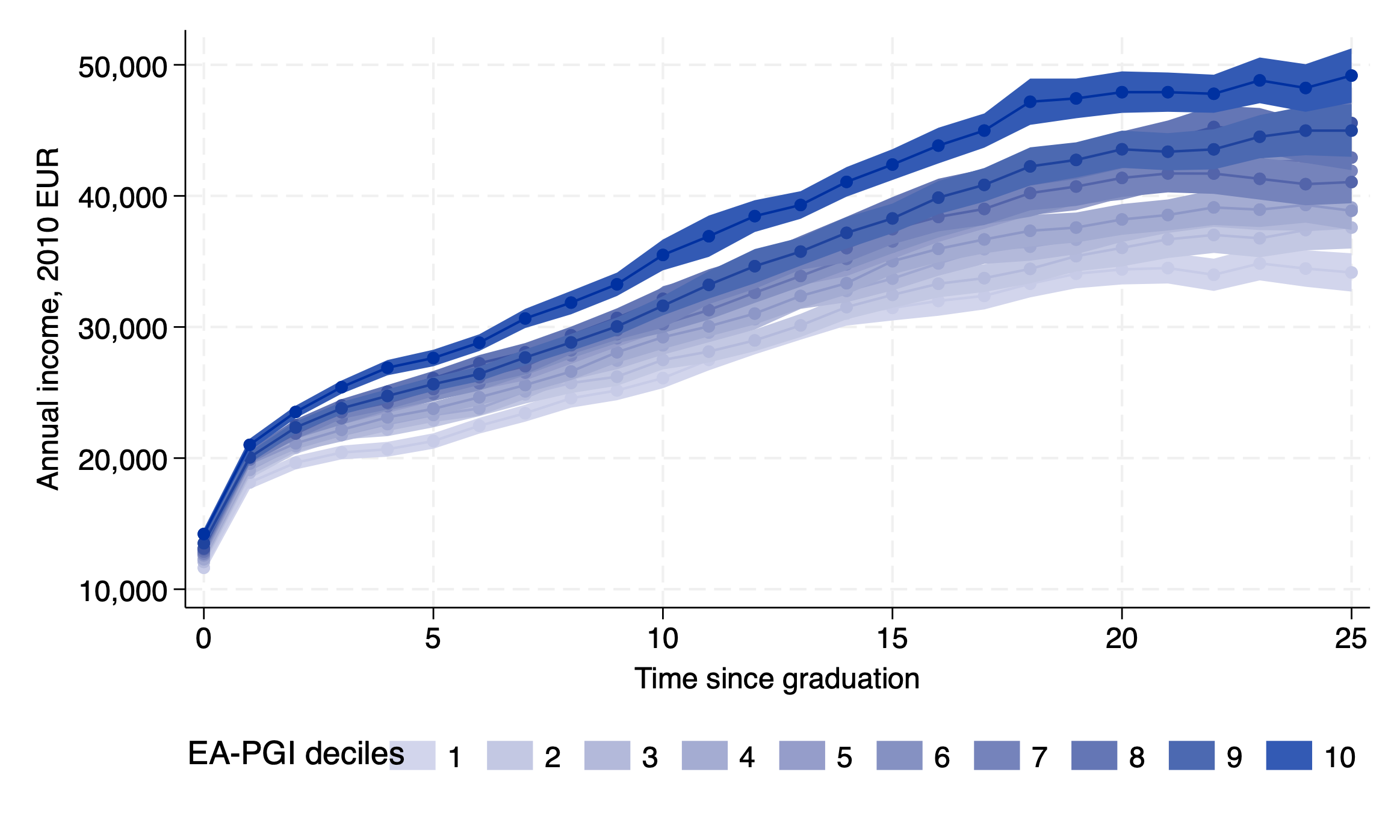}
    \caption{\textbf{Average annual income of tertiary-educated individuals by EA-PGI level, over time}\\\footnotesize The figure uses a subset of tertiary-educated individuals ($N =$ \num[round-precision=0]{\finalNindTer}). The lines correspond to ten deciles of the EA-PGI distribution. \SentenceIncomeReg\SentenceCI{95}}
    \label{fig:trajectory_10groups}
\end{figure}

\begin{figure}[htbp!]
\centering
\caption{\textbf{Employer mobility and quality of secondary-educated individuals by EA-PGI level, over time}\\\footnotesize Panel \subref{subfig:njobs-time-sec} plots the average number of employment spells, and Panel \subref{subfig:akm-fe-time-sec} - the average firm quality index over time since graduation. \SentenceBlueRed Both panels are estimated from a regression of respective outcome on EA-PGI fully interacted with indicators measuring years since graduation and controlling for first ten genetic principal components, gender, year of birth, calendar year, and biobank indicators. The estimation sample is restricted to secondary-educated workers with non-missing firm identifiers ($N=$ \num[round-precision=0]{\predMargEmpSpellNSec}) in panel \subref{subfig:njobs-time-sec} and non-missing firm quality index ($N=$ \num[round-precision=0]{\predMargPsiNSec}) in panel \subref{subfig:akm-fe-time-sec}. \SentenceCI{95}}
\label{fig:akm-fe-njobs-time-sec}
\begin{subfigure}{1\textwidth}
    \centering
    \subcaption{\textbf{Number of jobs since labor market entry}}
    \includegraphics[width = 0.8\linewidth]{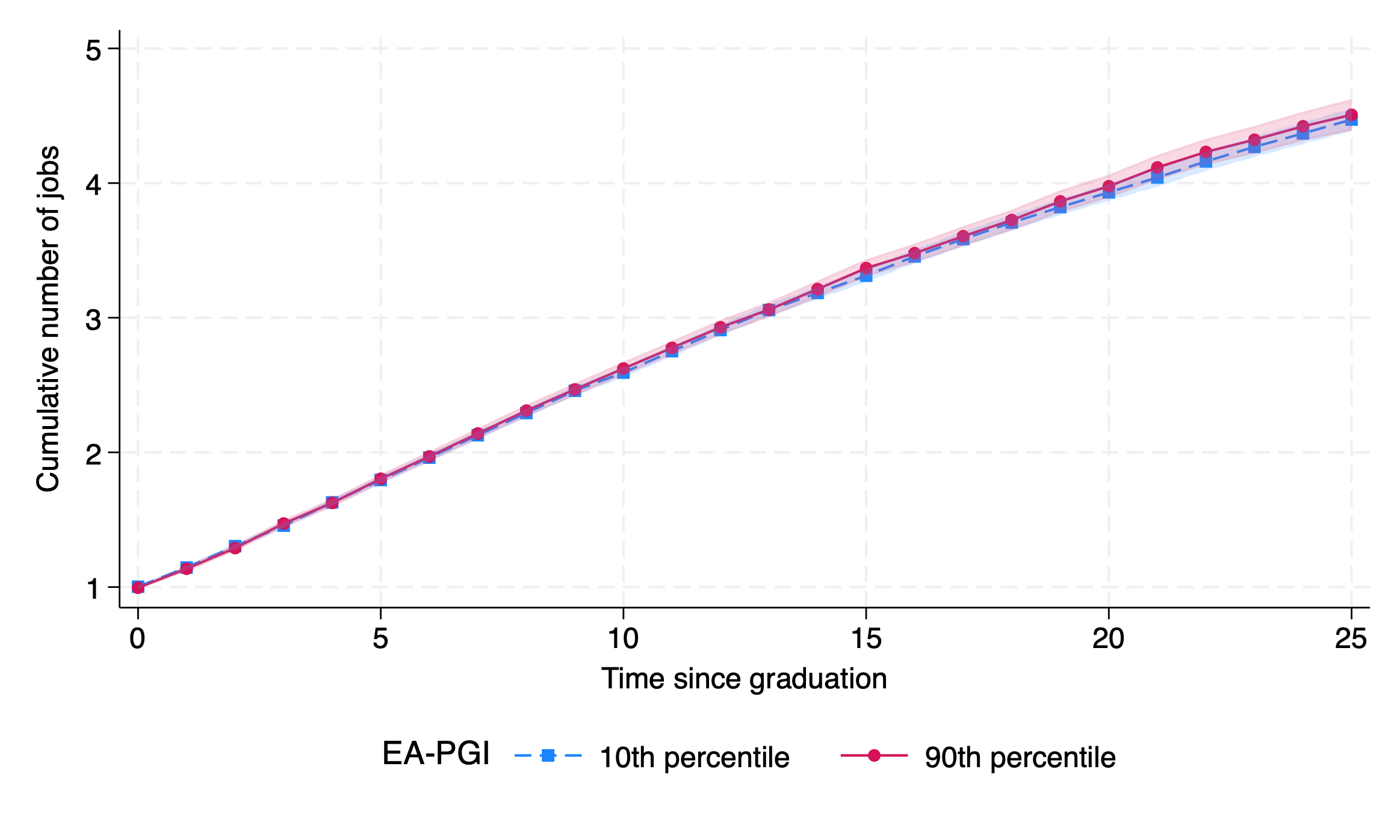}
    \label{subfig:njobs-time-sec}
\end{subfigure}
\begin{subfigure}{1\textwidth}
    \centering
    \subcaption{\textbf{Firm quality since labor market entry}}
    \includegraphics[width = 0.8\linewidth]{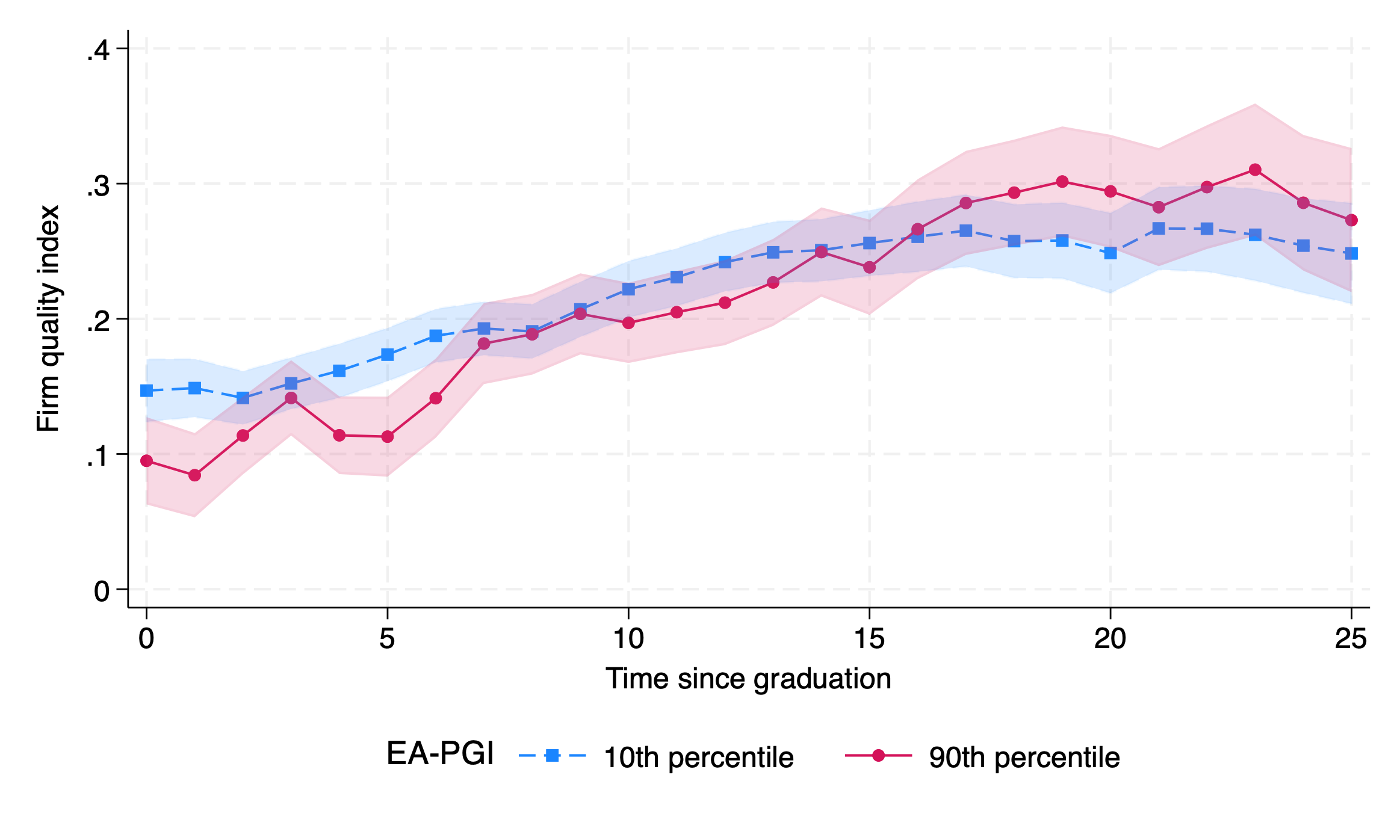}
    \label{subfig:akm-fe-time-sec}
\end{subfigure}
\end{figure}


\begin{figure}
    \centering
    \begin{subfigure}[b]{\linewidth}
        \centering
        \caption{\textbf{Cumulative contribution to log(earnings) growth}}
        \label{subfig:log-earn-growth-decomp-cumcontrib-detailed}
        \includegraphics[width = 0.9\linewidth]{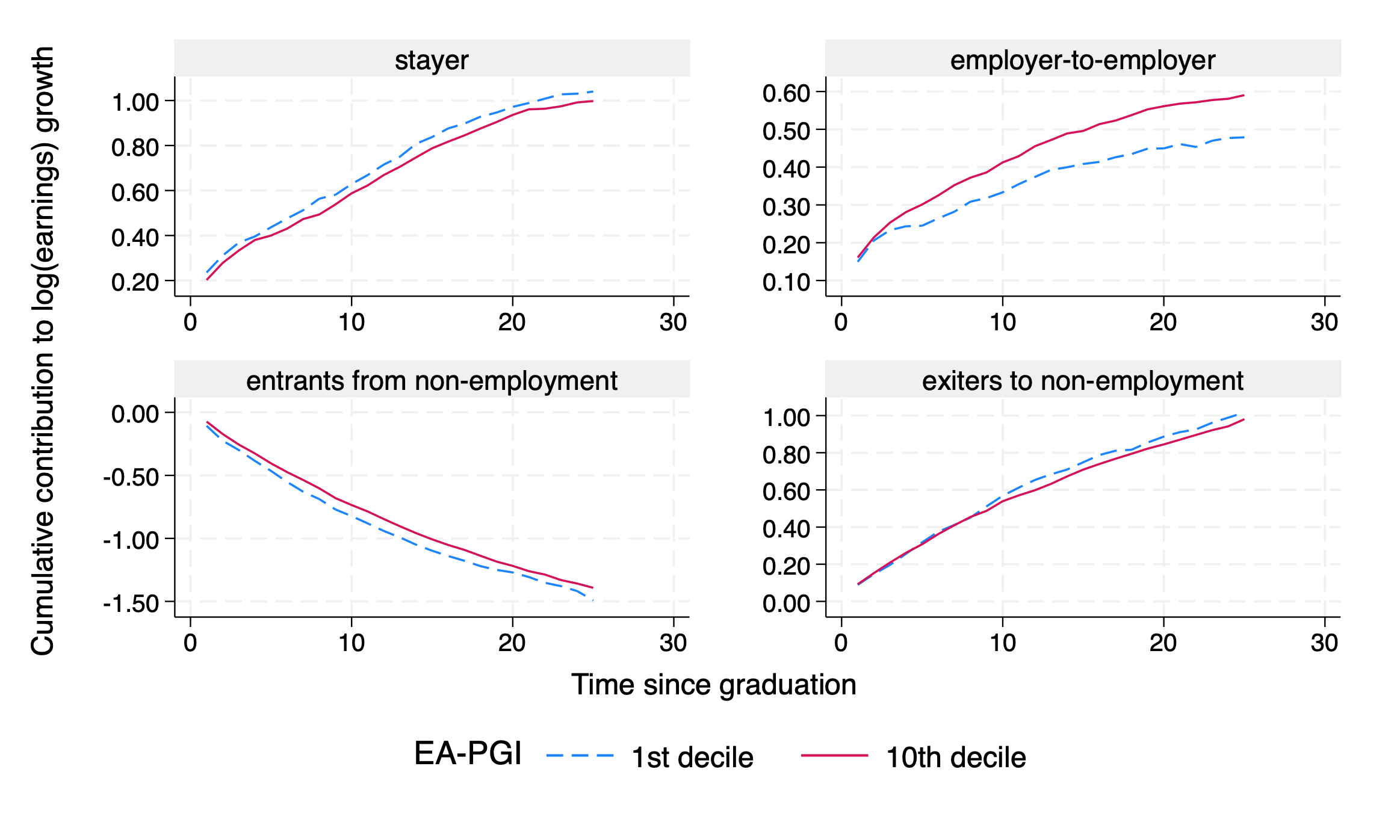}
    \end{subfigure}
    
    \begin{subfigure}[b]{\linewidth}
        \centering
        \caption{\textbf{Employment share}}
        \label{subfig:log-earn-growth-decomp-empshare}
        \includegraphics[width = 0.9\linewidth]{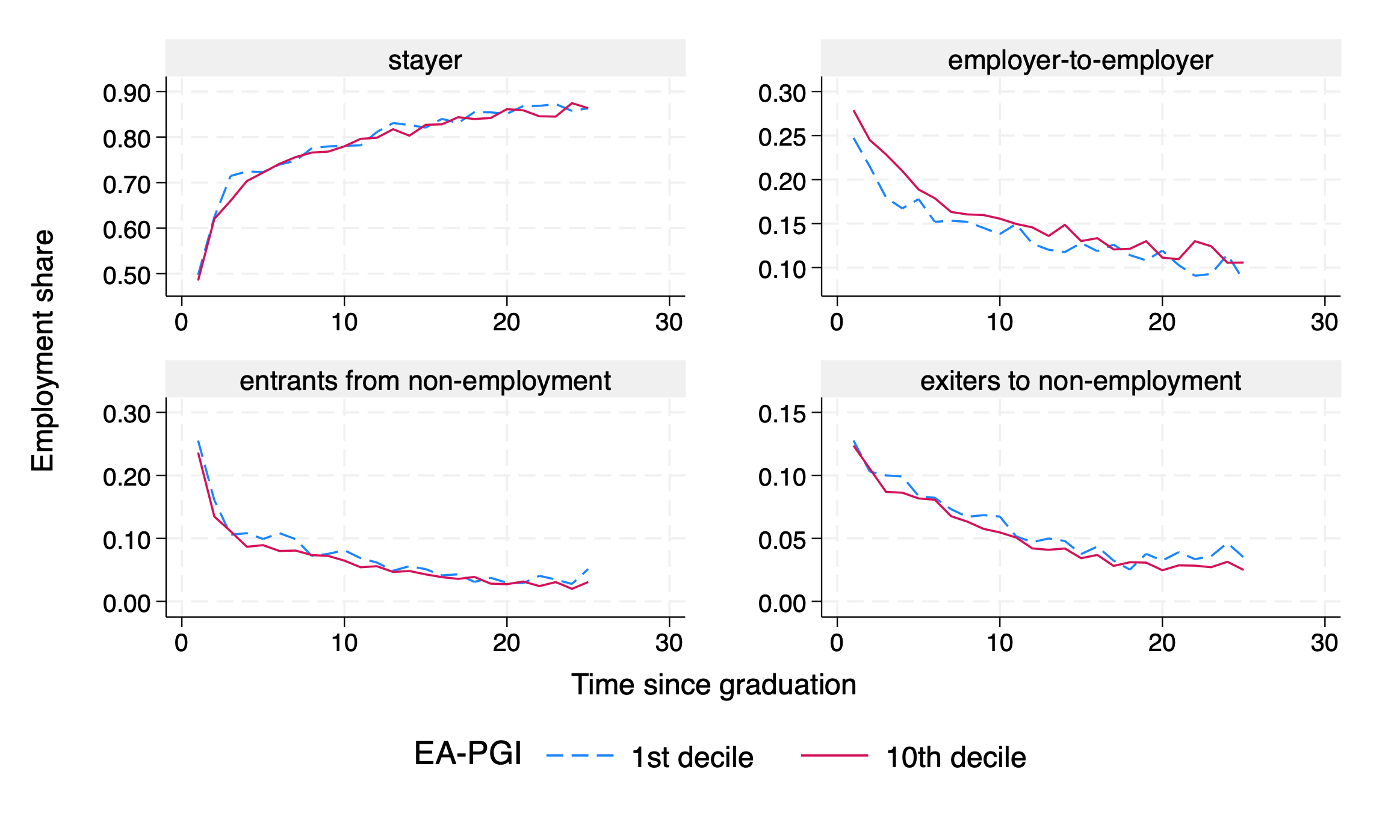}
    \end{subfigure}
    
    \caption{\textbf{Decomposition of cumulative growth in log annual income of tertiary-educated individuals by EA-PGI level, over time, including mobility into and out of non-employment}\\\footnotesize Panel \subref{subfig:log-earn-growth-decomp-cumcontrib-detailed} reports cumulative contributions of four mobility types to overall cumulative log earnings growth over time since graduation. Panel \subref{subfig:log-earn-growth-decomp-empshare} reports average employment shares of four mobility types over time since graduation. The blue line corresponds to 1st and the red - to 10th decile of EA-PGI distribution. The decomposition follows the algorithm in \cite{hahn2021}. The sample for the decomposition is restricted to tertiary-educated workers ($N =$ \num[round-precision=0]{\finalNindTer}).}
    \label{fig:log-earn-growth-decomp-detailed}
\end{figure}

\begin{figure}[htbp!]
    \centering
    \begin{subfigure}[b]{0.33\linewidth}
        \caption{\textbf{Pooled}}
        \includegraphics[width = \linewidth]{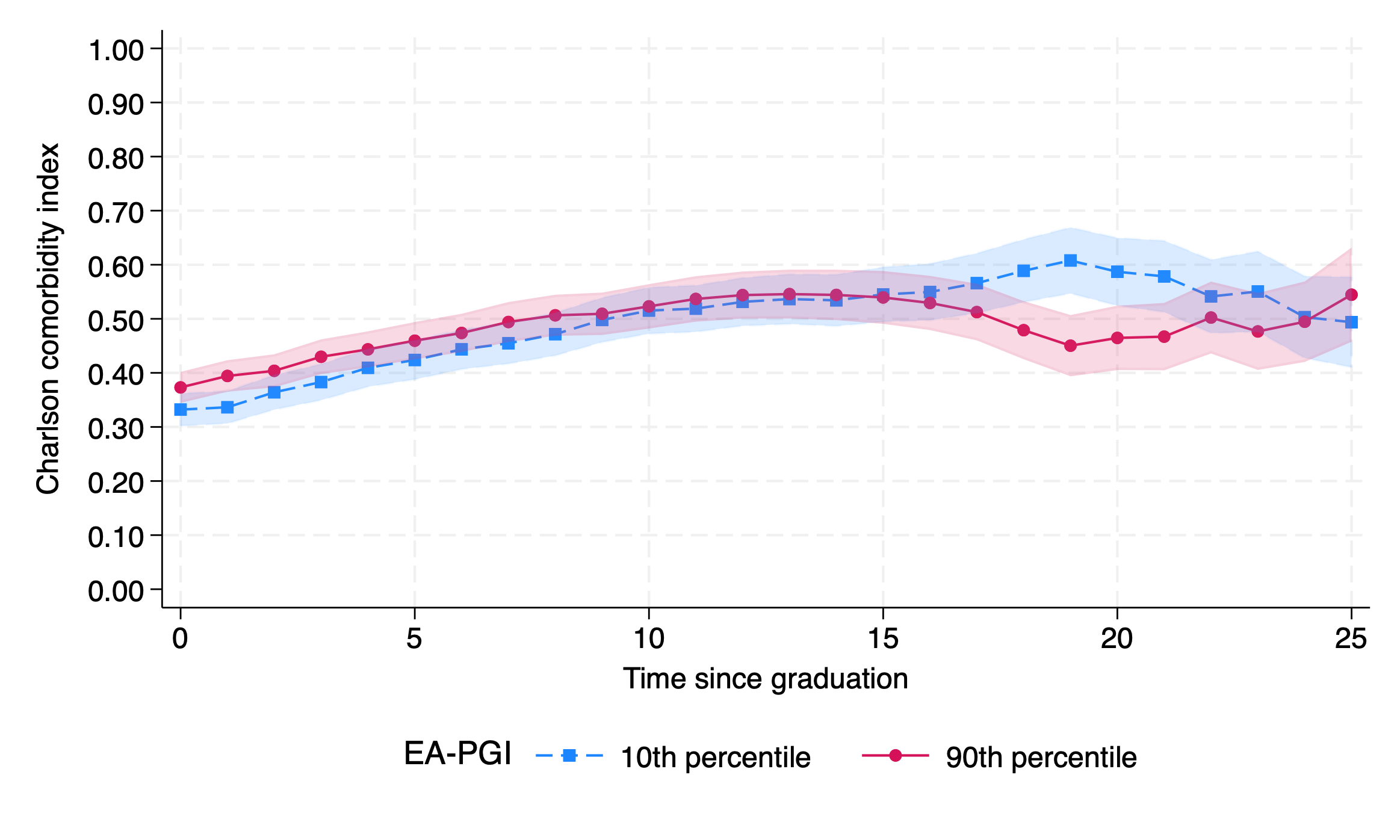}
        \label{subfig:predmarg-cci-pgiea-time-pld-famp}
    \end{subfigure}%
    \begin{subfigure}[b]{0.33\linewidth}
        \caption{\textbf{Secondary}}
        \includegraphics[width = \linewidth]{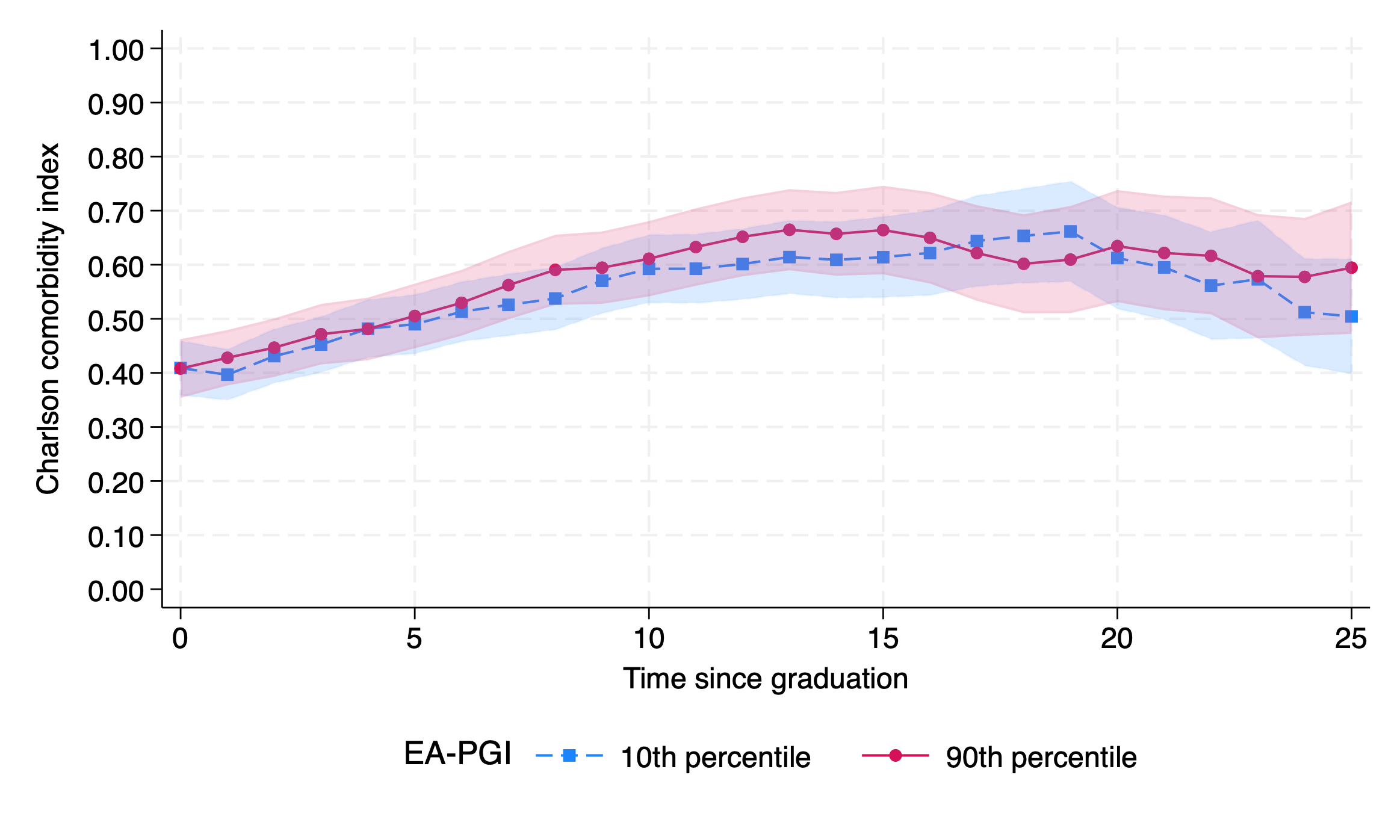}
        \label{subfig:predmarg-cci-pgiea-time-sec-famp}
    \end{subfigure}%
    \begin{subfigure}[b]{0.33\linewidth}
        \caption{\textbf{Tertiary}}
        \includegraphics[width = \linewidth]{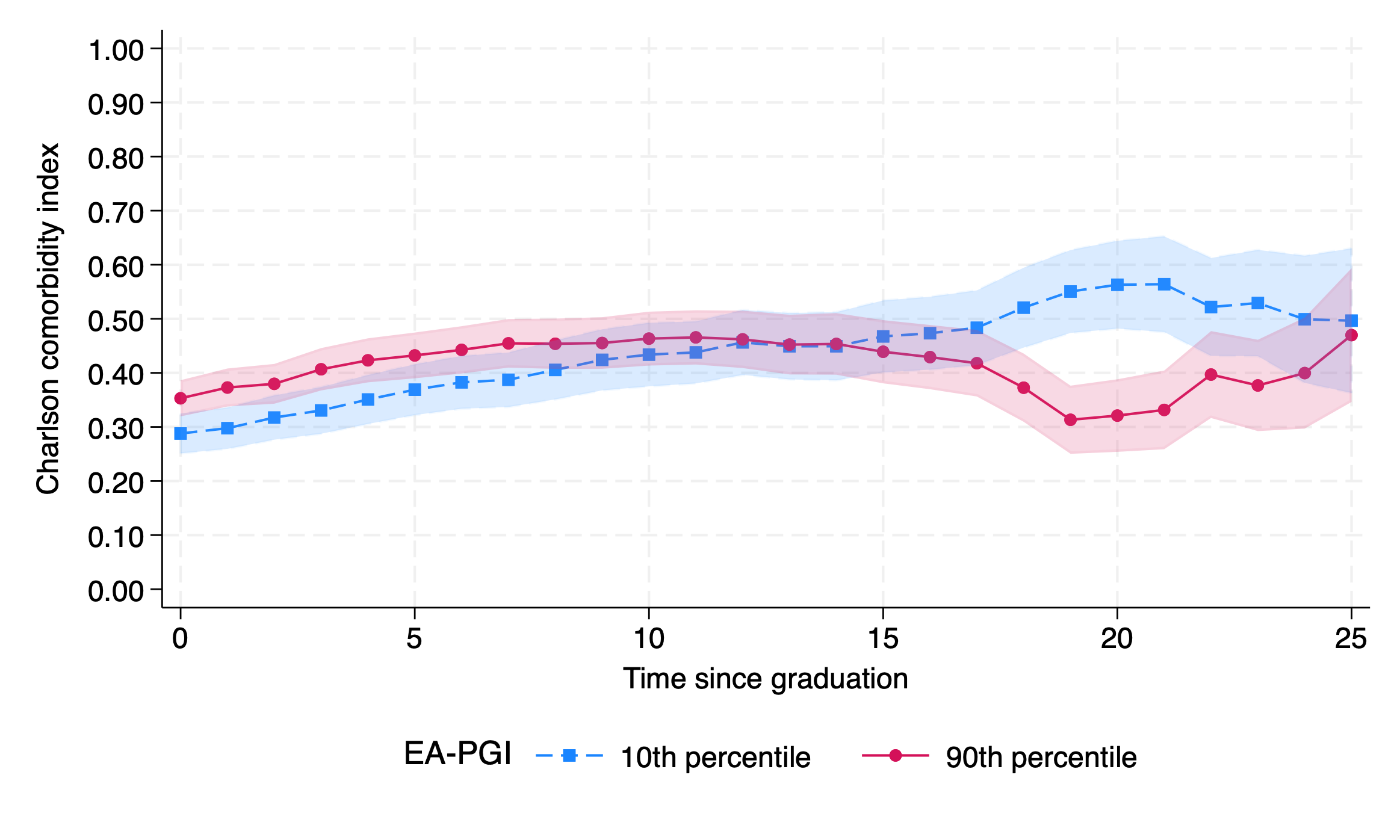}
        \label{subfig:predmarg-cci-pgiea-time-ter-famp}
    \end{subfigure}
    \caption{\textbf{Average health index by EA-PGI level, over time and by education, conditional on parental EA-PGI among parent-offspring trios}\\\footnotesize Panel \subref{subfig:predmarg-cci-pgiea-time-pld-famp} uses full sample of parent-offspring sample with non-missing Charlson Comorbidity Index ($N =$ \num[round-precision=0]{\predMargCCINPld}), while Panels \subref{subfig:predmarg-cci-pgiea-time-sec-famp} and \subref{subfig:predmarg-cci-pgiea-time-ter-famp} use subset of these workers based on their highest qualification being either secondary ($N =$ \num[round-precision=0]{\predMargCCINSec}) or tertiary degree ($N =$ \num[round-precision=0]{\predMargCCINTer}), respectively. \SentenceBlueRed\SentenceHealthRegFamp\SentenceCI{95}}
    \label{fig:predmarg-health-pgiea-age-byedu-famp}
\end{figure}

\begin{figure}[htbp!]
    \centering
    \begin{subfigure}[b]{0.33\linewidth}
        \caption{\textbf{Own EA-PGI}}
        \includegraphics[width = \linewidth]{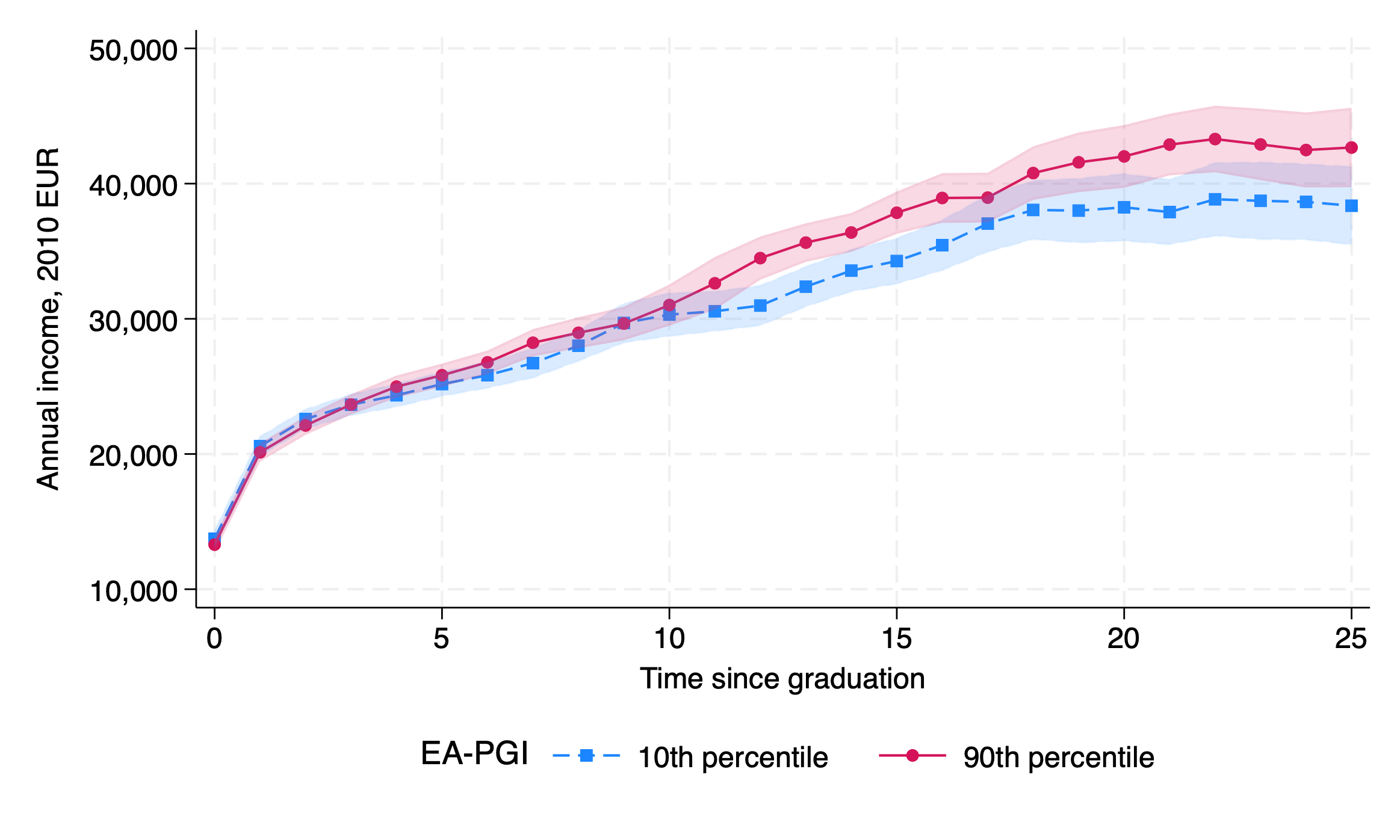}
        \label{subfig:predmarg-inc-pgiea-famp-ter-own}
    \end{subfigure}%
    \begin{subfigure}[b]{0.33\linewidth}
        \caption{\textbf{Father EA-PGI}}
        \includegraphics[width = \linewidth]{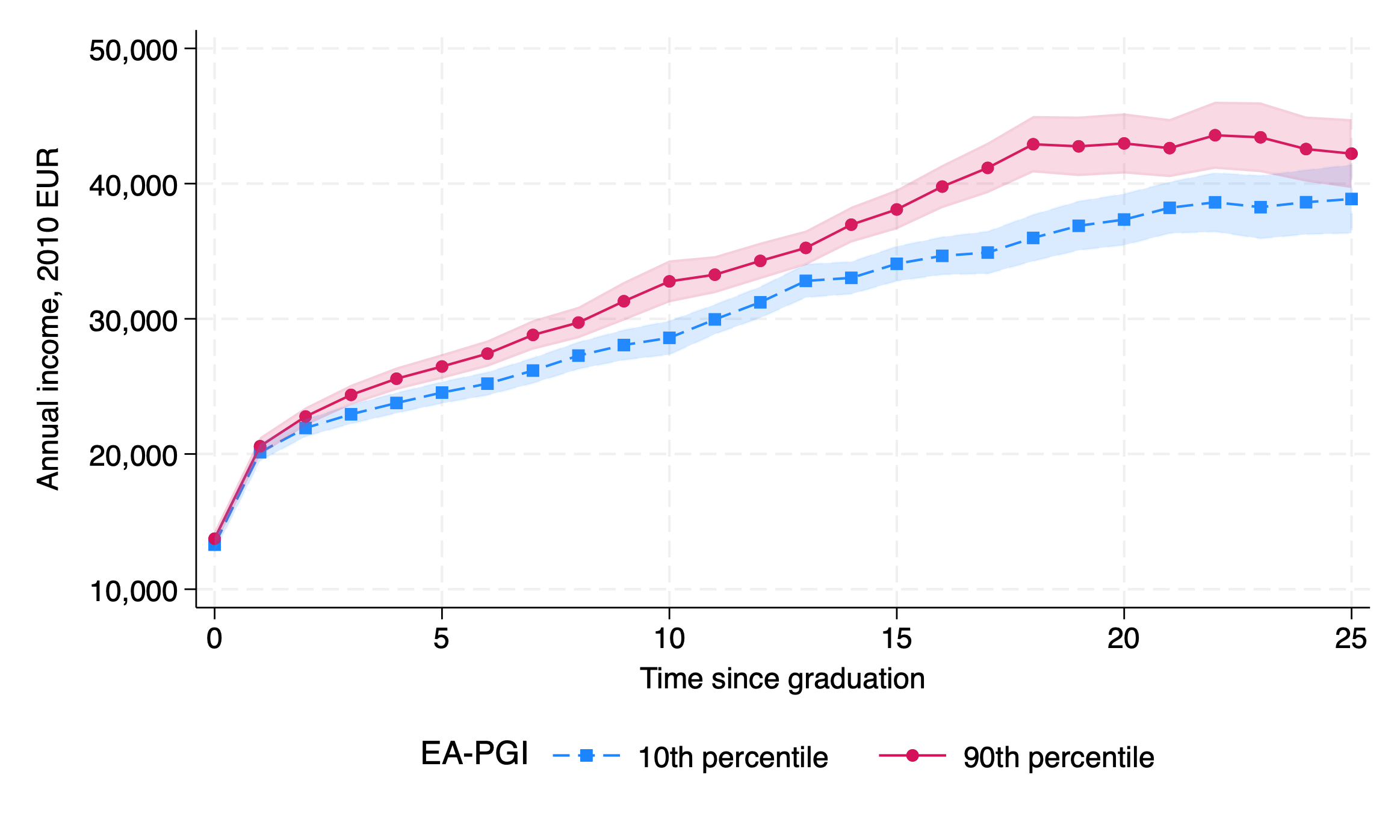}
        \label{subfig:predmarg-inc-pgiea-famp-ter-father}
    \end{subfigure}%
    \begin{subfigure}[b]{0.33\linewidth}
        \caption{\textbf{Mother EA-PGI}}
        \includegraphics[width = \linewidth]{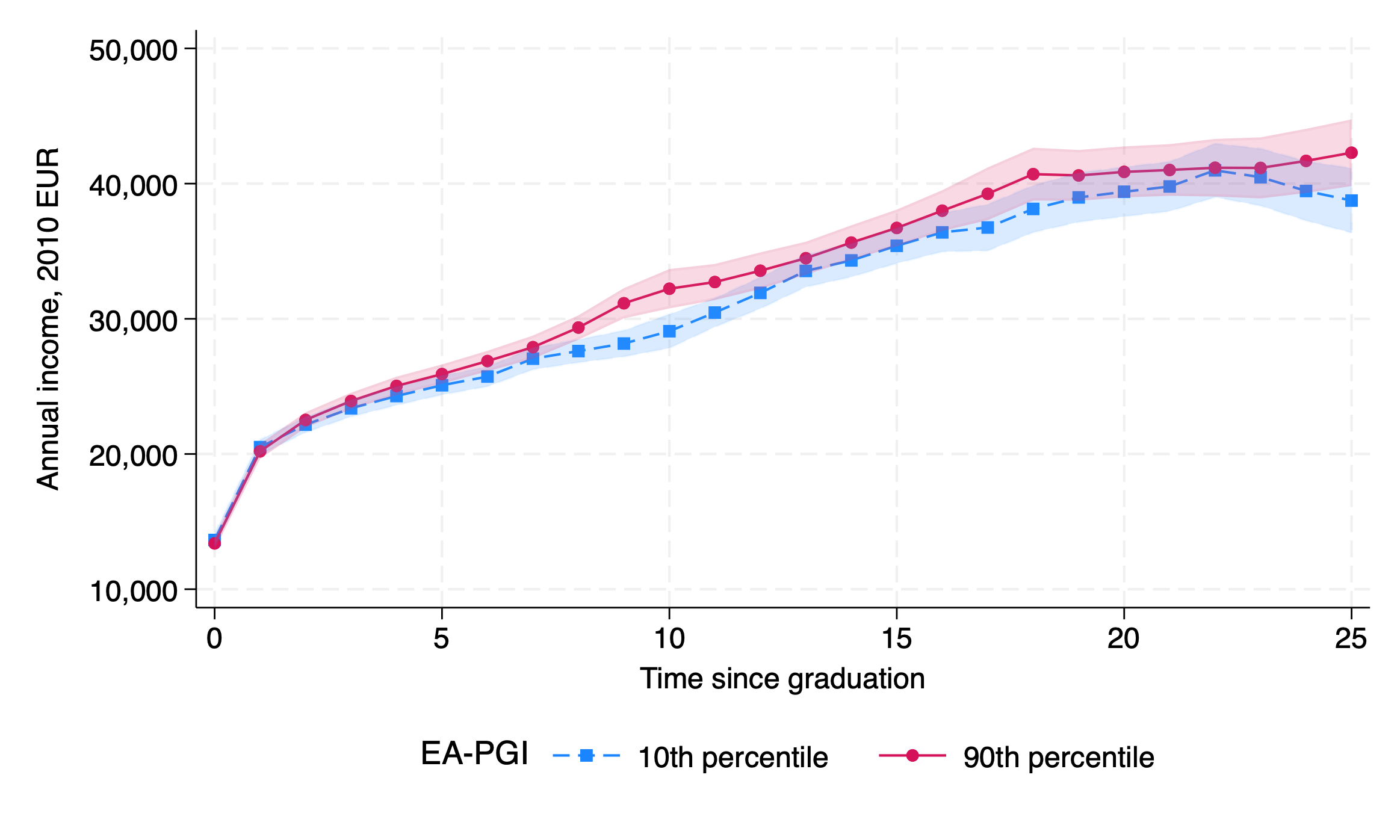}
        \label{subfig:predmarg-inc-pgiea-famp-ter-mother}
    \end{subfigure}
    \caption{\textbf{Average annual income of tertiary-educated individuals by own or parental EA-PGI levels, over time}\\\footnotesize The figure uses a subset of parent-offspring trios with highest offspring qualification being tertiary degree ($N= $ \num[round-precision=0]{\finalNindTrioTer}). The lines in Panel \subref{subfig:predmarg-inc-pgiea-famp-ter-own} correspond to 10th and 90th percentiles of offspring EA-PGI; in Panel \subref{subfig:predmarg-inc-pgiea-famp-ter-father} - 10th and 90th percentiles of paternal EA-PGI; and in Panel \subref{subfig:predmarg-inc-pgiea-famp-ter-mother} - 10th and 90th percentiles of maternal EA-PGI. \SentenceIncomeRegFamp\SentenceCI{95}}
    \label{fig:predmarg-inc-pgiea-famp-ter-own-par}
\end{figure}


\begin{figure}
    \centering
    \begin{subfigure}[b]{0.5\linewidth}
        \caption{\textbf{Worker productivity indices}}
        \includegraphics[width = \linewidth]{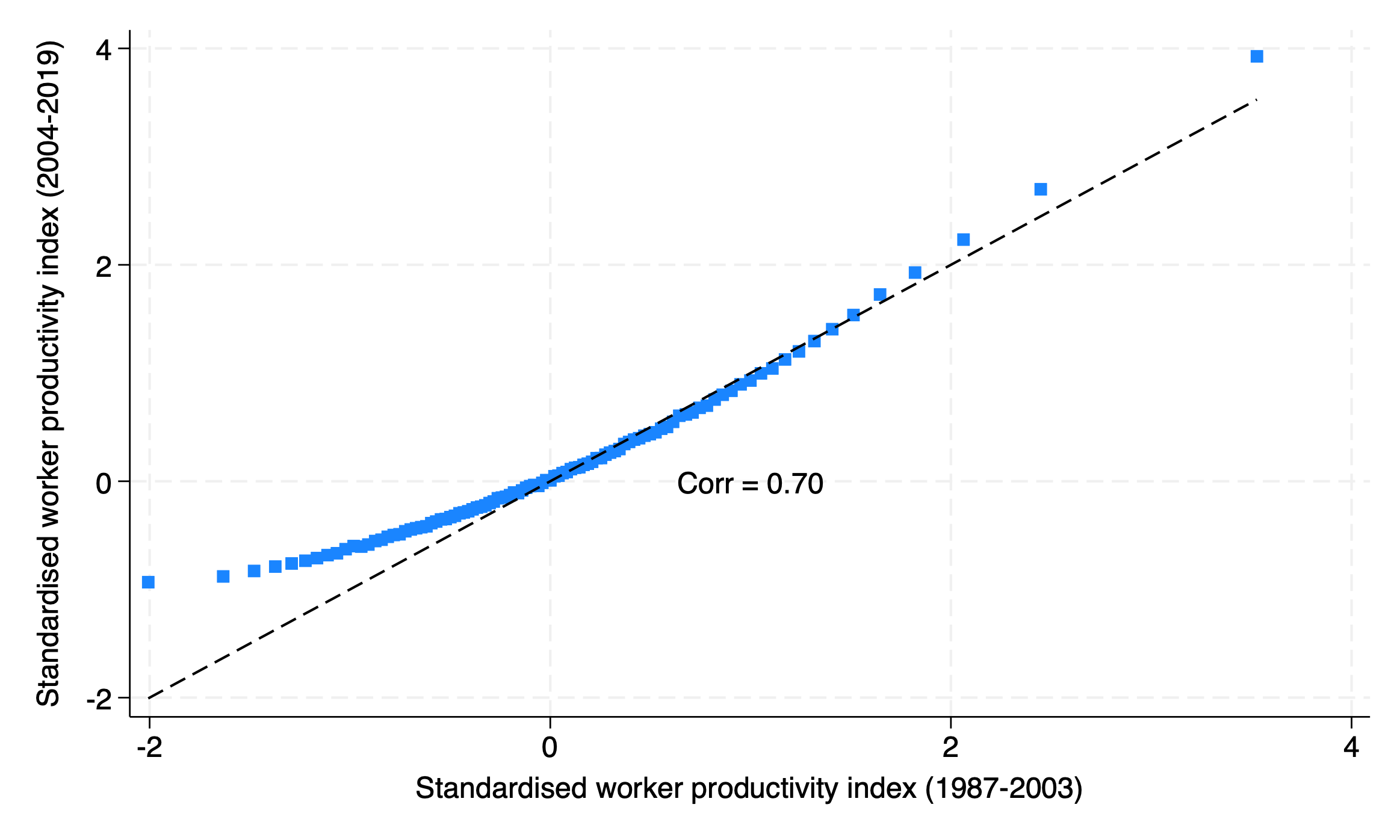}
        \label{subfig:akm-theta-8703-0419-binscatter}
    \end{subfigure}%
    \begin{subfigure}[b]{0.5\linewidth}
        \caption{\textbf{Firm quality indices}}
        \includegraphics[width = \linewidth]{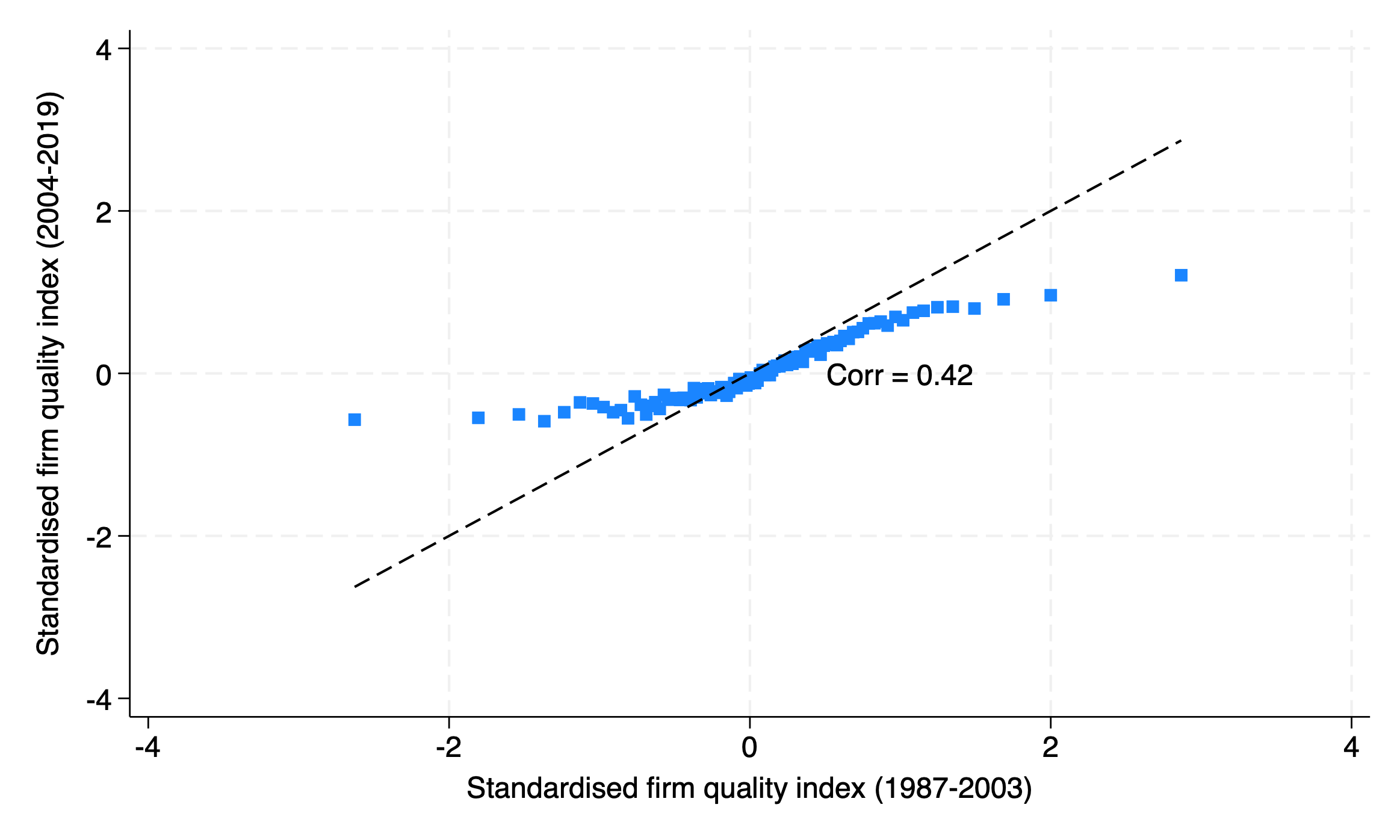}
        \label{subfig:akm-psi-8703-0419-binscatter}
    \end{subfigure}
    \caption{\textbf{AKM worker productivity and firm quality indices between two AKM estimation periods (1987-2003 and 2004-2019)}\\\footnotesize Panel \subref{subfig:akm-theta-8703-0419-binscatter} is a binscatter plot of worker productivity indices estimated using matched employer-employee data between 2004-2019 (on the y axis) and 1987-2003 (on the x axis). Panel \subref{subfig:akm-psi-8703-0419-binscatter} is a binscatter plot of firm quality indices estimated using matched employer-employee data between 2004-2019 (on the y axis) and 1987-2003 (on the x axis). The dashed black lines correspond to 45$^\circ$ line. The sample in Panel \subref{subfig:akm-theta-8703-0419-binscatter} are individuals with both worker productivity indices non-missing ($N = 13,715$); in Panel \subref{subfig:akm-psi-8703-0419-binscatter} - individuals with both firm quality indices non-missing ($N = 32,494$).}
    \label{fig:akm-8703-0419-binscatter}
\end{figure}

\clearpage
\newpage
\subsection{Supplementary Tables}\label{subsec:appendix_tbls}

\begin{table}[htbp!]
    \centering
    \scriptsize
    \begin{tabular}{lllllll}
\toprule
\multicolumn{1}{c}{} &
  \multicolumn{3}{c}{Person-year observations} &
  \multicolumn{3}{c}{Unique individuals} \\\cmidrule(lr){2-4}\cmidrule(lr){5-7}
\multicolumn{1}{c}{} &
  \multicolumn{1}{c}{All} &
  \multicolumn{1}{c}{THL} &
  \multicolumn{1}{c}{BDB} &
  \multicolumn{1}{c}{All} &
  \multicolumn{1}{c}{THL} &
  \multicolumn{1}{c}{BDB} \\
\multicolumn{1}{c}{} &
  \multicolumn{1}{c}{(1)} &
  \multicolumn{1}{c}{(2)} &
  \multicolumn{1}{c}{(3)} &
  \multicolumn{1}{c}{(4)} &
  \multicolumn{1}{c}{(5)} &
  \multicolumn{1}{c}{(6)} \\
\midrule
\multicolumn{1}{l}{Start} &
  \multicolumn{1}{r}{5,374,521} &
  \multicolumn{1}{r}{3,963,254} &
  \multicolumn{1}{r}{1,411,267} &
  \multicolumn{1}{r}{176,523} &
  \multicolumn{1}{r}{132,171} &
  \multicolumn{1}{r}{44,352} \\
\multicolumn{1}{l}{Keep individuals with non-missing genetic PCs} &
  \multicolumn{1}{r}{5,335,564} &
  \multicolumn{1}{r}{3,963,254} &
  \multicolumn{1}{r}{1,372,310} &
  \multicolumn{1}{r}{175,050} &
  \multicolumn{1}{r}{132,171} &
  \multicolumn{1}{r}{42,879} \\
\multicolumn{1}{l}{Keep graduates only} &
  \multicolumn{1}{r}{3,215,453} &
  \multicolumn{1}{r}{1,957,308} &
  \multicolumn{1}{r}{1,258,145} &
  \multicolumn{1}{r}{98,810} &
  \multicolumn{1}{r}{59,416} &
  \multicolumn{1}{r}{39,394} \\
\multicolumn{1}{l}{Keep graduates with non-missing graduation year} &
  \multicolumn{1}{r}{3,215,453} &
  \multicolumn{1}{r}{1,957,308} &
  \multicolumn{1}{r}{1,258,145} &
  \multicolumn{1}{r}{98,810} &
  \multicolumn{1}{r}{59,416} &
  \multicolumn{1}{r}{39,394} \\
\multicolumn{1}{l}{Graduated between 1970 and 2020} &
  \multicolumn{1}{r}{3,139,889} &
  \multicolumn{1}{r}{1,882,118} &
  \multicolumn{1}{r}{1,257,771} &
  \multicolumn{1}{r}{96,186} &
  \multicolumn{1}{r}{56,803} &
  \multicolumn{1}{r}{39,383} \\
\multicolumn{1}{l}{Observed between 0 and 25 years since graduation} &
  \multicolumn{1}{r}{1,692,473} &
  \multicolumn{1}{r}{1,065,113} &
  \multicolumn{1}{r}{627,360} &
  \multicolumn{1}{r}{96,166} &
  \multicolumn{1}{r}{56,786} &
  \multicolumn{1}{r}{39,380} \\
\multicolumn{1}{l}{Followed from 0 years since graduation} &
  \multicolumn{1}{r}{1,000,872} &
  \multicolumn{1}{r}{576,894} &
  \multicolumn{1}{r}{423,978} &
  \multicolumn{1}{r}{57,956} &
  \multicolumn{1}{r}{29,205} &
  \multicolumn{1}{r}{28,751} \\
\multicolumn{1}{l}{Followed at least up to age 30 (if secondary)} &
  \multicolumn{1}{r}{963,715} &
  \multicolumn{1}{r}{567,236} &
  \multicolumn{1}{r}{396,479} &
  \multicolumn{1}{r}{51,056} &
  \multicolumn{1}{r}{27,135} &
  \multicolumn{1}{r}{23,921} \\
\bottomrule
\end{tabular}

    \caption{\textbf{Working sample in trajectory analysis}\\\footnotesize The table reports sample counts by working sample restrictions. Columns (1)-(3) report person-year observation counts in full sample and by biobanks, respectively. Columns (4)-(6) report unique individual counts in full sample and by biobanks.}
    \label{tab:trajectory-restrictions-samplecounts}
\end{table}

\begin{table}[htbp!]
    \centering
    \begin{subtable}[t]{\linewidth}
    \caption{\textbf{Genotyped sample}}
    \label{subtab:descriptives-genotyped}
    \resizebox{\linewidth}{!}{%
    \begin{threeparttable}
        \begin{tabular}{llllllll}
\toprule
\multicolumn{1}{c}{} &
  \multicolumn{1}{c}{Population} &
  \multicolumn{1}{c}{Genotyped sample} &
  \multicolumn{1}{c}{Reweighted sample} &
  \multicolumn{1}{c}{$\Delta^{(1)}_{(2)}$} &
  \multicolumn{1}{c}{$\Delta^{(1)}_{(3)}$} &
  \multicolumn{1}{c}{$ N^{(1)}$} &
  \multicolumn{1}{c}{$ N^{(2)}$} \\
\multicolumn{1}{c}{} &
  \multicolumn{1}{c}{(1)} &
  \multicolumn{1}{c}{(2)} &
  \multicolumn{1}{c}{(3)} &
  \multicolumn{1}{c}{(4)} &
  \multicolumn{1}{c}{(5)} &
  \multicolumn{1}{c}{(6)} &
  \multicolumn{1}{c}{(7)} \\
\midrule
\multicolumn{1}{l}{Cohort: 1950-59} &
  \multicolumn{1}{r}{0.01} &
  \multicolumn{1}{r}{0.01} &
  \multicolumn{1}{r}{0.01} &
  \multicolumn{1}{r}{0.000} &
  \multicolumn{1}{r}{1.000} &
  \multicolumn{1}{r}{1,599,332} &
  \multicolumn{1}{r}{51,056} \\
\multicolumn{1}{l}{Cohort: 1960-69} &
  \multicolumn{1}{r}{0.17} &
  \multicolumn{1}{r}{0.22} &
  \multicolumn{1}{r}{0.16} &
  \multicolumn{1}{r}{0.000} &
  \multicolumn{1}{r}{1.000} &
  \multicolumn{1}{r}{1,599,332} &
  \multicolumn{1}{r}{51,056} \\
\multicolumn{1}{l}{Cohort: 1970-79} &
  \multicolumn{1}{r}{0.34} &
  \multicolumn{1}{r}{0.36} &
  \multicolumn{1}{r}{0.34} &
  \multicolumn{1}{r}{0.000} &
  \multicolumn{1}{r}{1.000} &
  \multicolumn{1}{r}{1,599,332} &
  \multicolumn{1}{r}{51,056} \\
\multicolumn{1}{l}{Cohort: 1980-89} &
  \multicolumn{1}{r}{0.36} &
  \multicolumn{1}{r}{0.29} &
  \multicolumn{1}{r}{0.37} &
  \multicolumn{1}{r}{0.000} &
  \multicolumn{1}{r}{1.000} &
  \multicolumn{1}{r}{1,599,332} &
  \multicolumn{1}{r}{51,056} \\
\multicolumn{1}{l}{Cohort: 1990-99} &
  \multicolumn{1}{r}{0.13} &
  \multicolumn{1}{r}{0.11} &
  \multicolumn{1}{r}{0.13} &
  \multicolumn{1}{r}{0.000} &
  \multicolumn{1}{r}{1.000} &
  \multicolumn{1}{r}{1,599,332} &
  \multicolumn{1}{r}{51,056} \\
\multicolumn{1}{l}{Graduation age: 16-20} &
  \multicolumn{1}{r}{0.36} &
  \multicolumn{1}{r}{0.31} &
  \multicolumn{1}{r}{0.36} &
  \multicolumn{1}{r}{0.000} &
  \multicolumn{1}{r}{1.000} &
  \multicolumn{1}{r}{1,599,332} &
  \multicolumn{1}{r}{51,056} \\
\multicolumn{1}{l}{Graduation age: 21-25} &
  \multicolumn{1}{r}{0.39} &
  \multicolumn{1}{r}{0.43} &
  \multicolumn{1}{r}{0.39} &
  \multicolumn{1}{r}{0.000} &
  \multicolumn{1}{r}{1.000} &
  \multicolumn{1}{r}{1,599,332} &
  \multicolumn{1}{r}{51,056} \\
\multicolumn{1}{l}{Graduation age: 26-30} &
  \multicolumn{1}{r}{0.25} &
  \multicolumn{1}{r}{0.25} &
  \multicolumn{1}{r}{0.24} &
  \multicolumn{1}{r}{0.001} &
  \multicolumn{1}{r}{1.000} &
  \multicolumn{1}{r}{1,599,332} &
  \multicolumn{1}{r}{51,056} \\
\multicolumn{1}{l}{Education: secondary} &
  \multicolumn{1}{r}{0.44} &
  \multicolumn{1}{r}{0.37} &
  \multicolumn{1}{r}{0.44} &
  \multicolumn{1}{r}{0.000} &
  \multicolumn{1}{r}{1.000} &
  \multicolumn{1}{r}{1,599,332} &
  \multicolumn{1}{r}{51,056} \\
\multicolumn{1}{l}{Education: tertiary} &
  \multicolumn{1}{r}{0.56} &
  \multicolumn{1}{r}{0.63} &
  \multicolumn{1}{r}{0.56} &
  \multicolumn{1}{r}{0.000} &
  \multicolumn{1}{r}{1.000} &
  \multicolumn{1}{r}{1,599,332} &
  \multicolumn{1}{r}{51,056} \\
\multicolumn{1}{l}{Male} &
  \multicolumn{1}{r}{0.48} &
  \multicolumn{1}{r}{0.39} &
  \multicolumn{1}{r}{0.48} &
  \multicolumn{1}{r}{0.000} &
  \multicolumn{1}{r}{1.000} &
  \multicolumn{1}{r}{1,599,332} &
  \multicolumn{1}{r}{51,056} \\
\multicolumn{1}{l}{Married} &
  \multicolumn{1}{r}{0.10} &
  \multicolumn{1}{r}{0.13} &
  \multicolumn{1}{r}{0.11} &
  \multicolumn{1}{r}{0.000} &
  \multicolumn{1}{r}{0.000} &
  \multicolumn{1}{r}{1,599,332} &
  \multicolumn{1}{r}{51,056} \\
\multicolumn{1}{l}{Rural} &
  \multicolumn{1}{r}{0.24} &
  \multicolumn{1}{r}{0.25} &
  \multicolumn{1}{r}{0.25} &
  \multicolumn{1}{r}{0.712} &
  \multicolumn{1}{r}{1.000} &
  \multicolumn{1}{r}{1,599,332} &
  \multicolumn{1}{r}{51,056} \\
\multicolumn{1}{l}{Cumulated income} &
  \multicolumn{1}{r}{279,534} &
  \multicolumn{1}{r}{302,077} &
  \multicolumn{1}{r}{277,373} &
  \multicolumn{1}{r}{0.000} &
  \multicolumn{1}{r}{0.692} &
  \multicolumn{1}{r}{1,599,332} &
  \multicolumn{1}{r}{51,056} \\
\multicolumn{1}{l}{Income at t=0} &
  \multicolumn{1}{r}{9,301} &
  \multicolumn{1}{r}{9,527} &
  \multicolumn{1}{r}{9,328} &
  \multicolumn{1}{r}{0.000} &
  \multicolumn{1}{r}{1.000} &
  \multicolumn{1}{r}{1,599,332} &
  \multicolumn{1}{r}{51,056} \\
\bottomrule
\end{tabular}

    \end{threeparttable}}
    \end{subtable}

    \vspace{12pt}
    
    \begin{subtable}[t]{\linewidth}
    \caption{\textbf{Family trio sample}}
    \label{subtab:descriptives-trio}
    \resizebox{\linewidth}{!}{%
    \begin{threeparttable}
        \begin{tabular}{llllllll}
\toprule
\multicolumn{1}{c}{} &
  \multicolumn{1}{c}{Population} &
  \multicolumn{1}{c}{Genotyped family trios} &
  \multicolumn{1}{c}{Reweighted family trios} &
  \multicolumn{1}{c}{$\Delta^{(1)}_{(2)}$} &
  \multicolumn{1}{c}{$\Delta^{(1)}_{(3)}$} &
  \multicolumn{1}{c}{$ N^{(1)}$} &
  \multicolumn{1}{c}{$ N^{(2)}$} \\
\multicolumn{1}{c}{} &
  \multicolumn{1}{c}{(1)} &
  \multicolumn{1}{c}{(2)} &
  \multicolumn{1}{c}{(3)} &
  \multicolumn{1}{c}{(4)} &
  \multicolumn{1}{c}{(5)} &
  \multicolumn{1}{c}{(6)} &
  \multicolumn{1}{c}{(7)} \\
\midrule
\multicolumn{1}{l}{Cohort: 1950-59} &
  \multicolumn{1}{r}{0.01} &
  \multicolumn{1}{r}{0.01} &
  \multicolumn{1}{r}{0.01} &
  \multicolumn{1}{r}{0.446} &
  \multicolumn{1}{r}{1.000} &
  \multicolumn{1}{r}{1,599,332} &
  \multicolumn{1}{r}{12,871} \\
\multicolumn{1}{l}{Cohort: 1960-69} &
  \multicolumn{1}{r}{0.17} &
  \multicolumn{1}{r}{0.15} &
  \multicolumn{1}{r}{0.17} &
  \multicolumn{1}{r}{0.001} &
  \multicolumn{1}{r}{1.000} &
  \multicolumn{1}{r}{1,599,332} &
  \multicolumn{1}{r}{12,871} \\
\multicolumn{1}{l}{Cohort: 1970-79} &
  \multicolumn{1}{r}{0.34} &
  \multicolumn{1}{r}{0.35} &
  \multicolumn{1}{r}{0.34} &
  \multicolumn{1}{r}{0.017} &
  \multicolumn{1}{r}{1.000} &
  \multicolumn{1}{r}{1,599,332} &
  \multicolumn{1}{r}{12,871} \\
\multicolumn{1}{l}{Cohort: 1980-89} &
  \multicolumn{1}{r}{0.36} &
  \multicolumn{1}{r}{0.38} &
  \multicolumn{1}{r}{0.36} &
  \multicolumn{1}{r}{0.000} &
  \multicolumn{1}{r}{1.000} &
  \multicolumn{1}{r}{1,599,332} &
  \multicolumn{1}{r}{12,871} \\
\multicolumn{1}{l}{Cohort: 1990-99} &
  \multicolumn{1}{r}{0.13} &
  \multicolumn{1}{r}{0.11} &
  \multicolumn{1}{r}{0.12} &
  \multicolumn{1}{r}{0.000} &
  \multicolumn{1}{r}{1.000} &
  \multicolumn{1}{r}{1,599,332} &
  \multicolumn{1}{r}{12,871} \\
\multicolumn{1}{l}{Graduation age: 16-20} &
  \multicolumn{1}{r}{0.36} &
  \multicolumn{1}{r}{0.34} &
  \multicolumn{1}{r}{0.36} &
  \multicolumn{1}{r}{0.000} &
  \multicolumn{1}{r}{1.000} &
  \multicolumn{1}{r}{1,599,332} &
  \multicolumn{1}{r}{12,871} \\
\multicolumn{1}{l}{Graduation age: 21-25} &
  \multicolumn{1}{r}{0.39} &
  \multicolumn{1}{r}{0.42} &
  \multicolumn{1}{r}{0.40} &
  \multicolumn{1}{r}{0.000} &
  \multicolumn{1}{r}{1.000} &
  \multicolumn{1}{r}{1,599,332} &
  \multicolumn{1}{r}{12,871} \\
\multicolumn{1}{l}{Graduation age: 26-30} &
  \multicolumn{1}{r}{0.25} &
  \multicolumn{1}{r}{0.24} &
  \multicolumn{1}{r}{0.24} &
  \multicolumn{1}{r}{0.086} &
  \multicolumn{1}{r}{1.000} &
  \multicolumn{1}{r}{1,599,332} &
  \multicolumn{1}{r}{12,871} \\
\multicolumn{1}{l}{Education: secondary} &
  \multicolumn{1}{r}{0.44} &
  \multicolumn{1}{r}{0.39} &
  \multicolumn{1}{r}{0.44} &
  \multicolumn{1}{r}{0.000} &
  \multicolumn{1}{r}{1.000} &
  \multicolumn{1}{r}{1,599,332} &
  \multicolumn{1}{r}{12,871} \\
\multicolumn{1}{l}{Education: tertiary} &
  \multicolumn{1}{r}{0.56} &
  \multicolumn{1}{r}{0.61} &
  \multicolumn{1}{r}{0.56} &
  \multicolumn{1}{r}{0.000} &
  \multicolumn{1}{r}{1.000} &
  \multicolumn{1}{r}{1,599,332} &
  \multicolumn{1}{r}{12,871} \\
\multicolumn{1}{l}{Male} &
  \multicolumn{1}{r}{0.48} &
  \multicolumn{1}{r}{0.40} &
  \multicolumn{1}{r}{0.48} &
  \multicolumn{1}{r}{0.000} &
  \multicolumn{1}{r}{1.000} &
  \multicolumn{1}{r}{1,599,332} &
  \multicolumn{1}{r}{12,871} \\
\multicolumn{1}{l}{Married} &
  \multicolumn{1}{r}{0.10} &
  \multicolumn{1}{r}{0.12} &
  \multicolumn{1}{r}{0.12} &
  \multicolumn{1}{r}{0.000} &
  \multicolumn{1}{r}{1.000} &
  \multicolumn{1}{r}{1,599,332} &
  \multicolumn{1}{r}{12,871} \\
\multicolumn{1}{l}{Rural} &
  \multicolumn{1}{r}{0.24} &
  \multicolumn{1}{r}{0.27} &
  \multicolumn{1}{r}{0.24} &
  \multicolumn{1}{r}{0.000} &
  \multicolumn{1}{r}{1.000} &
  \multicolumn{1}{r}{1,599,332} &
  \multicolumn{1}{r}{12,871} \\
\multicolumn{1}{l}{Cumulated income} &
  \multicolumn{1}{r}{279,534} &
  \multicolumn{1}{r}{282,107} &
  \multicolumn{1}{r}{283,464} &
  \multicolumn{1}{r}{0.446} &
  \multicolumn{1}{r}{1.000} &
  \multicolumn{1}{r}{1,599,332} &
  \multicolumn{1}{r}{12,871} \\
\multicolumn{1}{l}{Income at t=0} &
  \multicolumn{1}{r}{9,301} &
  \multicolumn{1}{r}{9,782} &
  \multicolumn{1}{r}{9,557} &
  \multicolumn{1}{r}{0.000} &
  \multicolumn{1}{r}{1.000} &
  \multicolumn{1}{r}{1,599,332} &
  \multicolumn{1}{r}{12,871} \\
\bottomrule
\end{tabular}

    \end{threeparttable}}
    \end{subtable}
    \caption{\textbf{Average population and sample characteristics of graduates}\\\footnotesize Panel \subref{subtab:descriptives-genotyped} compares full analysis sample ($N =$ \num[round-precision=0]{\finalNindAll}) and Panel \subref{subtab:descriptives-trio} compares subsample of parent-offspring trios ($N=$ \num[round-precision=0]{\finalNindTrioPld}) to the population of fresh graduates. The population of fresh graduates is selected from the full Finnish population with same criteria described in Table \ref{tab:trajectory-restrictions-samplecounts}, except for conditioning on genotypes and genetic principal components. Column (1) reports average characteristics in the population of fresh graduates. Columns (2) and (3) report average characteristics in the corresponding sample, before and after applying inverse probability weights, respectively. Columns (4) and (5) report \textit{p}-values for the equality of means between population and working sample, before and after applying inverse probability weights, respectively. All \textit{p}-values are adjusted for multiple hypotheses testing using Holm correction. Columns (6) and (7) report population and sample counts, respectively. Inverse probability weights are estimated to balance the sample according to year of birth and graduation year (fully interacted with highest education level and gender), and rural area indicator fully interacted with gender.}
    \label{tab:descriptives}
\end{table}

\begin{table}[htbp!]
    \centering
    \caption{\textbf{Descriptive statistics in AKM sample}\\\footnotesize The table reports descriptive statistics in full matched employee-employer panel used in AKM estimation. Column (1) reports sample mean, column (2) - standard deviation, and column (3) - person-year observation counts for the variables in rows.}
    \label{tab:sumstats-akm-gtp}
    \begin{threeparttable}
        \begin{tabular}{llll}
\toprule
\multicolumn{1}{c}{} &
  \multicolumn{1}{c}{Mean} &
  \multicolumn{1}{c}{SD} &
  \multicolumn{1}{c}{N} \\
\multicolumn{1}{c}{} &
  \multicolumn{1}{c}{(1)} &
  \multicolumn{1}{c}{(2)} &
  \multicolumn{1}{c}{(3)} \\
\midrule
\multicolumn{1}{l}{Cohort: 1946-1955} &
  \multicolumn{1}{r}{0.231} &
  \multicolumn{1}{r}{0.421} &
  \multicolumn{1}{r}{7,225,843} \\
\multicolumn{1}{l}{Cohort: 1956-1965} &
  \multicolumn{1}{r}{0.299} &
  \multicolumn{1}{r}{0.458} &
  \multicolumn{1}{r}{9,379,633} \\
\multicolumn{1}{l}{Cohort: 1966-1975} &
  \multicolumn{1}{r}{0.240} &
  \multicolumn{1}{r}{0.427} &
  \multicolumn{1}{r}{7,504,156} \\
\multicolumn{1}{l}{Cohort: 1976-1985} &
  \multicolumn{1}{r}{0.162} &
  \multicolumn{1}{r}{0.369} &
  \multicolumn{1}{r}{5,086,257} \\
\multicolumn{1}{l}{Cohort: 1986-1995} &
  \multicolumn{1}{r}{0.064} &
  \multicolumn{1}{r}{0.245} &
  \multicolumn{1}{r}{2,010,473} \\
\multicolumn{1}{l}{Cohort: 1996-2005} &
  \multicolumn{1}{r}{0.004} &
  \multicolumn{1}{r}{0.062} &
  \multicolumn{1}{r}{121,624} \\
\multicolumn{1}{l}{Age group: 20-29} &
  \multicolumn{1}{r}{0.200} &
  \multicolumn{1}{r}{0.400} &
  \multicolumn{1}{r}{6,875,154} \\
\multicolumn{1}{l}{Age group: 30-39} &
  \multicolumn{1}{r}{0.288} &
  \multicolumn{1}{r}{0.453} &
  \multicolumn{1}{r}{9,929,190} \\
\multicolumn{1}{l}{Age group: 40-49} &
  \multicolumn{1}{r}{0.288} &
  \multicolumn{1}{r}{0.453} &
  \multicolumn{1}{r}{9,926,283} \\
\multicolumn{1}{l}{Age group: 50-59} &
  \multicolumn{1}{r}{0.212} &
  \multicolumn{1}{r}{0.409} &
  \multicolumn{1}{r}{7,306,357} \\
\multicolumn{1}{l}{Age group: 60-69} &
  \multicolumn{1}{r}{0.011} &
  \multicolumn{1}{r}{0.106} &
  \multicolumn{1}{r}{391,987} \\
\multicolumn{1}{l}{Male} &
  \multicolumn{1}{r}{0.604} &
  \multicolumn{1}{r}{0.489} &
  \multicolumn{1}{r}{34,428,971} \\
\multicolumn{1}{l}{Education level: Compulsory} &
  \multicolumn{1}{r}{0.212} &
  \multicolumn{1}{r}{0.409} &
  \multicolumn{1}{r}{7,314,987} \\
\multicolumn{1}{l}{Education level: Secondary} &
  \multicolumn{1}{r}{0.447} &
  \multicolumn{1}{r}{0.497} &
  \multicolumn{1}{r}{15,406,896} \\
\multicolumn{1}{l}{Education level: Tertiary} &
  \multicolumn{1}{r}{0.340} &
  \multicolumn{1}{r}{0.474} &
  \multicolumn{1}{r}{11,707,088} \\
\multicolumn{1}{l}{Firm size} &
  \multicolumn{1}{r}{1,690} &
  \multicolumn{1}{r}{4,283} &
  \multicolumn{1}{r}{34,428,971} \\
\multicolumn{1}{l}{Annual total earning, 2015 EUR} &
  \multicolumn{1}{r}{27,319} &
  \multicolumn{1}{r}{30,515} &
  \multicolumn{1}{r}{34,428,936} \\
\multicolumn{1}{l}{Months worked in a year} &
  \multicolumn{1}{r}{11.680} &
  \multicolumn{1}{r}{1.241} &
  \multicolumn{1}{r}{34,428,971} \\
\multicolumn{1}{l}{Monthly total earning, 2015 EUR} &
  \multicolumn{1}{r}{2,330} &
  \multicolumn{1}{r}{2,585} &
  \multicolumn{1}{r}{34,428,936} \\
\multicolumn{1}{l}{Firm quality index} &
  \multicolumn{1}{r}{0.365} &
  \multicolumn{1}{r}{0.771} &
  \multicolumn{1}{r}{15,699,075} \\
\multicolumn{1}{l}{Worker productivity index} &
  \multicolumn{1}{r}{0.122} &
  \multicolumn{1}{r}{0.990} &
  \multicolumn{1}{r}{16,162,338} \\
\bottomrule
\end{tabular}

    \end{threeparttable}
\end{table}

\newpage
\begin{table}[htbp!]
    \centering
    \caption{\textbf{AKM summary statistics and variance decomposition}\\\footnotesize The table reports summary statistics and variance decomposition following AKM estimations. The dependent variable in AKM estimation is log monthly earnings calculated as the ratio of total annual earnings by number of months worked. The sample includes employees aged between 20 and 60, with monthly earnings above 50\% of national median monthly income, working in firms with at least 5 employees and for at least four months in a calendar year. The estimations control for calendar year indicators, education level, cubic polynomial in age, as well as interactions of calendar year and age polynomial with education level. For further details, see Section \ref{sec:measurement}.}
    \label{tab:akm-sumstats-vardec}
    \begin{threeparttable}
        \begin{tabular}{lll}
\toprule
& \multicolumn{2}{c}{Dependent variable: log monthly earnings}  \\ \cmidrule(lr){2-3}
& 1987-2003 & 2004-2019 \\ \midrule
Standard deviation of outcome & 0.5003 & 0.4614 \\
N largest connected set & 16 862 428 & 15 435 023 \\
N singletons & 275 680 & 374 028 \\
N estimation sample & 16 586 748 & 15 060 995 \\
\multicolumn{3}{l}{\itshape Panel A: Summary of parameter estimates} \\
\hspace{1em}N worker FE & 1 881 715 & 1 842 564 \\
\hspace{1em}N firm FE & 126 605 & 50 430 \\
\hspace{1em}Std. dev. of worker FE & 0.2969 & 0.3208 \\
\hspace{1em}Std. dev. of firm FE & 0.1067 & 0.1027 \\
\hspace{1em}Std. dev. of Xb & 0.3416 & 0.2437 \\
\hspace{1em}Std. dev. of residual & 0.1587 & 0.1561 \\
\hspace{1em}Corr(worker FE, firm FE) & 0.1054 & 0.2496 \\
\hspace{1em}RMSE & 0.1693 & 0.1669 \\
\hspace{1em}Adjusted R2 & 0.8846 & 0.8681 \\
\multicolumn{3}{l}{\itshape Panel B: Share of outcome variance attributed to} \\
\hspace{1em}Worker FE & 0.3547 & 0.4868 \\
\hspace{1em}Firm FE & 0.0458 & 0.0499 \\
\hspace{1em}Cov(worker FE, firm FE) & 0.0269 & 0.0778 \\
\hspace{1em}Xb and associated covariances & 0.4712 & 0.2703 \\
\hspace{1em}Residual & 0.1014 & 0.1153 \\
\bottomrule
\end{tabular}

    \end{threeparttable}
\end{table}

\begin{landscape}
    \begin{table}[htbp!]
        \centering
        \caption{\textbf{Descriptive statistics by EA-PGI level}\\\footnotesize The table reports descriptive statistics among tertiary-educated individuals ($N =$ \num[round-precision=0]{\finalNindTer}) by deciles of EA-PGI. Columns (1)-(3) report sample means in 1st, 2nd-9th and 10th decile of EA-PGI, respectively. Column (4) reports difference in sample means and standard error of the difference in parentheses for the 2nd-9th decile relative to 1st decile, respectively. Column (5) reports difference in sample means and standard error of the difference in parentheses for the 10th decile relative to 1st decile, respectively. Column (6) reports total sample count for each variable considered. All estimates control for first ten genetic principal components.}
        \label{tab:cross-sumstats-pgiea-dec}
        \begin{threeparttable}
            \small\begin{tabular}{lllllllll}
\toprule
\multicolumn{1}{c}{} &
  \multicolumn{3}{c}{Sample means} &
  \multicolumn{4}{c}{Diff.} &
  \multicolumn{1}{r}{} \\\cmidrule(lr){2-4}\cmidrule(lr){5-8}
\multicolumn{1}{c}{} &
  \multicolumn{1}{c}{1st decile} &
  \multicolumn{1}{c}{2nd-9th deciles} &
  \multicolumn{1}{c}{10th decile} &
  \multicolumn{2}{c}{2nd-9th deciles} &
  \multicolumn{2}{c}{10th decile} &
  \multicolumn{1}{c}{} \\
\multicolumn{1}{c}{} &
  \multicolumn{1}{c}{(1)} &
  \multicolumn{1}{c}{(2)} &
  \multicolumn{1}{c}{(3)} &
  \multicolumn{2}{c}{(4)} &
  \multicolumn{2}{c}{(5)} &
  \multicolumn{1}{c}{(6)} \\
\midrule
\multicolumn{1}{l}{Male} &
  \multicolumn{1}{r}{0.267} &
  \multicolumn{1}{r}{0.341} &
  \multicolumn{1}{r}{0.402} &
  \multicolumn{1}{r}{0.074***} &
  \multicolumn{1}{r}{(0.011)} &
  \multicolumn{1}{r}{0.135***} &
  \multicolumn{1}{r}{(0.013)} &
  \multicolumn{1}{r}{32,364} \\
\multicolumn{1}{l}{Birth year} &
  \multicolumn{1}{r}{1977.2} &
  \multicolumn{1}{r}{1977.3} &
  \multicolumn{1}{r}{1977.6} &
  \multicolumn{1}{r}{0.170\phantom{***}} &
  \multicolumn{1}{r}{(0.231)} &
  \multicolumn{1}{r}{0.455\phantom{***}} &
  \multicolumn{1}{r}{(0.272)} &
  \multicolumn{1}{r}{32,364} \\
\multicolumn{1}{l}{Mother edu: compulsory} &
  \multicolumn{1}{r}{0.332} &
  \multicolumn{1}{r}{0.250} &
  \multicolumn{1}{r}{0.175} &
  \multicolumn{1}{r}{-0.082***} &
  \multicolumn{1}{r}{(0.010)} &
  \multicolumn{1}{r}{-0.157***} &
  \multicolumn{1}{r}{(0.012)} &
  \multicolumn{1}{r}{32,364} \\
\multicolumn{1}{l}{Mother edu: secondary} &
  \multicolumn{1}{r}{0.413} &
  \multicolumn{1}{r}{0.355} &
  \multicolumn{1}{r}{0.258} &
  \multicolumn{1}{r}{-0.059***} &
  \multicolumn{1}{r}{(0.011)} &
  \multicolumn{1}{r}{-0.155***} &
  \multicolumn{1}{r}{(0.013)} &
  \multicolumn{1}{r}{32,364} \\
\multicolumn{1}{l}{Mother edu: tertiary} &
  \multicolumn{1}{r}{0.247} &
  \multicolumn{1}{r}{0.386} &
  \multicolumn{1}{r}{0.554} &
  \multicolumn{1}{r}{0.139***} &
  \multicolumn{1}{r}{(0.011)} &
  \multicolumn{1}{r}{0.307***} &
  \multicolumn{1}{r}{(0.013)} &
  \multicolumn{1}{r}{32,364} \\
\multicolumn{1}{l}{Mother edu: missing} &
  \multicolumn{1}{r}{0.008} &
  \multicolumn{1}{r}{0.009} &
  \multicolumn{1}{r}{0.012} &
  \multicolumn{1}{r}{0.001\phantom{***}} &
  \multicolumn{1}{r}{(0.002)} &
  \multicolumn{1}{r}{0.005\phantom{***}} &
  \multicolumn{1}{r}{(0.003)} &
  \multicolumn{1}{r}{32,364} \\
\multicolumn{1}{l}{Father edu: compulsory} &
  \multicolumn{1}{r}{0.351} &
  \multicolumn{1}{r}{0.278} &
  \multicolumn{1}{r}{0.196} &
  \multicolumn{1}{r}{-0.073***} &
  \multicolumn{1}{r}{(0.010)} &
  \multicolumn{1}{r}{-0.155***} &
  \multicolumn{1}{r}{(0.012)} &
  \multicolumn{1}{r}{32,364} \\
\multicolumn{1}{l}{Father edu: secondary} &
  \multicolumn{1}{r}{0.388} &
  \multicolumn{1}{r}{0.314} &
  \multicolumn{1}{r}{0.218} &
  \multicolumn{1}{r}{-0.074***} &
  \multicolumn{1}{r}{(0.011)} &
  \multicolumn{1}{r}{-0.170***} &
  \multicolumn{1}{r}{(0.012)} &
  \multicolumn{1}{r}{32,364} \\
\multicolumn{1}{l}{Father edu: tertiary} &
  \multicolumn{1}{r}{0.220} &
  \multicolumn{1}{r}{0.366} &
  \multicolumn{1}{r}{0.551} &
  \multicolumn{1}{r}{0.147***} &
  \multicolumn{1}{r}{(0.011)} &
  \multicolumn{1}{r}{0.332***} &
  \multicolumn{1}{r}{(0.013)} &
  \multicolumn{1}{r}{32,364} \\
\multicolumn{1}{l}{Father edu: missing} &
  \multicolumn{1}{r}{0.042} &
  \multicolumn{1}{r}{0.042} &
  \multicolumn{1}{r}{0.035} &
  \multicolumn{1}{r}{0.001\phantom{***}} &
  \multicolumn{1}{r}{(0.005)} &
  \multicolumn{1}{r}{-0.007\phantom{***}} &
  \multicolumn{1}{r}{(0.005)} &
  \multicolumn{1}{r}{32,364} \\
\multicolumn{1}{l}{Age at graduation} &
  \multicolumn{1}{r}{24.506} &
  \multicolumn{1}{r}{24.854} &
  \multicolumn{1}{r}{25.245} &
  \multicolumn{1}{r}{0.349***} &
  \multicolumn{1}{r}{(0.054)} &
  \multicolumn{1}{r}{0.740***} &
  \multicolumn{1}{r}{(0.063)} &
  \multicolumn{1}{r}{32,364} \\
\multicolumn{1}{l}{Graduation: years since predicted graduation} &
  \multicolumn{1}{r}{2.256} &
  \multicolumn{1}{r}{2.211} &
  \multicolumn{1}{r}{2.126} &
  \multicolumn{1}{r}{-0.045\phantom{***}} &
  \multicolumn{1}{r}{(0.049)} &
  \multicolumn{1}{r}{-0.131*\phantom{**}} &
  \multicolumn{1}{r}{(0.058)} &
  \multicolumn{1}{r}{32,364} \\
\multicolumn{1}{l}{Age at first job} &
  \multicolumn{1}{r}{26.332} &
  \multicolumn{1}{r}{26.448} &
  \multicolumn{1}{r}{26.739} &
  \multicolumn{1}{r}{0.116\phantom{***}} &
  \multicolumn{1}{r}{(0.072)} &
  \multicolumn{1}{r}{0.407***} &
  \multicolumn{1}{r}{(0.085)} &
  \multicolumn{1}{r}{31,015} \\
\multicolumn{1}{l}{First job: years since predicted graduation} &
  \multicolumn{1}{r}{3.588} &
  \multicolumn{1}{r}{3.302} &
  \multicolumn{1}{r}{3.092} &
  \multicolumn{1}{r}{-0.286***} &
  \multicolumn{1}{r}{(0.071)} &
  \multicolumn{1}{r}{-0.496***} &
  \multicolumn{1}{r}{(0.084)} &
  \multicolumn{1}{r}{31,015} \\
\multicolumn{1}{l}{Annual average income at first job} &
  \multicolumn{1}{r}{11,729} &
  \multicolumn{1}{r}{12,811} &
  \multicolumn{1}{r}{14,208} &
  \multicolumn{1}{r}{1,082***} &
  \multicolumn{1}{r}{(242)} &
  \multicolumn{1}{r}{2,478***} &
  \multicolumn{1}{r}{(285)} &
  \multicolumn{1}{r}{32,364} \\
\multicolumn{1}{l}{AKM firm FE of first job} &
  \multicolumn{1}{r}{0.217} &
  \multicolumn{1}{r}{0.236} &
  \multicolumn{1}{r}{0.268} &
  \multicolumn{1}{r}{0.019\phantom{***}} &
  \multicolumn{1}{r}{(0.022)} &
  \multicolumn{1}{r}{0.052*\phantom{**}} &
  \multicolumn{1}{r}{(0.026)} &
  \multicolumn{1}{r}{16,860} \\
\multicolumn{1}{l}{AKM firm FE at t=15} &
  \multicolumn{1}{r}{0.407} &
  \multicolumn{1}{r}{0.462} &
  \multicolumn{1}{r}{0.527} &
  \multicolumn{1}{r}{0.055\phantom{***}} &
  \multicolumn{1}{r}{(0.035)} &
  \multicolumn{1}{r}{0.120**\phantom{*}} &
  \multicolumn{1}{r}{(0.043)} &
  \multicolumn{1}{r}{8,004} \\
\bottomrule
\end{tabular}

        \end{threeparttable}
    \end{table}
\end{landscape}

\newpage
\begin{table}[htbp!]
    \centering
    \caption{\textbf{Worker productivity and EA-PGI}\\\footnotesize The table reports estimation results from a regression of of worker productivity index on EA-PGI of interacted with education level. All regressions control for first ten genetic PCs, gender, year of birth, calendar year and biobank indicators. Column (1) reports the baseline results. Column (2) additionally controls for detailed field of education (3-digit) indicators. Column (3) adds interaction between level and detailed field of education. Column (4) adds indicators for academic institution identifiers. \SentenceSE}
    \label{tab:workerFE_pgiEA_reg}
    \begin{threeparttable}
        {
\def\sym#1{\ifmmode^{#1}\else\(^{#1}\)\fi}
\begin{tabular}{l*{4}{c}}
\toprule
                &\multicolumn{4}{c}{Dependent variable: Worker productivity index}          \\
                &\multicolumn{1}{c}{(1)}         &\multicolumn{1}{c}{(2)}         &\multicolumn{1}{c}{(3)}         &\multicolumn{1}{c}{(4)}         \\
\midrule
Secondary education $\times$ EA-PGI&    0.017\sym{*}  &    0.010         &    0.011         &    0.007         \\
                &  (0.007)         &  (0.007)         &  (0.007)         &  (0.007)         \\
Tertiary education $\times$ EA-PGI&    0.115\sym{***}&    0.086\sym{***}&    0.082\sym{***}&    0.016\sym{*}  \\
                &  (0.007)         &  (0.007)         &  (0.007)         &  (0.006)         \\
\midrule Level  &      Yes         &      Yes         &      Yes         &      Yes         \\
Field           &       No         &      Yes         &      Yes         &      Yes         \\
Level $\times$ Field&       No         &       No         &      Yes         &       No         \\
Institution ID  &       No         &       No         &       No         &      Yes         \\
\midrule
Obs.            &   31,866         &   31,709         &   31,702         &   31,574         \\
Avg. obs. per cell&   15,933         &      352         &      145         &       23         \\
Adj R2          &    0.198         &    0.251         &    0.256         &    0.330         \\
RMSE            &    0.836         &    0.808         &    0.806         &    0.765         \\
\bottomrule
\multicolumn{5}{l}{\footnotesize \sym{*} \(p<0.05\), \sym{**} \(p<0.01\), \sym{***} \(p<0.001\)}\\
\end{tabular}
}

    \end{threeparttable}
\end{table}



\begin{table}[htbp!]
    \centering
    \caption{\textbf{Years of education and EA-PGI, unconditional and conditional on parental EA-PGI among parent-offspring trios}\\\footnotesize The table reports estimation results from regression of predicted years of education (given highest qualification) on EA-PGI, unconditional and conditional on parental EA-PGI in parent-offspring trios. The top panel reports baseline estimations without controlling for parental EA-PGI. The bottom panel reports estimates conditional on parental EA-PGI. Column (1) uses full sample of parent-offspring trios ($N =$ \num[round-precision=0]{\finalNindTrioPld}), including the cases where parental genotypes were imputed from parent-offspring duos and siblings. Column (2) uses subset of directly genotyped parent-offspring trios ($N =$ \num[round-precision=0]{\finalNindTrioGtpPld}). All estimations control for first ten genetic principal components, gender, year of birth, and biobank indicators. \SentenceSE}
    \label{tab:reg-yedu-predicted-pgiea-byspec-pld}
    \begin{threeparttable}
        \begin{tabular}{lll}\toprule
& \multicolumn{2}{c}{Dependent var: predicted years of education} \\\cmidrule(lr){2-3}
& \multicolumn{1}{c}{All family trios} & \multicolumn{1}{c}{Directly genotyped trios} \\
& \multicolumn{1}{c}{(1)} & \multicolumn{1}{c}{(2)} \\\midrule
\multicolumn{3}{l}{Baseline without parental EA-PGI} \\
\hspace{1em}Own EA-PGI & 0.553*** & 0.570*** \\
& (0.016) & (0.027) \\
\hspace{1em}Constant & 14.691*** & 13.741*** \\
& (0.491) & (1.058) \\\cmidrule(lr){2-3}
\hspace{1em}Obs. & 12 871 & 4 586 \\
\multicolumn{3}{l}{Controlling for parental EA-PGI} \\
\hspace{1em}Own EA-PGI & 0.413*** & 0.441*** \\
& (0.026) & (0.040) \\
\hspace{1em}Mother EA-PGI & 0.128*** & 0.110*** \\
& (0.021) & (0.030) \\
\hspace{1em}Father EA-PGI & 0.093*** & 0.095*** \\
& (0.021) & (0.030) \\
\hspace{1em}Constant & 14.641*** & 13.717*** \\
& (0.482) & (1.028) \\\cmidrule(lr){2-3}
\hspace{1em}Obs. & 12 871 & 4 586 \\\bottomrule
\end{tabular}

    \end{threeparttable}
\end{table}

\begin{table}[htbp!]
    \centering
    \begin{tabular}{lllllll}
\toprule
& \multicolumn{6}{c}{Dependent variable: Cumulated income} \\\cmidrule(lr){2-7}
& \multicolumn{3}{c}{Baseline} & \multicolumn{3}{c}{Controlling for parental EA-PGI} \\\cmidrule(lr){2-4}\cmidrule(lr){5-7}
& \multicolumn{1}{c}{Pooled} & \multicolumn{1}{c}{Secondary} & \multicolumn{1}{c}{Tertiary} & \multicolumn{1}{c}{Pooled}  & \multicolumn{1}{c}{Secondary}  & \multicolumn{1}{c}{Tertiary} \\
& \multicolumn{1}{c}{(1)} & \multicolumn{1}{c}{(2)} & \multicolumn{1}{c}{(3)} & \multicolumn{1}{c}{(4)} & \multicolumn{1}{c}{(5)} & \multicolumn{1}{c}{(6)} \\\midrule
Own EA-PGI & 13 724 & -7 721 & 15 602 & 7 398 & -3 613 & 5 048 \\
& (  1 804) & (  2 132) & (  2 546) & (  2 746) & (  3 478) & (  3 833) \\
Mother EA-PGI &  &  &  & 1 769 & -5 002 & 4 658 \\
&  &  &  & (  2 163) & (  2 680) & (  3 016) \\
Father EA-PGI &  &  &  & 8 211 & -1 702 & 12 113 \\
&  &  &  & (  2 375) & (  2 795) & (  3 370) \\
Constant & 414 682 & 270 156 & 507 419 & 411 300 & 270 147 & 501 314 \\
& ( 61 295) & ( 58 545) & ( 84 159) & ( 61 302) & ( 57 824) & ( 84 234) \\\midrule
Obs. & 12 871 & 5 063 & 7 808 & 12 871 & 5 063 & 7 808 \\\bottomrule
\end{tabular}

    \caption{\textbf{Cumulated lifetime income, own and parental EA-PGI, regression results}\\\footnotesize The table reports estimation results from regression of cumulated lifetime income on own and parental EA-PGI. Columns (1)-(3) report the baseline estimates without controlling for parental EA-PGI. Columns (4)-(6) report estimates conditional on parental EA-PGI. Columns (1) and (4) use full sample of parent-offspring trios ($N =$ \num[round-precision=0]{\finalNindTrioPld}), columns (2) and (5) - subset of trios with secondary-educated offspring ($N =$ \num[round-precision=0]{\finalNindTrioSec}) and columns (3) and (6) - subset of trios with tertiary-educated offspring ($N =$ \num[round-precision=0]{\finalNindTrioTer}). All estimates control for first ten genetic principal components, gender, birth year, calendar year, and biobank indicators. \SentenceDPVMethod\SentenceSE}
    \label{tab:reg-dpvinc-pgiea-family}
\end{table}

\begin{table}[htbp!]
    \centering
    \resizebox{\linewidth}{!}{\begin{tabular}{lllllll}
\toprule
& \multicolumn{6}{c}{Dependent variable: Cumulated income} \\\cmidrule(lr){2-7}
& \multicolumn{3}{c}{All family trios} & \multicolumn{3}{c}{Directly genotyped trios} \\\cmidrule(lr){2-4}\cmidrule(lr){5-7}
& \multicolumn{1}{c}{Own EA-PGI} & \multicolumn{1}{c}{Mother EA-PGI} & \multicolumn{1}{c}{Father EA-PGI} & \multicolumn{1}{c}{Own EA-PGI}  & \multicolumn{1}{c}{Mother EA-PGI}  & \multicolumn{1}{c}{Father EA-PGI}  \\
& \multicolumn{1}{c}{(1)} & \multicolumn{1}{c}{(2)} & \multicolumn{1}{c}{(3)} & \multicolumn{1}{c}{(4)} & \multicolumn{1}{c}{(5)} & \multicolumn{1}{c}{(6)} \\\midrule
10th percentile & 339 181 & 339 879 & 330 699 & 350 648 & 342 498 & 336 264 \\
& (5 690) & (4 311) & (4 459) & ( 9 782) & ( 6 842) & ( 6 792) \\
90th percentile & 352 172 & 351 466 & 360 946 & 350 086 & 358 149 & 364 690 \\
& (5 198) & (4 500) & (5 126) & ( 8 547) & ( 6 833) & ( 7 444) \\
90-10 gap & 12 991 & 11 587 & 30 247 & -562 & 15 650 & 28 426 \\
& (9 866) & (7 502) & (8 414) & (16 426) & (10 933) & (11 639) \\\midrule
Obs. & 7 808 & 7 808 & 7 808 & 2 704 & 2 704 & 2 704 \\\bottomrule
\end{tabular}
}
    \caption{\textbf{Cumulated lifetime income by own and parental EA-PGI levels}\\\footnotesize The table reports adjusted lifetime income (up to \num[round-precision=0]{\maxTime} years since graduation) by EA-PGI percentiles, conditional on parental EA-PGI. Columns (1)-(3) use full sample of parent-offspring trios with tertiary-educated offspring ($N =$ \num[round-precision=0]{\finalNindTrioTer}). Columns (4)-(6) use subset of directly genotyped parent-offspring trios with tertiary-educated offspring ($N =$ \num[round-precision=0]{\finalNindTrioGtpTer}). Columns (1) and (4) report cumulated lifetime income at 10th and 90th percentiles of own EA-PGI, holding maternal and paternal EA-PGI constant. Columns (2) and (5) report cumulated lifetime income at 10th and 90th percentiles of maternal EA-PGI, holding own and paternal EA-PGI constant.  Columns (3) and (6) report cumulated lifetime income at 10th and 90th percentiles of paternal EA-PGI, holding own and maternal EA-PGI constant. \SentenceDPVRegFamp\SentenceDPVMethod\SentenceSE}
    \label{tab:predmarg-dpvinc-pgiea-family}
\end{table}

\begin{table}[htbp!]
    \centering
    \begin{tabular}{lrrrr}
\toprule
& \multicolumn{4}{c}{Dependent variable: Cumulated income} \\\cmidrule(lr){2-5}
& Baseline & \multicolumn{3}{c}{Controlling for parental EA-PGI}  \\\cmidrule(lr){2-2}\cmidrule(lr){3-5}
& Own EA-PGI & Own EA-PGI  & Mother EA-PGI & Father EA-PGI \\
 & \multicolumn{1}{c}{(1)} & \multicolumn{1}{c}{(2)} & \multicolumn{1}{c}{(3)} & \multicolumn{1}{c}{(4)} \\ \midrule
\multicolumn{5}{l}{\textit{EA-PGI percentiles}} \\
\hspace{1em}10th & 265 492 & 259 961 & 261 546 & 257 400 \\
& (2 782) & (4 621) & (3 938) & (3 954) \\
\hspace{1em}20th & 262 054 & 258 352 & 259 384 & 256 677 \\
& (2 229) & (3 323) & (3 036) & (3 029) \\
\hspace{1em}30th & 259 582 & 257 195 & 257 866 & 256 156 \\
& (2 022) & (2 568) & (2 532) & (2 503) \\
\hspace{1em}40th & 257 496 & 256 219 & 256 508 & 255 710 \\
& (2 017) & (2 184) & (2 240) & (2 223) \\
\hspace{1em}50th & 255 614 & 255 338 & 255 320 & 255 311 \\
& (2 150) & (2 154) & (2 161) & (2 159) \\
\hspace{1em}60th & 253 639 & 254 413 & 254 078 & 254 906 \\
& (2 405) & (2 459) & (2 274) & (2 294) \\
\hspace{1em}70th & 251 508 & 253 416 & 252 817 & 254 471 \\
& (2 775) & (3 057) & (2 566) & (2 624) \\
\hspace{1em}80th & 248 921 & 252 206 & 251 177 & 253 929 \\
& (3 311) & (3 978) & (3 130) & (3 216) \\
\hspace{1em}90th & 245 621 & 250 661 & 249 103 & 253 149 \\
& (4 074) & (5 294) & (4 008) & (4 253) \\
\midrule Obs. & 5 063 & 5 063 & 5 063 & 5 063 \\\bottomrule
\end{tabular}

    \caption{\textbf{Cumulated lifetime income of secondary-educated individuals by EA-PGI level, conditional on parental EA-PGI among parent-offspring trios}\\\footnotesize The table reports adjusted lifetime income (up to \num[round-precision=0]{\maxTime} years since graduation) by EA-PGI percentiles. The estimation sample is restricted to parent-offspring trios with secondary-educated offspring ($N =$ \num[round-precision=0]{\predMargDPVFamNSec}). Column (1) reports the estimates in the baseline specification without controlling for parental EA-PGI. Columns (2)-(4) report average cumulated lifetime income by own, maternal and paternal EA-PGI percentiles, respectively, conditional on parental EA-PGI. \SentenceDPVReg\SentenceDPVMethod\SentenceSE}
    \label{tab:predmarg-dpvinc-pgiea-fam-sec}
\end{table}

\begin{table}[htbp!]
    \centering
    \resizebox{\linewidth}{!}{\begin{tabular}{lllllll}
\toprule
& \multicolumn{6}{c}{Dependent variable: Cumulated income} \\\cmidrule(lr){2-7}
& \multicolumn{2}{c}{Pooled} & \multicolumn{2}{c}{Secondary} & \multicolumn{2}{c}{Tertiary} \\\cmidrule(lr){2-3}\cmidrule(lr){4-5}\cmidrule(lr){6-7}
& \multicolumn{1}{c}{Unweighted} & \multicolumn{1}{c}{Weighted} & \multicolumn{1}{c}{Unweighted}  & \multicolumn{1}{c}{Weighted}  & \multicolumn{1}{c}{Unweighted}   & \multicolumn{1}{c}{Weighted}   \\
& \multicolumn{1}{c}{(1)} & \multicolumn{1}{c}{(2)} & \multicolumn{1}{c}{(3)}  & \multicolumn{1}{c}{(4)}  & \multicolumn{1}{c}{(5)}   & \multicolumn{1}{c}{(6)}   \\
\midrule
\multicolumn{7}{l}{\textbf{Panel A: Genotyped sample}} \\
\hspace{1em}10th percentile & 309 659 & 291 728 & 262 386 & 257 996 & 346 194 & 331 362 \\
& (1 306) & (1 286) & (1 429) & (1 501) & (1 944) & (1 920) \\
\hspace{1em}50th percentile & 329 893 & 308 756 & 255 422 & 249 549 & 368 728 & 350 947 \\
& (  857) & (  832) & (1 116) & (1 157) & (1 137) & (1 105) \\
\hspace{1em}90th percentile & 350 418 & 325 930 & 248 358 & 241 029 & 391 585 & 370 700 \\
& (1 591) & (1 525) & (2 120) & (2 185) & (2 006) & (1 938) \\\cmidrule(lr){2-7}
\hspace{1em}Obs. & 51 056 & 51 056 & 18 692 & 18 692 & 32 364 & 32 364 \\
\multicolumn{7}{l}{\textbf{Panel B: Family trio sample (conditional on parental EA-PGI)}} \\
\hspace{1em}10th percentile & 303 725 & 304 546 & 259 961 & 263 787 & 339 181 & 345 745 \\
& (3 906) & (4 417) & (4 621) & (5 261) & (5 690) & (6 559) \\
\hspace{1em}50th percentile & 313 190 & 313 886 & 255 338 & 258 422 & 345 639 & 351 212 \\
& (1 697) & (1 981) & (2 154) & (2 388) & (2 320) & (2 755) \\
\hspace{1em}90th percentile & 322 764 & 323 347 & 250 661 & 252 986 & 352 172 & 356 750 \\
& (3 934) & (4 502) & (5 294) & (5 717) & (5 198) & (6 114) \\\cmidrule(lr){2-7}
\hspace{1em}Obs. & 12 871 & 12 871 & 5 063 & 5 063 & 7 808 & 7 808 \\\bottomrule
\end{tabular}
}
    \caption{\textbf{Cumulated lifetime income by own EA-PGI level, education, genotype sample and weighting scheme}\\\footnotesize The table reports adjusted lifetime income (up to \num[round-precision=0]{\maxTime} years since graduation) by own EA-PGI percentiles. Panel A uses full analysis sample ($N =$ \num[round-precision=0]{\finalNindAll} in pooled, $N =$ \num[round-precision=0]{\finalNindSec} in secondary-educated and $N =$ \num[round-precision=0]{\finalNindTer} in tertiary-educated subsamples), while Panel B uses sample of parent-offspring trios ($N =$ \num[round-precision=0]{\finalNindTrioPld} in pooled, $N =$ \num[round-precision=0]{\finalNindTrioSec} in secondary-educated and $N =$ \num[round-precision=0]{\finalNindTrioTer} in tertiary-educated subsamples). Columns (1), (3) and (5) report unweighted estimates. Columns (2), (4) and (6) report weighted estimates using inverse probability weights. \SentenceDPVReg Panel B additionally controls for maternal and paternal EA-PGI. \SentenceDPVMethod Inverse probability weights are estimated to balance the sample according to year of birth and graduation year (fully interacted with highest education level and gender), and rural area indicator fully interacted with gender. \SentenceSE}
    \label{tab:predmarg-dpvinc-pgiea-weighted}
\end{table}

\begin{table}[htbp!]
    \centering
    {
\def\sym#1{\ifmmode^{#1}\else\(^{#1}\)\fi}
\begin{tabular}{l*{2}{c}}
\toprule
            &\multicolumn{2}{c}{Dependent variable: predicted years of education}\\
            &\multicolumn{1}{c}{(1)}         &\multicolumn{1}{c}{(2)}         \\
\midrule
EA-PGI      &                     &       0.530\sym{***}\\
            &                     &     (0.008)         \\
Constant    &      14.984\sym{***}&      14.979\sym{***}\\
            &     (0.202)         &     (0.194)         \\
\midrule
Obs.        &      51,056         &      51,056         \\
$ R^2$      &       0.086         &       0.158         \\
Adj. $ R^2$ &       0.085         &       0.157         \\
Incremental $ R^2$&                     &       0.071         \\
\bottomrule
\multicolumn{3}{l}{\footnotesize \sym{*} \(p<0.05\), \sym{**} \(p<0.01\), \sym{***} \(p<0.001\)}\\
\end{tabular}
}

    \caption{\textbf{Predictive power of EA-PGI for years of education}\\\footnotesize The table reports estimation results from regression of predicted years of education (given highest qualification) on EA-PGI. Column (1) reports estimates without EA-PGI, and column (2) - with EA-PGI. All estimations controls for gender fully interacted with year of birth indicators, biobank indicator and first ten genetic principal components. \SentenceSE}
    \label{tab:predicted-ea-pgi-increm-r2}
\end{table}






\end{document}